\documentclass[%
 reprint,
 amsmath,amssymb,
 aps,
superscriptaddress
]{revtex4-1}

\usepackage{graphicx}% Include figure files
\usepackage{dcolumn}% Align table columns on decimal point
\usepackage{bm}% bold math
\usepackage{bbding}
\usepackage{braket}
\usepackage{physics}
\usepackage{pifont}
\usepackage[colorlinks,
            linkcolor=blue,       
            anchorcolor=blue,  %%修改此处为你想要的颜色
            citecolor=blue,
            urlcolor=blue,
            ]{hyperref}
\allowdisplaybreaks[4]

\begin{document}

\title{Moir\'{e} optical phonons dancing with heavy electrons in magic-angle twisted bilayer graphene}
\author{Hao Shi}
\affiliation{Department of Physics, The Hong Kong University of Science and Technology, Clear Water Bay, Hong Kong, China}
\author{Wangqian Miao}
\affiliation{Materials Department, University of California, Santa Barbara, California 93106-5050, USA}

\author{Xi Dai}
\email{daix@ust.hk}
\affiliation{Department of Physics, The Hong Kong University of Science and Technology, Clear Water Bay, Hong Kong, China}

\date{\today}
\begin{abstract}
Electron-phonon coupling in magic-angle twisted bilayer graphene is an important but difficult topic. We propose a scheme to simplify and understand this problem. Weighted by the coupling strength with the low-energy heavy electrons ($f$ orbitals), several moir\'{e} optical phonons are singled out which strongly couple to the flat bands. These modes have localized envelopes in the moir\'{e} scale, while in the atomic scale they inherit the monolayer oscillations like the Kekul\'{e} pattern. They flip the flavor of $f$ orbitals, helping stabilize some symmetry-breaking orders. Such electron-phonon couplings are incorporated into an effective extended Holstein model, where both phonons and electrons are written as moir\'{e} scale basis. We hope this model will inspire some insights guiding further studies about the superconductivity and other correlated effects in this system.
\end{abstract}

\maketitle

%\tableofcontents

\section{\label{sec:level1}Introduction}

Magic angle twisted bilayer graphene (MATBG) hosts eight flat bands near the charge neutrality point (CNP) and exhibits captivating properties in transport and optical experiments \cite{18nature_cao_SC, 18nature_cao_CI, 19nature_Xie_SC_STM, 19nature_jiang_charge_order, 19nature_Lu_SC, 19science_Yankowitz_SC, 20science_Serline_QAH, 20n_cascades,21_nm_Wu2021,21s_nematicity_sc,22np_hofstadter_exp}. Numerous studies have been dedicated to understanding the ground states and the associated symmetry-breaking mechanisms \cite{18prx_IVC_Mott_sc, 19prb_ZLL_jianpeng, 19prl_filling2_from_wannier, 19prl_ferromagnetic_Mott_state, 19prx_JT_TB_Fabrizio, 20prl_HF_by_MacDonald, 20prb_HF_by_Fuchun, 20prx_KIVC_ashvin, 20prl_QAH_from_hBN, 20prb_HF_by_Guinea, 21prx_liao_ivc, 21prb_HF_by_XiDai, 21prr_nematic_order, 21prl_strain_induced_transition, 21prx_kekule_order_kwan, 22prl_kekule_phase_wagner, 22prl_optical_response, 22prl_spectroscopy_theory_bab,22prl_stm_theory_mpz,22prb_hofstadter_xiaoyu,22prx_MC_KIVC_SM,23prb_qmc_xu,23prr_cluster_theory,23prb_dmrg_filling3}. The recent topological heavy fermion representation \cite{22prl_heavy_fermion_Zhida,22prb_ZLL_OPW_representation,23ltp_THF2,23prb_THF_for_TTG} accelerates the study of strongly correlated physics in moir\'{e} graphene systems \cite{23prl_kondo_sarma1,23a_kondo_song,23prl_kondo_haoyu1,23prl_kondo_haoyu2,21prl_cascades_kang,23nc_cascades_dmft,23a_kondo_coleman,23a_mixed_valence_yantao,23a_transition_ziyang,23a_hofstadter_thf,23prb_kondo_sarma2,23a_dmft_bernevig,23prb_kondo_shankar,23a_tunable_kondo,23_liu_ek_phonon,21prl_kondo_TTG,23prb_Dirac_kondo_effect,23a_vacancy_kondo,24a_TTG_exp,24a_nodal_sc_onsite}.

As experimental techniques continue to develop, more hidden stories of this system are gradually revealed. Recent high-resolution STM experiments demonstrate that the atomic-scale Kekul\'{e} pattern is a salient feature of the correlated states \cite{23_nuckolls_quantum_texture,23a_IKS_kim}, suggesting an intervalley coherent nature. At filling $\nu = \pm 2$, the ground state can be attributed to either the time-reversal symmetric intervalley coherent order (TIVC) \cite{19prl_filling2_from_wannier, 19prl_ferromagnetic_Mott_state, 21prx_liao_ivc, 22prl_spectroscopy_theory_bab,18prb_kekule_xiaoyan} or the incommensurate Kekul\'{e} spiral order (IKS) \cite{21prx_kekule_order_kwan,22prl_kekule_phase_wagner,23a_EPC_vs_IKS_kwan,23a_kekule_TTG}, depending on the strength of strain. Such Kekul\'{e} pattern could easily arise from graphene's inherent vibrational modes, an aspect that has been largely neglected until recently. Two additional experiments, one using the nano-Raman technique \cite{21n_local_phonon_exp} and the other using the $\mu$-ARPES technique \cite{23_chen_replica_band}, provide some hints that optical phonons may play important roles in MATBG. In the nano-Raman experiment, two strong optical signals appear at 1584 cm$^{-1}$ (196 meV) and 2640 cm$^{-1}$ (2 $\times$ 164 meV), corresponding to an in-plane longitudinal optical (iLO) phonon process ($G$ band) and an in-plane transverse optical (iTO) double phonon process ($G'$ band) \cite{05pr_dresselhaus_raman,08prb_graphene_symmetry,17_book_graphene_phonon,23prl_phonon_exp}. The phonons carry moir\'{e} structures so that the vibrations are strongest at the AA-stacking area, indicating the enhanced electron-phonon coupling (EPC) there. In the $\mu$-ARPES experiment, phonon-induced replica bands with equal energy spacing ($150\pm 15$ meV) are observed in superconducting samples, while no such phenomena appear in non-superconducting samples with aligned hBN substrates. All these signs indicate that EPC is quite important, not only limited to the discussion of superconductivity \cite{18prl_SC_from_phonon, 18prb_SC_meanfield, 19prl_lian_epc_sc, 19prb_wu_epc_sc, 21pnas_cea_coulomb, 21prl_epc_flat_band_choi,23prb_pair_critical_field,23prb_preformed_pair,23_liu_ek_phonon,23a_off_diagonal_sc,24a_nodal_sc_onsite}.

However, a strict and thorough theoretical treatment of moir\'{e} phonons is notoriously challenging \cite{19prx_JT_TB_Fabrizio,19prb_phason_epc,202Dm_soliton_phonon_TBG,22nl_phonon_by_Jianpeng,22prb_csc_model,23prb_phonon_continuum_tb,23prb_continuum_phonon_Girotto,23prb_optical_flat_phonon,17prb_phason}. In MATBG there are more than $30,000$ moir\'{e} phonon branches. Even if we diagonalize all phonons accurately, dealing with the subsequent EPC can again lead to complex data processing, obscuring a clear physical understanding. The physical picture is deeply hidden by the complexity of both electron and phonon structures. Thus, a more comprehensive model is needed.

In this study we attempt to simplify the EPC model in MATBG by constructing the optical moir\'{e} phonons that strongly couple to the flat bands. The basic idea is the following. On the one hand, the folded phonons originating from the monolayer can be considered as a suitable basis set to start with. Previous studies have shown that the phonon density of states (DOS) is nearly independent of the stacking-style or twist angle \cite{13prb_phonon_in_TBG,18prb_epc_tbg_choi,19prx_JT_TB_Fabrizio}. The phonon DOS features two peaks that coincide with those of the Eliashberg spectrum \cite{18prb_epc_tbg_choi,21prl_epc_flat_band_choi}, which suggests that the interlayer force field has a minimal impact on high-frequency lattice dynamics. Focusing on the phonons folded from the atomic $K(K')$ and $\Gamma$ valleys (the two DOS peaks), we still have many [$O(10^{2})$] phonons. These phonon mini-bands couple with the electron flat bands with very different coupling strength \cite{19prb_electron_filter}. Only a few of them strongly couple with flat bands and the rest of them couple very weakly (at least one order smaller). On the other hand, the heavy-fermion picture of MATBG \cite{22prl_heavy_fermion_Zhida,22prb_ZLL_OPW_representation} suggests that the localized $f$ orbitals centered at AA stacking-area play the significant role for the formation of the flat bands. In heavy-fermion theories, the $f$ orbitals also dominate the strongly correlated aspects of the system. As a much simpler indicator of the flat bands, the $f$ orbitals have been applied to study the Coulomb interaction \cite{19prb_ZLL_jianpeng,22prb_ZLL_OPW_representation}, disorder \cite{22prb_disorder_koshino}, and impurity \cite{23c_impurity} effects. As such, we could instead focus on the couplings with $f$ orbitals. Several moir\'{e} phonons can then be constructed in real-space from folded modes, weighted by their coupling strength with $f$ orbitals. These manually constructed modes will by construction strongly couple with the flat bands.

Symmetry analysis shows that in the $K(K')$ and $\Gamma$ valleys only 9 moir\'{e} phonons will strongly couple to the flat bands, which are listed in Table \ref{tab:epc_const_lam} and Fig. \ref{fig:All_distortion}. In the present study, we identify them numerically using two different methods. The first one starts from an approximated layer-decoupled continuum EPC model \cite{08prb_graphene_symmetry,18prl_SC_from_phonon} and the projection of phonons is straightforward. The other method uses a more realistic tight-binding (TB) model \cite{12prb_Koshino_Moon,23prb_tapw,23prb_tapw_shengjun}, based on which a frozen-phonon scheme is specially developed to effectively determine the EPC matrix for moir\'{e} systems. The resulting moir\'{e} phonons have typical features in real space. In moir\'{e} scale, their distortion fields distribute mainly near the AA-stacking area, although the shapes are distinct depending on their specific symmetries. In atomic scale, they inherit the monolayer vibrations such as the Kekul\'{e} pattern. Moreover, two dominant modes with $A_1$ and $B_1$ symmetries are found qualitatively consistent with those proposed in Refs. \cite{19prx_JT_TB_Fabrizio,20epjp_JT_BM_Fabrizio,22prb_Kekule_Fabrizio}. 

With only 9 local modes involved, a clear picture of real-space EPC can be established, which helps gain a deeper understanding of the underlying physics at play. Their unique patterns near the AA-stacking domain suggest that they might be responsible for the strong vibrational and IVC signals in experiments. To investigate the impact of them on the ground states, we perform the mean-field calculation at fillings $\nu=0,2$. The phonon-mediated attractive channel favors these phonon-stabilized orders but the stabilization energy is really sensitive to the EPC strength. With realistic parameters, these states become competitive with Kramers-IVC states and the flavor-polarized states. To have a more complete model for future studies, we finally express it as an extended moir\'{e}-scale Holstein lattice model, using the 8-band model proposed in Ref. \cite{19prr_8orbital_model}.

The rest of the paper is organized as follows. Section \ref{Sec2} shows the projection theory of moir\'{e} phonons. In Section \ref{Sec3} we present the phonon structures and outline the frozen-phonon method to determine them. In Section \ref{Sec4} we incorporate the phonon-mediated attraction into a mean-field study. In Section \ref{section:lattice_epc}, the effective lattice EPC model is given. We summarize and discuss their implications in Section \ref{Sec5}. The main text is followed by a series of appendices, where readers can find additional details and complementary information about this work.

\section{Projection theory of moir\'{e} optical phonons \label{Sec2}}

\begin{figure}
\includegraphics[width=0.48\textwidth]{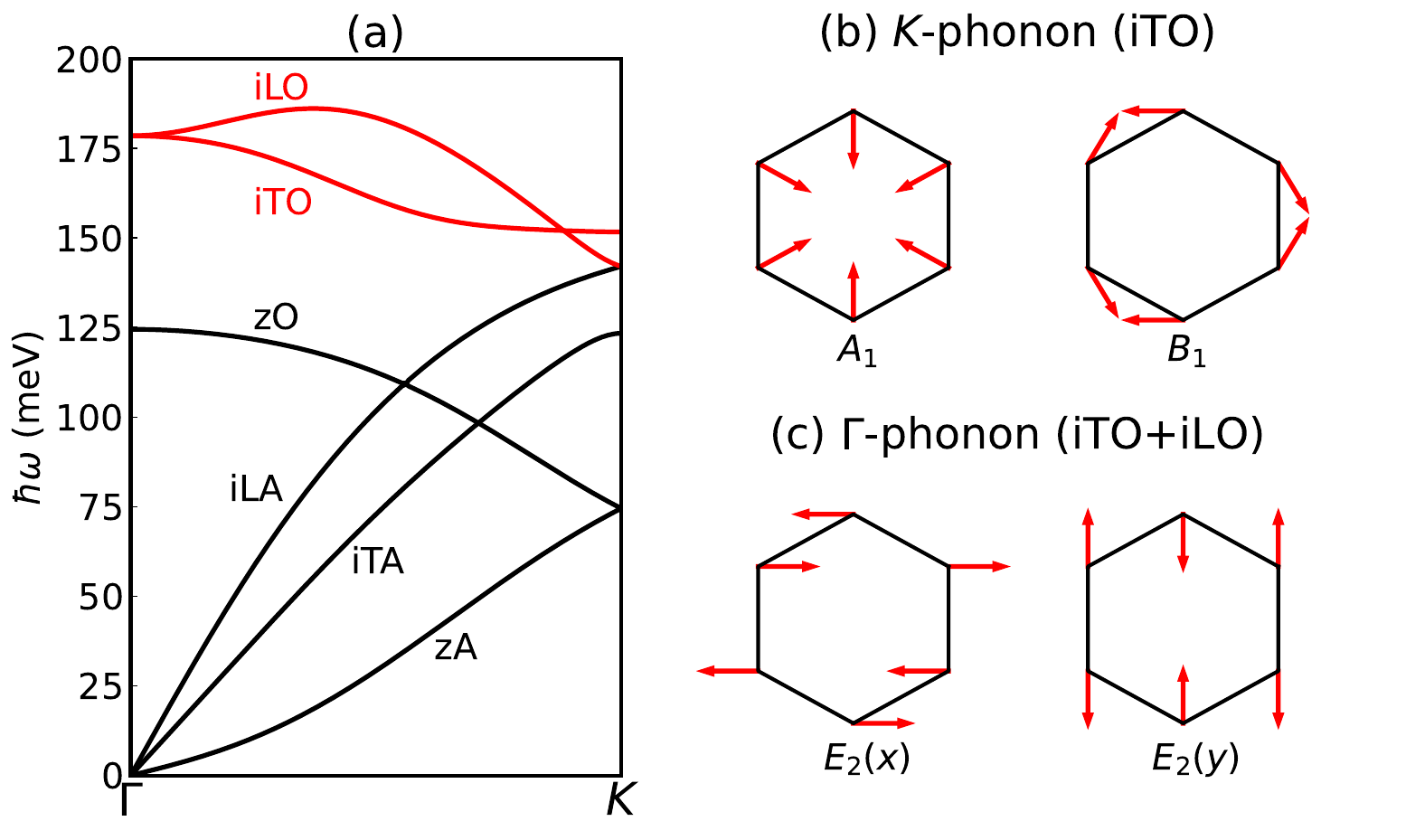}
\caption{\label{fig:monolayer_phonon}(a) The phonon bands of graphene. The iTO and iLO branches (red) will be projected into moir\'{e} phonons. (b) and (c) show the atomic displacement patterns of $K$-phonons (iTO near $\bm{K}$ and $\bm{K}'$ with $\hbar\omega_K\approx 150$ meV) and $\Gamma$-phonons (iLO and iTO near $\bm{\Gamma}$ with $\hbar\omega_{\Gamma}\approx 180$ meV).}
\end{figure}

\subsection{Electronic structure of MATBG}
As pointed out in Refs. \cite{22prl_heavy_fermion_Zhida,22prb_ZLL_OPW_representation}, the single-particle Hamiltonian of MATBG is a hybridized $f+c$ model 
\begin{align}
H_0 = \sum_{\bar{\bm{k}}}(f_{\bar{\bm{k}}}^{\dagger},c_{\bar{\bm{k}}}^{\dagger})\left(\begin{matrix}\mathcal{H}_{f}(\bar{\bm{k}})&\mathcal{H}_{fc}(\bar{\bm{k}})\\ \mathcal{H}^{\dagger}_{fc}(\bar{\bm{k}}) & \mathcal{H}_c(\bar{\bm{k}})
\end{matrix}\right)\left(\begin{matrix}f_{\bar{\bm{k}}}\\c_{\bar{\bm{k}}}\end{matrix}\right),
\label{eq:H0}
\end{align}
where $\bar{\bm{k}}$ resides within the moir\'{e} Brillouin zone (mBZ). The kernels $\mathcal{H}_f$, $\mathcal{H}_c$, and $\mathcal{H}_{fc}$ are diagonal with respect to spin ($s=\uparrow,\downarrow$) and valley ($\eta=K,K'$) indexes.

The $f$ electrons are $p_{\pm}$ orbitals localized around AA-stacking sites $\bm{R}$ with the wave function $\Phi_{s\eta t}({\bm{r}-\bm{R}})$, where $t=\pm1$ denotes the angular momentum (eigenvalue of $C_{3z}$). The $f$ fermion in Eq. (\ref{eq:H0}) is the Bloch sum of local orbitals (suppose we have $N_m$ supercells)
\begin{align}
\begin{split}
f_{\bar{\bm{k}}}=&\frac{1}{\sqrt{N_m}}\sum_{\bm{R}}e^{-i\bar{\bm{k}}\cdot\bm{R}}f_{\bm{R}},\\
f_{\bm{R}}=&\left(f_{\uparrow K +,\bm{R}}, f_{\uparrow K -,\bm{R}},f_{\uparrow K'+,\bm{R}},..., f_{\downarrow K'-,\bm{R}} \right)^T.
\end{split}
\label{eq:f_orbital}
\end{align}
These orbitals are obtained in Refs. \cite{22prl_heavy_fermion_Zhida,19prr_8orbital_model} as Wannier orbitals using active bands. In Ref. \cite{22prb_ZLL_OPW_representation} they have been identified as the pseudo zeroth Landau levels trapped by moir\'{e} potential. In the present study we obtain them using the Wannierization proposed in Ref. \cite{19prr_8orbital_model}, starting from a TB model \cite{12prb_Koshino_Moon,23prb_tapw}, see Appendices \ref{Appendix_TAPW}, \ref{Appendix_BM_ZLL}. The overlap between the $f$ orbitals and the flat bands are optimized to be $94.34\%$. %which is relatively high but can hardly be increased further because near $\bar{\bm{\Gamma}}$ the flat bands are occupied by $c$ orbitals. 

\begin{table*}[!htb]
\caption{\label{tab:table1_symmetry}%
Irreducible representations (irrep) of the group $D_6$ ($\phi=2\pi/3$), and the classification of $16$ spinless generators $\tau_i\sigma_j$ ($\tau$: valley, $\sigma$: angular momentum) of $f$ orbitals according to the real irreps. The generator $\tau_i\sigma_j$ acted by a generic symmetry operator $g$ will transform to $g(\tau_i\sigma_j)=D^{f}(g)\tau_i\sigma_j[D^f(g)]^{-1}$. The symbol \ding{55} labels generators forbidden by time reversal $\mathcal{T}=\tau_x\sigma_x\mathcal{K}$.}
\begin{ruledtabular}
\begin{tabular}{cccccc}
Irrep  & $C_{3z} (e^{i\phi\sigma_z}) $  &  $C_{2x}(\sigma_x)$  & $C_{2z} (\tau_x)$ & intervalley ($K\&K'$)     & intravalley ($\Gamma$) \\
\hline
$A_1$       &$+1$      &   $+1$      &   $+1$     & $\tau_x$                       & $\tau_0\sigma_0$\\
$B_1$       &$+1$      &   $+1$      &   $-1$     & $\tau_y$                       & $\tau_z $  (\ding{55})\\
$A_2$       &$+1$      &   $-1$      &   $+1$     & $\tau_x\sigma_z$ (\ding{55})   & $\sigma_z$ (\ding{55})\\
$B_2$       &$+1$      &   $-1$      &   $-1$     & $\tau_y\sigma_z$ (\ding{55})   & $\tau_z\sigma_z$ \\
$E_1$       & 
$ \begin{pmatrix}  \cos\phi & -\sin\phi \\\sin\phi & \cos\phi \end{pmatrix}$
& $ \begin{pmatrix}  1 &0  \\ 0 &-1 \end{pmatrix} $ & $
\begin{pmatrix}  -1 &0  \\ 0  &-1 \end{pmatrix}
$ & $\tau_y\sigma_x, \tau_y\sigma_y$  & $\tau_z\sigma_x, \tau_z \sigma_y$ (\ding{55}) \\ $E_2$ & $
\begin{pmatrix}  \cos\phi &-\sin\phi \\ \sin\phi&\cos\phi \end{pmatrix}$ 
& $ \begin{pmatrix} 1 &0  \\ 0 &-1 \end{pmatrix}$ & $
\begin{pmatrix} 1 &0  \\ 0 &1
\end{pmatrix}$ & $\tau_x\sigma_x, \tau_x\sigma_y$ & $\sigma_x,\sigma_y$ \\
\end{tabular}
\end{ruledtabular}
\end{table*}

The $c$ orbitals shuttle itinerantly in the system, bringing topology to the flat bands near the mBZ center through hybridization with $f$ orbitals \cite{19prb_10band_model,19prr_8orbital_model,19prr_10band_dft,22prl_heavy_fermion_Zhida}. In Eq. (\ref{eq:H0}) the $k$-space $c$ operator is
\begin{align}
c_{\bar{\bm{k}}}=\left( c_{\uparrow K 1\bar{\bm{k}}}, c_{\uparrow K 2\bar{\bm{k}}},...,c_{\uparrow K N_c\bar{\bm{k}}},..., c_{\downarrow K' N_c\bar{\bm{k}}} \right)^T.
\end{align}
Here $c_{s\eta a\bar{\bm{k}}}$ annihilates the $c$ electron with momentum $\bar{\bm{k}}$, spin $s$, valley $\eta$, and orbital index $a=1,2,..,N_c$, $N_c$ is the number of $c$ orbitals within the cutoff. Previous models treat them differently \cite{22prl_heavy_fermion_Zhida,22prb_ZLL_OPW_representation,19prb_10band_model,19prr_8orbital_model,20prb_8orbital_interaction,19prr_10band_dft}. %In Refs. \cite{22prl_heavy_fermion_Zhida,22prb_ZLL_OPW_representation} the $c$ electrons resemble plane waves and $N_c\sim O(10^2)$, while in earlier models \cite{19prb_10band_model,19prr_8orbital_model,20prb_8orbital_interaction,19prr_10band_dft} they are also fitted as Wannier orbitals and $N_c\leq 8$. In the following projection and mean-field study, we adopt them as orthogonalized plane waves as in Ref. \cite{22prb_ZLL_OPW_representation}. But finally we will write a complete lattice EPC model using the 8-band model \cite{19prr_8orbital_model} once the moir\'{e} phonons have been determined.

\subsection{Constructing moir\'{e} optical phonons by projection}
Our starting point is the phonons directly folded from the monolayer ones. Let us first focus on the $K$-phonons folded by iTO modes near the Dirac points $\bm{K}_l^{\eta}$. The bosonic operator $a_{l\eta\bm{q}}$ ($a_{l\eta\bm{q}}^{\dagger}$) is used to annihilate (create) the phonon with momentum $\bm{q}+\bm{K}_l^{\eta}$ in the layer $l$ and valley $\eta$. The corresponding polarization field (phonon wave function) has the form of ``plane waves'',
\begin{align}
\bm{u}_{l\eta\bm{q}}(\bm{r})=\frac{1}{\sqrt{N_m N_a}}\sum_{\bm{R}_l \alpha}e^{i(\bm{q}+\bm{K}_l^{\eta})\cdot \bm{r}}\delta_{\bm{r},\bm{R}_l+\bm{\tau}_{l\alpha}}\bm{\epsilon}^{\eta}_{l\alpha}(\bm{q}), \label{eq:u_field}
\end{align}
where $\bm{R}_l$ denotes the lattice vector of the $l$-th monolayer, $\bm{\epsilon}^{\eta}_{l}$ is the displacement eigenvector [see Fig. \ref{fig:monolayer_phonon}(b) and Appendix \ref{appendix_phonon}] telling which direction the sublattice $\alpha=A,B$ (located at $\bm{\tau}_{l\alpha}$) oscillates, $N_a=11164/4$ is the number of atomic cells in each layer of the supercell. Assuming the Einstein approximation $\omega_{K}\approx 150$ meV, the Hilbert space spanned by Eq. (\ref{eq:u_field}) is highly degenerate, allowing for arbitrary superposition of these modes.

The $K$-phonons induce the EPC Hamiltonian
\begin{align}
H_{\text{epc}}=\frac{1}{\sqrt{2N_m}}\sum_{l\eta\bm{q}}h_{l\eta}(\bm{q})a_{l\eta\bm{q}}+\text{H.c.}, \label{eq:H_epc_full}
\end{align}
where $h_{l\eta}(\bm{q})$ describes the electron scattering associated with the distortion $\bm{u}_{l\eta\bm{q}}$. Using the $f+c$ basis, and we pick out the onsite scatterings among $f$ orbitals,
\begin{align}
H^f_{\text{epc}} =& \frac{1}{\sqrt{2N_m}}\sum_{l\eta\bm{q}}\sum_{\bm{R}}f^{\dagger}_{\bm{R}}\Gamma^{l\eta}_{\bm{q}}(\bm{R})f_{\bm{R}}a_{l\eta\bm{q}} + \text{H.c.},
\label{eq:H_epc}
\end{align}
which should be the dominant EPC term governing the splitting of flat bands. We can decompose 
\begin{align}
\bm{q}+\bm{K}_l^{\eta}=\bar{\bm{q}}+\bm{Q}_l^{\eta}+\bm{K}_l^{\eta} \label{eq:aBZ_to_mBZ}
\end{align}
so that $\bar{\bm{q}}\in$mBZ and $\bm{Q}_l^{\eta}+\bm{K}_l^{\eta}$ locates at some moir\'{e} $\bar{\bm{\Gamma}}$ point. Using such notation the $8\times 8$ onsite scattering matrix can be expressed as (flavor index is hidden)
\begin{align}
\Gamma^{l\eta}_{\bm{q}}(\bm{R})=&\langle \Phi(\bm{r}-\bm{R})|h_{l\eta}(\bm{q})|\Phi(\bm{r}-\bm{R})\rangle =e^{i\bar{\bm{q}}\cdot\bm{R}}\Gamma^{l\eta}_{\bm{q}},\label{eq:scattering_matrix1} \\
\Gamma^{l\eta}_{\bm{q}}=&\frac{1}{N_m}\sum_{\bar{\bm{k}}}\langle \Phi_{\bar{\bm{k}}}|h_{l\eta}(\bm{q})|\Phi_{[\bar{\bm{k}}-\bar{\bm{q}}]}\rangle, \label{eq:scattering_matrix2}
\end{align}
which is a result of lattice momentum conservation ($[\bm{k}]$ represents the residue part of $\bm{k}$ in mBZ). All the coupling information between the mode $a_{l\eta\bm{q}}$ and $f$ electrons are contained in the form factor $\Gamma_{\bm{q}}^{l\eta}$. When folded into mBZ, similar to the formation of electronic flat-band states, the moir\'{e} phonon with momentum $\bar{\bm{q}}$ will in general be composed of modes with momentum $\bar{\bm{q}}+\bm{Q}_l^{\eta}$ [see Eq. (\ref{eq:aBZ_to_mBZ})]. If we directly superpose the monolayer phonons with different $\bm{Q}_l^{\eta}$ by assigning the superposition coefficients as indicated by $\Gamma^{l\eta}_{\bar{\bm{q}}+\bm{Q}_l^{\eta}}$, the resulting moir\'{e} modes will by construction maximally couple with $f$ electrons, while all other modes are definitely decoupled with $f$ electrons due to orthogonality, in the present onsite limit. It can also be expected that $\Gamma^{l\eta}_{\bm{q}}$ will exponentially decay with $\bm{q}$, because it resembles the Fourier transform of a Gaussian function (the phonon field is a plane wave, and $f$ orbitals are exponentially localized). 

We then decompose Eq. (\ref{eq:scattering_matrix2}) by Pauli matrices,
\begin{align}
\Gamma^{l\eta}_{\bm{q}}=\sum_{ij}\lambda^{l\eta}_{ij}(\bm{q})s_0\tau_i\sigma_j, \label{eq:Pauli_decomposition}
\end{align}
where $s_i$, $\tau_j$, and $\sigma_k$ are Pauli matrices defined for the spin, valley, and angular momentum. Phonons cause no spin flip, thus only $s_0$ appears. Putting all altogether, the onsite EPC can be written as a multi-flavor Holstein Hamiltonian \cite{59ap_holstein_model1,95prb_holstein_model2} (but in the moir\'{e} scale),
\begin{align}
H^{f}_{\text{epc}}=\frac{1}{\sqrt{2}}\sum_{ij}\lambda_{ij}\sum_{\bm{R}}f^{\dagger}_{\bm{R}}\tau_i\sigma_j f_{\bm{R}}\left(a_{ij\bm{R}}+a_{ij\bm{R}}^{\dagger}\right),\label{eq:onsite_epc}
\end{align}
and we have defined the moir\'{e} phonon operator
\begin{align}
a_{ij\bm{R}}=\frac{1}{\sqrt{N_m}\lambda_{ij}}\sum_{l\eta\bm{q}}e^{i\bar{\bm{q}}\cdot\bm{R}}\lambda_{ij}^{l\eta}(\bm{q})a_{l\eta\bm{q}}. \label{eq:moire_phonon_operator}
\end{align}
From Eq. (\ref{eq:onsite_epc}), we see that, for each specific channel $ij$ ($ij$ denotes the channel's symmetry, see below), the local orbitals only couple to a unique combination Eq. (\ref{eq:moire_phonon_operator}) of different branches. The effective EPC constant $\lambda_{ij}$ of the $ij$-th mode is just the normalization factor
\begin{align}
\lambda_{ij}=\sqrt{\frac{1}{N_m}\sum_{l\eta\bm{q}}|\lambda_{ij}^{l\eta}(\bm{q})|^2}.\label{eq:lambda_normalized}
\end{align}
The above expression shows explicitly that the effective coupling with $f$ orbitals (and flat bands) is enhanced by umklapp scatterings (large momentum transfers). Similar effects also arise from low-frequency phonons and plasmons \cite{21prb_umklapp_pairing,21nl_umklapp_purcell,23a_umklapp_cooling}. Corresponding to the operator (\ref{eq:moire_phonon_operator}), the moir\'{e} polarization field centered at $\bm{R}$ reads
\begin{align}
\begin{split}
\bm{u}_{ij\bm{R}}&(\bm{r})=\frac{1}{N_m\sqrt{N_a}\lambda_{ij}}\sum_{l\eta\bm{q}}\sum_{\bm{R}_l\alpha}\lambda^{l\eta *}_{ij}(\bm{q})\\
& \times e^{i(\bm{q}+\bm{K}_l^{\eta})\cdot(\bm{r-\bm{R}})}\delta_{\bm{r},\bm{R}_l+\bm{\tau}_{l\alpha}}\bm{\epsilon}^{\eta}_{l\alpha}(\bm{q}).
\end{split}
\label{eq:moire_phonon_distortion}
\end{align}
One may easily check $\bm{u}_{ij\bm{R}}(\bm{r})=\bm{u}_{ij\bm{0}}(\bm{r}-\bm{R})$.

The symmetries of these moir\'{e} phonons can be identified as below. The $f$ orbitals at $\bm{R}=\bm{0}$ [Eq. (\ref{eq:f_orbital}), totally 8 states] form the following representation of the $D_6$ group and time reversal $\mathcal{T}$ ($\mathcal{K}$ takes complex conjugation)
\begin{align}
\begin{split}
&D^f(C_{3z})=e^{i\frac{2\pi}{3}\sigma_z},\quad D^f(C_{2x})=\sigma_x,\\
&D^f(C_{2z})=\tau_x, \quad D^f(\mathcal{T})=\tau_x\sigma_x\mathcal{K}.
\end{split}
\label{eq:Df}
\end{align}
Table \ref{tab:table1_symmetry} classifies the $16$ spinless generators $\tau_i\sigma_j$ according to the irreducible representations (irreps) of the $D_6$ group. Each term $\tau_i\sigma_j$ locks one specific symmetry. Therefore, modes with different indexes $(ij)$ are automatically orthogonal. The $K$-phonons induce purely intervalley scatterings, so only terms $\tau_{x,y}\sigma_j$ exist. From Table \ref{tab:table1_symmetry} we see 6 moir\'{e} $K$-phonons are allowed (the others are forbidden by time reversal): two 1D irreps $A_1$, $B_1$ and two pairs of 2D irreps $E_1$, $E_2$.

A definition of moir\'{e} $\Gamma$-phonons can be similarly done. Unlike (monolayer) $K$-phonons, $\Gamma$-phonons (with $\hbar\omega_{\Gamma}\approx 180$ meV) have the twofold degeneracy characterized by the oscillation directions $\mu=x,y$, see Fig. \ref{fig:monolayer_phonon}(c). The local EPC has exactly the same form as Eq. (\ref{eq:onsite_epc}), but this time $\tau_i$ can only take $\tau_{0}$ or $\tau_z$ due to the intravalley nature. The moir\'{e} phonon operator, distortion field, and EPC constant are parallel with Eqs. (\ref{eq:moire_phonon_operator}), (\ref{eq:moire_phonon_distortion}), and (\ref{eq:lambda_normalized}), respectively (the valley $\eta$ is replaced by the direction $\mu$ in the summation). Among the 4 allowable moir\'{e} $\Gamma$-phonons shown in Table \ref{tab:table1_symmetry}, the $A_1(\tau_0\sigma_0)$ mode is unrealistic, otherwise it will cause an overall shift of $f$ orbitals that contradicts the energy conservation. Therefore, the $\Gamma$ valley contributes 3 modes only: a pair of $E_2$ mode and a $B_2$ mode.

\begin{figure*}
\includegraphics[width=0.7\textwidth]{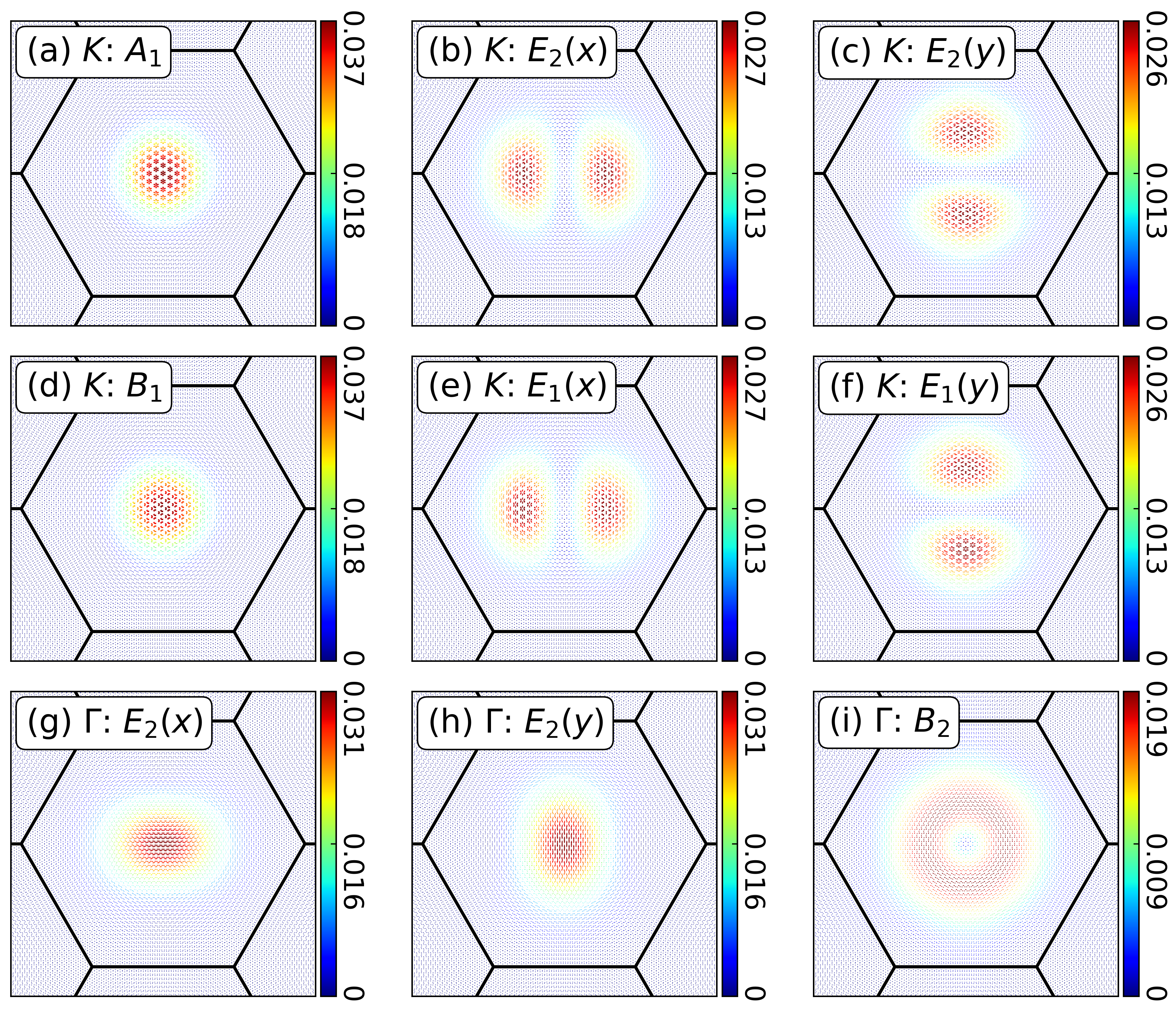}% Here is how to import EPS art
\caption{\label{fig:All_distortion}
The polarization fields $\bm{u}(\bm{r})$ of the 9 moir\'{e} phonons at $\bm{R}=0$, obtained by performing the numerical projection using the Wannierized $f$ orbitals and layer-decoupled EPC model Eq. (\ref{eq:Hepc_ld}). The $\bm{u}(\bm{r})$ fields are normalized so that $\sum_{l\bm{R}_l\alpha}|\bm{u}(\bm{R}_l+\bm{\tau}_{l\alpha})|^2=1$. The color and arrow represent the displacement strength and direction, respectively. (a)-(f) are the moir\'{e} $K$-phonons while (g)-(i) are the moir\'{e} $\Gamma$-phonons.}
\end{figure*}

\section{Numerical implementation \label{Sec3}}

\subsection{Moir\'{e} phonons from the layer-decoupled EPC}
To practically determine the moir\'{e} phonons, the simplest EPC model one can start with is \cite{08prb_graphene_symmetry,18prl_SC_from_phonon,23_liu_ek_phonon}
\begin{align}
H_{\text{epc}}=&H_{\text{epc}}^{K}+H_{\text{epc}}^{\Gamma}, \label{eq:Hepc_ld}\\
H_{\text{epc}}^K =& \frac{\gamma_{K}}{\sqrt{2N_t}}\sum_{l\eta\bm{q}\bm{k}}C^{\dagger}_{l,\bm{k}+\bm{q}}F^K_{\eta}C_{l\bm{k}}(a_{l\eta\bm{q}}+a_{l\bar{\eta},-\bm{q}}^{\dagger}),\label{eq:Hepc_K_plwv} \\
H_{\text{epc}}^{\Gamma} =& \frac{\gamma_{\Gamma}}{\sqrt{2N_t}}\sum_{l\mu\bm{q}\bm{k}}C^{\dagger}_{l,\bm{k}+\bm{q}}F^{\Gamma}_{\mu}C_{l\bm{k}}(a_{l\mu\bm{q}}+a_{l\mu,-\bm{q}}^{\dagger}), \label{eq:Hepc_G_plwv}
\end{align}
where $N_t=N_mN_a$, the $l$-th layer plane wave basis $C_{l\bm{k}}=(C_{KlA,\bm{k}+\bm{K}_l},C_{KlB,\bm{k}+\bm{K}_l},C_{K'lA,\bm{k}+\bm{K}_l'},C_{K'lB,\bm{k}+\bm{K}_l'})^T$, and the form factors $F_{\eta}^{K}=(\tau_x-i\eta \tau_y)\sigma_x/2$ for $K$-phonons, $F^{\Gamma}_{x}=-\tau_z\sigma_y$, $F^{\Gamma}_y=\sigma_x$ for $\Gamma$-phonons, with Pauli matrices $\tau_i\sigma_j$ defined in the valley-sublattice space. The coupling constants $\gamma_K\approx814$ meV, $\gamma_G\approx528$ meV, estimated using $\hbar\omega_K\approx150$ meV, $\hbar\omega_{\Gamma}\approx180$ meV. In this approximated model the dynamics of the two layers are decoupled, in the sense that the oscillation of one layer has no influence on the other.

\begin{table}[!htb]
\caption{\label{tab:epc_const_lam}%
The effective onsite coupling constants $\lambda_b$ ($\lambda_{ij}$), and the envelope shapes of the moir\'{e} phonons in each layer.}
\begin{ruledtabular}
\begin{tabular}{cccc}
Irrep & Order  &  $\lambda_b$ (meV) & Envelope \\
\hline
$K$: $A_1,B_1$       & $\tau_x$, $\tau_y$                    &   $7.364$   & $s$ \\
$K$: $E_1,E_2$       &$\tau_y(\sigma_x,\sigma_y)$, $\tau_x(\sigma_x,\sigma_y)$      &   $4.780$    & $(p_x,p_y)$  \\
$\Gamma$: $E_2$      &$(\sigma_x,\sigma_y)$                  &   $8.027$   & $s+d_{x^2-y^2}+d_{xy}$  \\
$\Gamma$: $B_2$      &$\tau_z\sigma_z$                       &   $6.166$   & $p_x+p_y$   \\
\end{tabular}
\end{ruledtabular}
\end{table}

Then the EPC matrix Eq. (\ref{eq:scattering_matrix2}) can be directly calculated and projected. We leave the numerical details in Appendix \ref{Appendix_epc_continuum}. The calculation is performed on a $27\times27$ $\bar{\bm{k}}$ sample mesh with $61$ reciprocal $\bm{G}$ vectors. The integrated coupling constants Eq. (\ref{eq:lambda_normalized}) of these 9 modes are listed in Table \ref{tab:epc_const_lam}. The corresponding polarization fields are shown in Fig. \ref{fig:All_distortion}.

The moir\'{e} phonons, by construction, share the typical property that the polarization strength is modulated in the moir\'{e} scale (which is the envelope function), while in the atomic scale the local distortion remains the monolayer pattern. For instance, the oscillating of moir\'{e} $A_1$ mode [Fig. \ref{fig:All_distortion}(a)] is strongest at AA-stacking, while in the graphene-scale the parent $\sqrt{3}\times\sqrt{3}$ Kekul\'{e} pattern is locally retained [Fig. \ref{fig:A1_distortion}(b)]. These features partially coincide with the experiment \cite{21n_local_phonon_exp} which is previously explained by classical lattice dynamics of the reconstructed moir\'{e} superlattice \cite{22prb_nearly_free_phonon_soliton,202Dm_soliton_phonon_TBG}. The envelope shapes of these modes are purely determined by the shape of $f$ orbitals, which can be made analytical if we approximate $f$ orbitals by Gaussian functions (Appendix \ref{Appendix_epc_continuum}). In that way the moir\'{e} polarization fields are explicitly expressed as atomic scale displacements modulated by Gaussian envelopes. The symmetry of the envelopes are summarized in Table \ref{tab:epc_const_lam}. Besides, for all the 6 moir\'{e} $K$-phonons, the two layers have in-phase envelopes, meaning that the two layers oscillate toward the same direction despite the tiny twisting angle, while for the 3 moir\'{e} $\Gamma$-phonons, the two layers' envelopes are opposite in phase.

\subsection{Frozen-phonon method}
Instead of using Eq. (\ref{eq:Hepc_ld}), we can start from more realistic ones. In this subsection we show how the frozen-phonon method \cite{70prl_modulated_hopping_frozen_phonon,80prb_frozen_phonon_method,85prl_frozen_phonon} can be applied to determine the moir\'{e} phonons as well. The logic follows the first-principle spirit: for a specific mode $b$ with the distortion field $\bm{u}_b$, the EPC Hamiltonian $h_{b}$ is by definition
\begin{align}
h_{b} = \bm{u}_{b}\cdot\nabla_{\bm{u}} H_0(\bm{u}_b)|_{\bm{u}_b=\bm{0}}\approx H_0(\bm{u}_{b})-H_0(0), \label{eq:frozen_phonon}
\end{align}
where $H_0(\bm{u}_b)$ is the single-particle Hamiltonian distorted by $\bm{u}_b$. In accordance with Eq. (\ref{eq:H_epc_full}), $\bm{u}_{b}$ should be carefully normalized for phonon quantization, see Appendix \ref{Appendix_frozen_phonon}. We then adopt the TB model to calculate $H_0(\bm{u}_b)$, but truncate it within the low-energy window \cite{23prb_tapw}. 

Some approximations help simplify calculations. First we replace the eigenmode vector $\bm{\epsilon}^b_{l}(\bm{q})$ by its values at $K(K')$ or $\Gamma$ points. This gets us free from the annoying random phases \cite{07prb_constant_plrz} and is actually an excellent approximation (Appendix \ref{appendix_phonon}). The EPC matrix Eq. (\ref{eq:scattering_matrix2}) ($K$-phonon as an example) is continuous about $\bm{q}=\bar{\bm{q}}+\bm{Q}_l^{\eta}$, so we neglect the $\bar{\bm{q}}$-dependence and keep only the $\bm{Q}_l^{\eta}$-dependence, in view of the tiny size of mBZ compared to the momentum cutoff. In other words, we focus exclusively on (monolayer) phonons with moir\'{e} translation symmetry ($\bar{\bm{q}}=\bar{\bm{0}}$), which can feasibly be integrated into the TB program. This approximation leads to
\begin{align}
\Gamma^{l\eta}_{\bar{\bm{q}}+\bm{Q}_l^{\eta}}\approx \Gamma^{l\eta}_{\bm{Q}_l^{\eta}}, \quad
\lambda_{b} \approx \sqrt{\sum_{l\eta\bm{Q}_l^{\eta}}|\lambda_{b}^{l\eta}(\bm{Q}_l^{\eta})|^2}.
\label{eq:scattering_approx}
\end{align}
The polarization field Eq. (\ref{eq:moire_phonon_distortion}) then reads
\begin{align}
\begin{split}
\bm{u}_{b\bm{R}}(\bm{r})\approx &\frac{1}{N_m\lambda_{b}}\sum_{\bar{\bm{q}}} e^{i\bar{\bm{q}}\cdot(\bm{r-\bm{R}})} \sum_{l\eta \bm{Q}_l^{\eta}}\sum_{\bm{R}_l\alpha}\lambda^{l\eta *}_{b}(\bm{Q}_l^{\eta})\\
& e^{i(\bm{Q}_l^{\eta}+\bm{K}_l^{\eta})\cdot(\bm{r-\bm{R}})}\delta_{\bm{r},\bm{R}_l+\bm{\tau}_{l\alpha}}\frac{\bm{\epsilon}^{\eta}_{l\alpha}}{\sqrt{N_a}},
\end{split}
\label{eq:moire_distortion_approx}
\end{align}
i.e., an envelope term $\sim\int_{\text{mBZ}}d\bar{\bm{q}}\exp(i\bar{\bm{q}}\cdot \bm{r})$ is factored out ($\bar{\bm{q}}$ should better be symmetrically sampled). Equivalently, the phonon version of the ``Bloch sum'', defined as
\begin{align}
\bm{u}_{b\bar{\bm{q}}}(\bm{r})=\frac{1}{\sqrt{N_m}}\sum_{\bm{R}}e^{i\bar{\bm{q}}\cdot\bm{R}}\bm{u}_{b\bm{R}}(\bm{r})=e^{i\bar{\bm{q}}\cdot\bm{r}}\bm{u}_{b}(\bm{r}), \label{eq:moire_phonon_q_space}
\end{align}
will have a $\bar{\bm{q}}$-independent periodic part \cite{22prb_Kekule_Fabrizio}
\begin{align}
\bm{u}_{b}(\bm{r})=\frac{1}{\lambda_{b}}\sum_{l\eta\bm{Q}_l^{\eta}}\lambda_{b}^{l\eta*}(\bm{Q}_l^{\eta})\bm{u}_{l\eta\bm{Q}_l^{\eta}}(\bm{r}).
\label{eq:moire_phonon_peridic}
\end{align}

The calculation is more elegant if $\bm{u}_{l\eta\bm{Q}_l^{\eta}}$ is symmetrized before distorting the lattice. This reduces to the standard problem of decomposing a reducible representation space of $D_6$ into its (real) irreps. The decomposition is given in Appendix \ref{appendix_frozen_phonon_symmetrized}. Here we take the $A_1$ and $B_1$ modes as an illustrative example. The $A_1$ mode is invariant in $D_6$, and $B_1$ is only odd under $C_{2z}$ ($C_{2z}$ flips the valley and reverses the momentum). As a result, $\lambda_{A_1}^{l\eta}(\bm{Q}_l^{\eta})$ should be the same for all 12 $\bm{u}_{l\eta\bm{Q}_l^{\eta}}$ (3 from each layer/valley) connected by $D_6$, while $\lambda_{B_1}^{l\eta}(\bm{Q}_l^{\eta})=\lambda_{B_1}^{l\bar{\eta}}(-\bm{Q}_l^{\eta})$. More specifically, focusing on the leading term nearest to $\bm{K}_l^{\eta}$, one may identify $\bm{Q}_{1}^K,\bm{Q}_2^{K'}\in\{\bm{q}_1,\bm{q}_2,\bm{q}_3\}=\{\bm{q}_j\}$ and $\bm{Q}_{1}^{K'},\bm{Q}_{2}^{K}\in \{-\bm{q}_j\}$, where $\bm{q}_1=(0,-k_{\theta})^T$, $\bm{q}_{j}=C_{3z}^{j-1}\bm{q}_1$, and $k_{\theta}$ is the mBZ sidelength. The $A_1$ and $B_1$ fields composed by them respectively read (not normalized)
\begin{subequations}
\begin{align}
&\sum_{j}\left(\bm{u}_{1K\bm{q}_j}+\bm{u}_{1K',-\bm{q}_j}+\bm{u}_{2K,-\bm{q}_j}+\bm{u}_{2K'\bm{q}_j}\right),\label{eq:leading_order_a}\\
&i\sum_{j}\left(\bm{u}_{1K\bm{q}_j}-\bm{u}_{1K',-\bm{q}_j}+\bm{u}_{2K,-\bm{q}_j}-\bm{u}_{2K'\bm{q}_j}\right). \label{eq:leading_order_b}
\end{align}
\end{subequations}
They are real and periodic, thus can be easily used to distort the lattice after suitable normalization. The electron bands of such frozen-phonon method provide a means to visualize the EPC effects. The resulting EPC matrix Eq. (\ref{eq:scattering_matrix2}) will no doubt form as $\tau_{x,y}$, which is now simply calculated in $k$-space. The higher-order modes with larger $|\bm{Q}_{l}^{\eta}|>k_{\theta}$ can be symmetrized in the same way.

\begin{figure}
\includegraphics[width=0.48\textwidth]{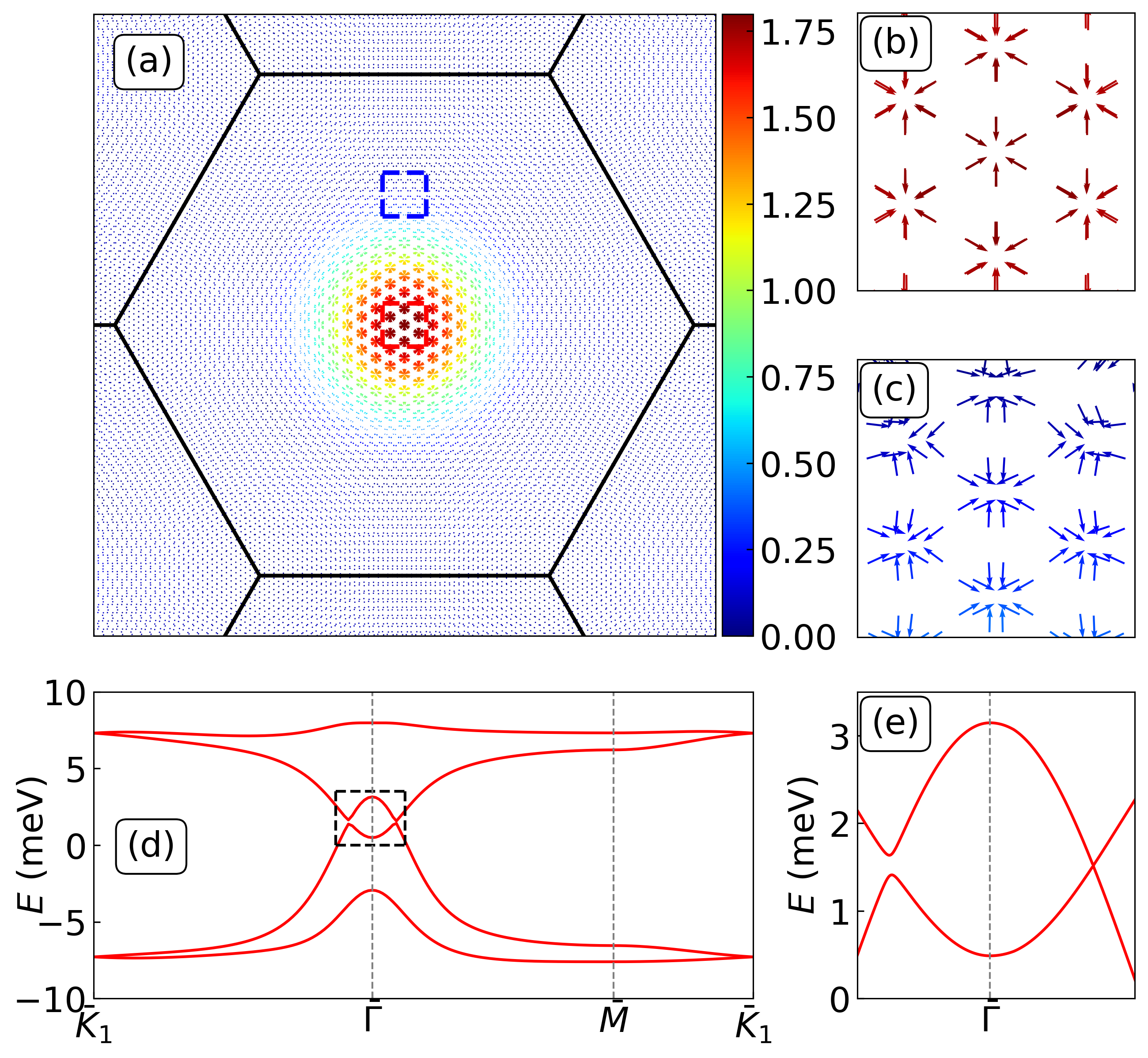}% Here is how to import EPS art
\caption{\label{fig:A1_distortion}
(a) The (periodic) distortion field of moir\'{e} $A_1$ mode, obtained using the frozen-phonon method. The color (arrow) shows the strength (direction) of the displacement, in the unit of m$\text{\AA}$. The field is normalized to have the mean displacement $\approx 0.25$ m$\text{\AA}/$atom. (b) and (c) show the local displacement in the AA [red square in (a)] and domain wall between AB and BA regions [blue square in (a)], respectively. (d) The frozen-phonon flat bands with the distortion in (a). The flat bands have a crossing point along $\bar{\bm{\Gamma}}-\bar{\bm{M}}$, and an anti-crossing point along $\bar{\bm{K}}-\bar{\bm{\Gamma}}$, as shown in (e). The bands will completely split with a stronger distortion. All these characters are consistent with the $A_1$ mode discovered in Ref. \cite{19prx_JT_TB_Fabrizio}.}
\end{figure}

If we stop at the leading terms (\ref{eq:leading_order_a}), (\ref{eq:leading_order_b}), we obtain the same moir\'{e} phonons used in Refs. \cite{20epjp_JT_BM_Fabrizio,22prb_disorder_koshino} (Appendix \ref{Appendix_leading_order}). They indeed strongly couple with the flat bands due to the tiny momentum transfer. However, modes with larger momentum (the same irrep) are found essential as well, although the coupling with $f$ orbitals decreases as $\bm{Q}_l^{\eta}$ goes to outer shells, see Appendix \ref{appendix_frozen_phonon_symmetrized}. The periodic moir\'{e} field (\ref{eq:moire_phonon_peridic}) is the superposition of all these symmetrized basis. We plot such obtained $A_1$ mode and the corresponding frozen-phonon bands in Fig. \ref{fig:A1_distortion}. The coupling is so efficient that a mean displacement $0.25$ m$\text{\AA}$ per atom causes a $15$ meV of flat bands splitting at $\bar{\bm{K}}_{l}$. Such band reshape induced by tiny distortions is akin to the strain, disorder, and pressure effects magnified in moir\'{e} systems \cite{14FD_strain_magnified,20prr_disorder_sarma,18prb_pressure,23prb_defect}. Our $A_1$ phonon shares the same characters with the one in Ref. \cite{19prx_JT_TB_Fabrizio} obtained by brutal diagonalization. The strong overlap between the phonon envelope and the charge density of $f$ orbitals, now naturally appears as a result of the projection theory.

\section{Phonon-mediated mean-field states \label{Sec4}}
\subsection{Mean-field scheme}
Phonons induce effective interactions among electrons. We then check the influence of some moir\'{e} phonons on various symmetry-breaking states, within the framework of mean-field theory. Integrating out the phonons gives the interaction \cite{22prb_Kekule_Fabrizio,23prb_pair_critical_field}
\begin{align}
H_{\text{P}} \approx -\frac{1}{2N_m}\sum_{b\bar{\bm{q}}}\frac{1}{\hbar \omega_b}h^{\dagger}_b(\bar{\bm{q}}) h_b(\bar{\bm{q}}), \label{eq:EPC_attraction}
\end{align}
where the complete EPC matrix $h_b(\bar{\bm{q}})$ encodes all complicated umklapp scatterings involving both $f$ and $c$ orbitals associated with the mode $\bm{u}_{\bar{\bm{q}}b}$. All complicated umklapp scatterings are encoded in $h_b(\bar{\bm{q}})$. For simplicity we consider only 4 modes: the $K$-phonon $A_1(\tau_x),B_1(\tau_y)$ modes and the $\Gamma$-phonon $E_2(\sigma_x,\sigma_y)$ mode. The Coulomb term is taken as \cite{22prb_ZLL_OPW_representation}
\begin{align}
H_{\text{C}}=\frac{1}{2N_m}\sum_{\xi\xi'}\sum_{\bm{q}\bm{k}\bm{k}'}v(\bm{q})C^{\dagger}_{\xi,\bm{k}+\bm{q}}C^{\dagger}_{\xi',\bm{k}'-\bm{q}}C_{\xi'\bm{k}'}C_{\xi\bm{k}},
\end{align}
where the plane wave $C_{\xi\bm{k}}$ has the index $\xi=(s,\eta,l,\alpha)$ composed by spin, valley, layer, and sublattice. We take the dielectric constant $\epsilon_s=10$ and screening length $d_s=25$ nm in the double-gated potential $v_{\bm{q}}=e^2\tanh(d_s |\bm{q}|)/(2\epsilon_0\epsilon_s S_m|\bm{q}|)$, $S_m$ is the area of supercell. 

In our calculation the total energy 
\begin{align}
e=\frac{1}{N_m}\langle H_{0}+H_{\text{C}}+H_{\text{P}}\rangle \label{eq:total_energy}
\end{align}
is self-consistently obtained using a $9\times 9$ sample mesh of $\bar{\bm{k}}$ points. To eliminate the double counting, in calculating interactions the non-symmetry-breaking density matrix is deducted \cite{22prb_Kekule_Fabrizio,22prb_ZLL_OPW_representation} (different subtraction schemes lead to different results at $\nu=0$, as argued in Ref. \cite{23a_EPC_vs_IKS_kwan}). The calculation is made more realistic using the TB model \cite{23prb_tapw} as the non-interacting $H_0$, and the EPC matrices $h_b$ are directly read from the frozen-phonon method. We neglect the $\bar{\bm{q}}$-dependence of $h_b$, which is justified since the dominant attractive Hartree channel can already be accurately captured by $\bar{\bm{q}}=\bar{\bm{0}}$ terms. 

\begin{figure}
\includegraphics[width=0.48\textwidth]{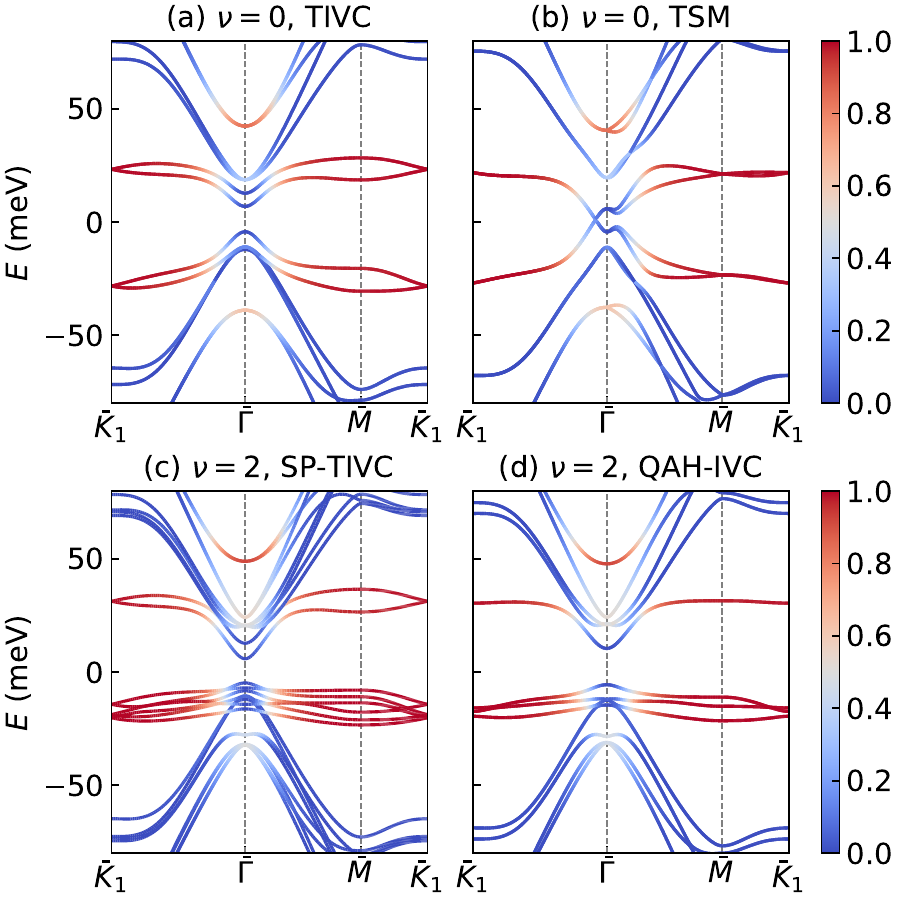}
\caption{\label{fig:HFband} (a) and (b) are the mean-field bands of TIVC and TSM states at filling $\nu=0$, respectively. (c) and (d) are the bands of SP-TIVC and QAH-IVC states at filling $\nu=2$, respectively. These bands are self-consistently obtained by minimizing the total energy Eq. (\ref{eq:total_energy}) with realistic EPC matrices ($\alpha=1$) obtained by the frozen-phonon method. The color represents the occupation of $f$ orbitals in each band.}
\end{figure}

\subsection{Numerical results}

\begin{figure}
\includegraphics[width=0.43\textwidth]{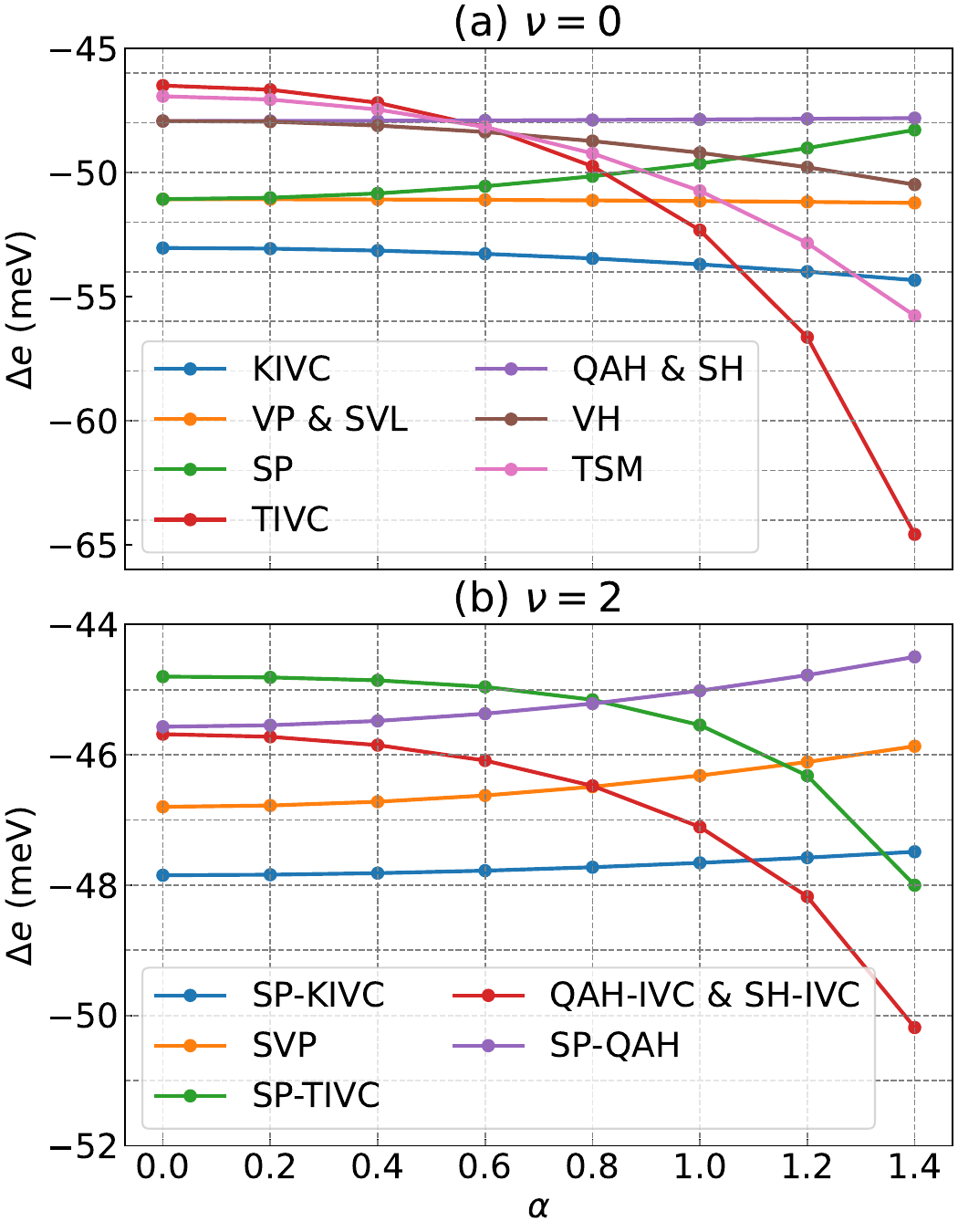}
\caption{\label{fig:Evsa} The self-consistent condensation energy per supercell of symmetry-breaking states at fillings (a) $\nu=0$ and (b) $\nu=2$, by gradually increasing the magnitude of EPC ($\alpha=1$ corresponds to the realistic case in the present study). In calculations only the four dominant phonon branches are incorporated, which are $A_1$ and $B_1$ modes of $K$-phonons and $E_2$ mode of $\Gamma$-phonons.}
\end{figure}

Previous studies using the continuum model suggest two groups of candidates for the insulating ground states if no other effect exists. The first group is the Kramers intervalley coherent states (KIVC) with the order parameter $\cos\phi\tau_x\sigma_z+\sin\phi\tau_y\sigma_z$ (using $f$ basis). The second group contains various flavor-polarized states, including the valley-polarized (VP: $\tau_z$), spin-polarized (SP: $s_z$), and spin-valley-locked (SVL: $s_z\tau_z$) states. Even using the particle-hole-asymmetric $H_0$ from TB model, they are still the most competitive states, with KIVC order being slightly more favorable (Fig. \ref{fig:Evsa}, $\alpha=0$). However, these candidates contradict some recent experiments where the Kekul\'{e} order in atomic-scale are observed \cite{23_nuckolls_quantum_texture,23a_IKS_kim,22prl_spectroscopy_theory_bab,22prl_stm_theory_mpz}.

We focus on even fillings. At the neutrality case $\nu=0$, the phonon-mediated $A_1,B_1$ channels favor the order $\cos\phi\tau_x+\sin\phi\tau_y$, which is the $\mathcal{T}$-invariant intervalley coherent (TIVC) state. This state has been pointed out as the candidate for the Kekul\'{e} charge order \cite{22prb_Kekule_Fabrizio,23a_EPC_vs_IKS_kwan,23_nuckolls_quantum_texture}. In addition, the $E_2$ channel spoils the intravalley order $\cos\phi\sigma_x+\sin\phi\sigma_y$, which preserves $\mathcal{T}$, $C_{2z}$ but breaks $C_{3z}$ \cite{20prx_KIVC_ashvin}. These states are in general semimetals with tiny gaps (depends on $\phi$) so we call them the $\mathcal{T}$-invariant semimetal (TSM) states. All TSM states have very close condensation energies. In Fig. \ref{fig:HFband}(b) we show a typical semimetal band when $\phi\approx\pi/3$. When we gradually enhance the EPC by tuning $\alpha h_b$ from $\alpha=0$ to $\alpha=1.4$, as shown in Fig. \ref{fig:Evsa}(a), the TIVC and TSM states indeed become more stabilized. This is mainly due to the energy decreasing through the Hartree channel. The other states, including the KIVC, VP, SP, SVL, quantum anomalous Hall (QAH: $\sigma_z$), spin Hall (SH: $s_z\sigma_z$), and valley Hall (VH: $\tau_z\sigma_z$) states, are only weakly affected. In realistic EPC range $\alpha\approx 1$, it is subtle to determine which is the actual ground state. 

More varieties are brought by moir\'{e} phonons at the half filling $\nu=\pm 2$. The spin-polarized versions of the KIVC (SP-KIVC), VP (SVP), and TIVC (SP-TIVC) states show the same energy evolution trends when the EPC strength increases, as shown in Fig. \ref{fig:Evsa}(b). It is then attempting to guess that the ground state are still among these spin-polarized states. However, the TIVC state becomes more stable if the spin polarization is broken, corresponding to two new candidates. One is the QAH-IVC state ($\tau_x+\sigma_z$) carrying Chern number $C=\pm 2$, while the other is the topologically trivial spin-Hall SH-IVC state ($\tau_x+s_z\sigma_z$). These two states are found numerically degenerate and are more promising than SP-TIVC state, consistent with the study in Ref. \cite{23a_EPC_vs_IKS_kwan}. A qualitative analysis is give in Appendix \ref{appendix_phonon_interaction}.

According to our calculation, at $\nu=\pm2$ the TIVC states overcome the KIVC states at $\alpha\approx1.10$, which is larger than the value predicted in Ref. \cite{22prb_Kekule_Fabrizio} ($\alpha\approx 0.50$) and close to the result in Ref. \cite{23a_EPC_vs_IKS_kwan} ($\alpha\approx 1.0$). The phonon model proposed in Refs. \cite{20epjp_JT_BM_Fabrizio,22prb_Kekule_Fabrizio} attributes all couplings to the leading orders (\ref{eq:leading_order_a}), (\ref{eq:leading_order_b}), which might lead to an overestimation of the coupling once the phonon is normalized (see Appendix \ref{Appendix_leading_order}). It is reasonable to expect that the TIVC and TSM states can be more stable than predicted here. Our moir\'{e} phonons are specially designed to maximize the couplings with $f$ orbitals. The huge amount of phonons thrown away could more or less couple to $c$ orbitals and alter the present results (e.g., the band gap near $\bar{\bm{\Gamma}}$ might become larger), making the TIVC states more competitive. Remarkably enough, although we incorporate only two intervalley phonons, the results are quantitatively in the same order as those containing all modes \cite{23a_EPC_vs_IKS_kwan}. The moir\'{e} phonons presented here should have captured the main effects.

\section{An effective lattice model \label{section:lattice_epc}}
\begin{figure}
\includegraphics[width=0.45\textwidth]{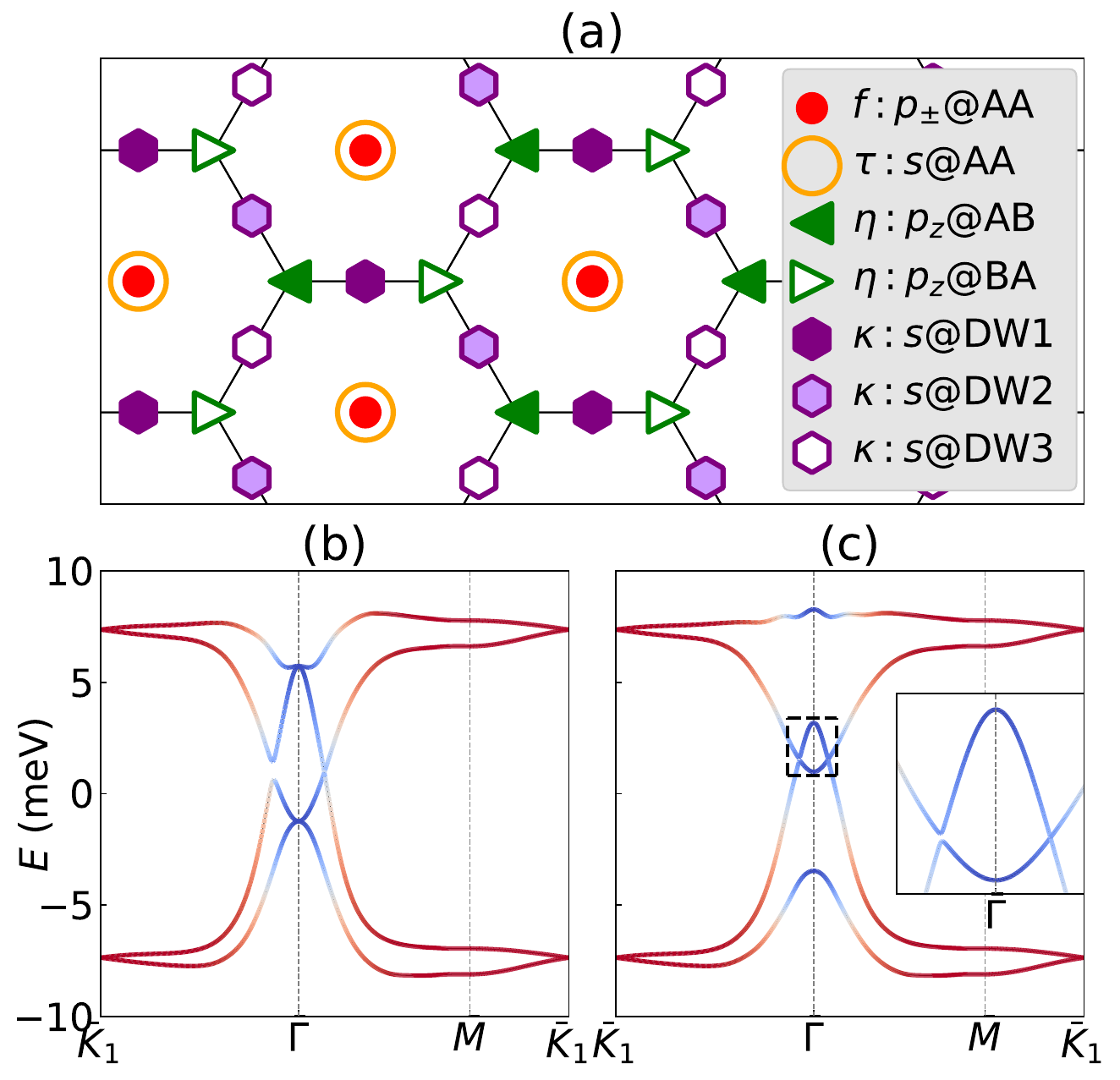}
\caption{\label{fig:lattice_band} (a) The lattice structure of the 8-band electron model. (b) and (c) show the frozen-phonon bands obtained using the lattice EPC model under the moir\'{e} $A_1$ distortion  shown in Fig. \ref{fig:A1_distortion}(a), but with different cutoffs. The color indicates the ratio of $f$ orbitals. In (b) we keep only the coupling with $f$ orbitals Eq. (\ref{eq:A1_epc_lattice_ff}), which has no influence on the flat bands near $\bar{\bm{\Gamma}}$. The gaps never open no matter how strong the distortion is. In (c) the couplings with both $f$ and $c$ orbitals Eqs. (\ref{eq:A1_epc_lattice_tt}), (\ref{eq:A1_epc_lattice_hh}) are included, which qualitatively recovers the bands with full couplings in Fig. \ref{fig:A1_distortion}(d).}
\end{figure}

If the couplings between phonons and $f$ orbitals tell the whole story, the simple Holstein model Eq. (\ref{eq:onsite_epc}) would be adequate to describe all relative physics. However, it is not that trivial in reality: the couplings with $c$ orbitals indeed play some roles. From the frozen-phonon viewpoint, a strong enough $A_1$ distortion can split the flat bands completely into two sets \cite{19prx_JT_TB_Fabrizio}, which never happens if only couplings with $f$ orbitals exist, see Fig. \ref{fig:lattice_band}(b). This suggests us retrieve the other couplings, at least the leading terms, for a more realistic EPC model.

Therefore, we first apply the multi-step Wannierization as in Ref. \cite{19prr_8orbital_model} to construct 6 additional Wannier $c$ orbitals (in each valley/spin), and write down the EPC as an extended 8-band Holstein model. These $c$ orbitals, together with two $f$ orbitals, are denoted by the spinless operator ($\xi$ is the valley)
\begin{align}
\begin{split}
d_{\xi\bm{R}}=(&f_{\xi+,\bm{R}},f_{\xi-,\bm{R}},\tau_{\xi\bm{R}},\eta_{\xi,\bm{R}+\bm{\tau}_{\text{AB}}},\eta_{\xi,\bm{R}+\bm{\tau}_{\text{BA}}},\\
&\kappa_{\xi,\bm{R}+\bm{\tau}_{\text{DW1}}},\kappa_{\xi,\bm{R}+\bm{\tau}_{\text{DW2}}},\kappa_{\xi,\bm{R}+\bm{\tau}_{\text{DW3}}})^T.
\end{split}
\end{align} 
The lattice formed by them is shown in Fig. \ref{fig:lattice_band}(a), and the Wannier orbitals are detailed in Appendix \ref{appendix_8band_model}. The Coulomb interaction among them have been thoroughly studied in Ref. \cite{20prb_8orbital_interaction}. In this paper we focus on their couplings with the moir\'{e} phonons. The general lattice EPC Hamiltonian reads
\begin{align}
\begin{split}
H_{\text{epc}}=&\frac{1}{\sqrt{2}}\sum_{b\bm{R}}\sum_{\xi'\xi}\sum_{\bm{L}\bm{T}}(a_{b\bm{R}}+a_{b\bm{R}}^{\dagger})\\
&\times d_{\xi',\bm{R}+\bm{L}}^{\dagger}\Lambda^{b}_{\xi'\bm{L},\xi\bm{T}}d_{\xi,\bm{R}+\bm{T}},
\end{split}
\label{eq:lattice_EPC_general}
\end{align}
where $\Lambda^{b}_{\xi'\bm{L},\xi\bm{T}}$ encodes all scatterings of electrons (mediated by the phonon $b$ centered at $\bm{R}$) from site $\bm{R}+\bm{T}$ in the valley $\xi$, to the site $\bm{R}+\bm{L}$ in the valley $\xi'$. Because both phonons and electrons are localized, the scattering should decay fast when $|\bm{L}|$ or $|\bm{T}|$ becomes larger. 

Our numerical examination indicates that, a nearest-neighbor cutoff already captures the important features compared with the frozen-phonon bands. Take the moir\'{e} $A_1$ mode of $K$-phonons as an example, the intervalley EPC is expanded as
\begin{align}
H_{\text{epc}}^{A_1}=&\frac{1}{\sqrt{2}}\sum_{\bm{R}}(a_{A_1\bm{R}}+a_{A_1\bm{R}}^{\dagger})(h^{ff}_{\bm{R}}+h^{\tau\tau}_{\bm{R}}+h^{\eta\eta}_{\bm{R}}+...). \label{eq:Hepc_A1_main}
\end{align}
The three leading channels listed above are
\begin{subequations}
\begin{align}
h_{\bm{R}}^{ff}=&\sum_{\xi t}\lambda_{A_1}^{ff}f^{\dagger}_{\bar{\xi}t \bm{R}}f_{\xi t\bm{R}}, \label{eq:A1_epc_lattice_ff}\\
h_{\bm{R}}^{\tau\tau}=&\sum_{\xi}\lambda_{A_1}^{\tau\tau}\tau^{\dagger}_{\bar{\xi}\bm{R}}\tau_{\xi\bm{R}},\label{eq:A1_epc_lattice_tt}\\
h_{\bm{R}}^{\eta\eta}=&\sum_{\xi}\sum_{g}^{0,1,2}\lambda_{A_1}^{\eta\eta}\left(\eta^{\dagger}_{\bar{\xi},\bm{R}+C_{3z}^{g+1}\bm{\tau}_{\text{BA}}}+\eta^{\dagger}_{\bar{\xi},\bm{R}+C_{3z}^{g-1}\bm{\tau}_{\text{BA}}}\right)\nonumber \\
&\times \eta_{\xi,\bar{\bm{R}}+C_{3z}^g\bm{\tau}_{\text{AB}}} +\left(\text{AB}\leftrightarrow\text{BA}\right), \label{eq:A1_epc_lattice_hh}
\end{align}
\end{subequations}
with $\lambda_{A_1}^{ff}=\lambda_{A_1}\approx 7.364$ meV, $\lambda_{A_1}^{\tau\tau}\approx -2.684$ meV, and $\lambda_{A_1}^{\eta\eta}\approx -0.370$ meV. The first term Eq. (\ref{eq:A1_epc_lattice_ff}) is just the onsite scattering among $f$ orbitals, which governs the gap $\Delta\approx 2\lambda_{A_1}^{ff}$ of the frozen-phonon bands at $\bar{\bm{K}}_{l}$. The second term Eq. (\ref{eq:A1_epc_lattice_tt}) is the onsite scattering between the ring-shape $c$ orbitals. The third term Eq. (\ref{eq:A1_epc_lattice_hh}) causes mutual scatterings of $c$ orbitals in the neighboring AB and BA sites. The second and third terms control respectively the splittings of the top and bottom two frozen-phonon bands at $\bar{\bm{\Gamma}}$ [Figs. \ref{fig:lattice_band}(c)]. The more complete lattice EPC Hamiltonians of the 9 moir\'{e} phonons are given in Appendix \ref{appendix_epc_lattice_model}.

We notice that, in such projection scheme, the couplings between phonons and $c$ orbitals are inevitably underestimated. One should treat them more strictly if they are proved essential in the future.

\section{Summary and Discussion \label{Sec5}}

In this paper, we propose a projection theory to simplify optical phonons and their couplings with electrons in MATBG. Instead of diagonalizing the formidable dynamical matrix, we obtain the moir\'{e} phonons by symmetrically superposing the folded monolayer modes, according to their couplings with the active heavy $f$ electrons. Only 9 moir\'{e} modes survive out of a huge amount of optical phonons, greatly simplifying the EPC model for further studies. Such obtained moir\'{e} phonons have well-localized distortion peaks near the AA-stacking area, while in the atomic scale they retain the typical monolayer oscillating characters.

These moir\'{e} phonons strongly couple with the flat bands, as proved by both the frozen-phonon and mean-field calculations. Moreover, we have demonstrated how the frozen-phonon method can be generalized to determine the EPC in a moir\'{e} system. The advantage is that all model parameters are derived from the reliable electron model, without fitting or adopting other empirical values. We believe this is crucial for moir\'{e} systems that depend quite sensitively on external effects.

Before ending, we would like to highlight several potential areas for improvement and raise open questions that warrant further investigation. The strength of the phonon-mediated attraction is estimated to be relatively weak (e.g., $\lambda_b^2/(\hbar\omega_b)=0.363$ meV for the $A_1$ mode), especially when compared to the Hubbard repulsion among $f$ orbitals ($>40$ meV \cite{22prl_heavy_fermion_Zhida,22prb_ZLL_OPW_representation,20prb_8orbital_interaction}). Therefore, in order to attribute superconductivity to these phonon channels, it is necessary to develop a theoretical mechanism that can explain the strong screening of the Coulomb interaction \cite{23a_kondo_song,23a_K_unflatten,23prb_kondo_shankar,23prl_kondo_haoyu2,24a_nodal_sc_onsite}. The lattice relaxation \cite{19natm_yoo_atomic,22prb_leconte_relax,20prb_koshino_epc,24a_analytical_relaxtion} and substrates \cite{12jpcl_substrate_raman,20prb_band_hBN_tbg,23prb_hbn_phonon} effects are not considered in this preliminary study, but exploring their influence on phonon properties would be of nontrivial importance. Additionally, the treatment of phonon couplings with $c$ orbitals could benefit from a more rigorous approach which may offer corrections to the obtained results, particularly regarding topological effects \cite{19prx_phonon_Weyl,23a_quantum_geometry_phonon,24prl_quantum_geometry_GL,23prx_3DTI_phonon,23prl_phonon_exp}.On a parallel note, it would also be intriguing to investigate how the presence of flat bands affects the softening or stiffening of these moir\'{e} modes \cite{08prl_phonon_soft_exp_yan,07nm_stiffening}. For example, the moir\'{e} modes of MATBG can be directly mapped to those of alternately twisted trilayer graphene, as these two systems share many similarities in their theoretical frameworks \cite{21prl_kondo_TTG,23prb_THF_for_TTG}.

\begin{acknowledgments}
We acknowledge the helpful discussions with T.-Y. Qiao, Z.-D. Song, C.-X. Liu, and Y.-L. Chen. H. Shi thanks A. Blason and Y. H. Kwan for their feedback in the early stage of this work. X. Dai is supported by a fellowship awarded from the Research Grants Council of the Hong Kong Special Administrative Region, China (Project No. HKUST SRFS2324-6S01).
\end{acknowledgments}

\appendix
\begin{widetext}
\section{Lattice configuration \label{Appendix_lattice}}
This appendix specifies the lattice configuration used throughout this study. Before rotatation the monolayer basis vectors are $\bm{a}_1 = a_0\left(1/2,\sqrt{3}/2\right)^T$, $\bm{a}_2 = a_0\left(-1/2,\sqrt{3}/2\right)^T$, where $a_0=0.246$ nm is the lattice constant. The sublattice $A$ and $B$ are located at $\bm{\tau}_{A,B}=\pm (\bm{a}_1+\bm{a}_2)/3$. The reciprocal basis vectors are $\bm{b}_1=4\pi/(\sqrt{3}a_0)\left(\sqrt{3}/2,1/2\right)^T$, $\bm{b}_2=4\pi/(\sqrt{3}a_0)\left(-\sqrt{3}/2,1/2\right)^T$. The two Dirac points living in atomic Brillouin zone (aBZ) are $\bm{K}^{\eta}=\eta(\bm{b}_1-\bm{b}_2)/3$ (if $\eta$ enters in calculations, it takes $+1$ for $K$ and $-1$ for $K'$). 

The hexagon center $\bm{r}=\bm{0}$ is chosen as the axis to rotate clockwisely the top ($l=1$) and bottom ($l=2$) layers by $\pm \theta/2$: $\bm{a}^{l}_{1,2}=R\left[(-1)^l \theta/2\right]\bm{a}_{1,2}$, $\bm{K}^{\eta}_{l}=R\left[(-1)^l \theta/2\right]\bm{K}^{\eta}$, etc. Such configuration has a $D_6$ point group consistent with the emergent symmetries of the BM model \cite{18prb_emergent_D6,18prb_Adrian_Po}. A different setup does not change the physics in moir\'{e} scale. For MATBG we define the commensurate moir\'{e} basis vectors
\begin{align}
\bm{L}_1^m = R \left( \frac{\theta}{2} \right) [n_m\bm{a}_1 - (2 n_m + 1) \bm{a}_2] = L_{\theta} \left( \frac{\sqrt{3}}{2}, -\frac{1}{2} \right)^T,\quad
  \bm{L}_2^m = R \left( \frac{\pi}{6} \right) \bm{L}_1^m = L_{\theta} \left(\frac{\sqrt{3}}{2}, \frac{1}{2} \right)^T, 
\end{align}
where $n_m=30$, which gives the magic angle $\theta = \arcsin \left[ \sqrt{3}(2n_m+1)/(6 n_m^2+6 n_m + 2) \right]\approx 1.085^{\circ}$. The moir\'{e} lattice constant $L_{\theta} = a_0\sqrt{3 n_m^2+3n_m+1}\approx 12.996$ nm. In each supercell there are $4N_a=4(3n_m^2+3n_m+1)=11164$ atoms, each can be labeled by index $(i,l,\alpha)$ representing the $\alpha\ (=A,B)$ sublattice in the $i$-th ($i=1,2,..,N_a$) atomic cell, $l$-th layer of the supercell. We use $\bm{R}_{Iil\alpha}=\bm{L}_I+\bm{\tau}_{il\alpha}$ to denote the position of sublattice ($i,l,\alpha$) in the $I$-th supercell.

The moir\'{e} reciprocal commensurate vectors are
\begin{align}
  \bm{G}_1^m = \bm{b}_1^{1} -\bm{b}_1^{2} = \sqrt{3} k_{\theta} \left( \frac{1}{2}, - \frac{\sqrt{3}}{2} \right)^T, \quad
  \bm{G}_2^m = \bm{b}_2^{1} -\bm{b}_2^{2} = \sqrt{3} k_{\theta} \left( \frac{1}{2}, \frac{\sqrt{3}}{2} \right)^T, 
\end{align}
where $k_{\theta}=4\pi/(3L_{\theta})$ is the edge length of the hexagonal moir\'{e} Brillouin zone (mBZ). Each vector $\bm{k}$ can be written as $\bm{k}=\bar{\bm{k}}+n_1\bm{G}_1^m+n_2\bm{G}_2^m$, where $\bar{\bm{k}}\in $mBZ. We define the high-symmetry points in mBZ as $\bar{\bm{\Gamma}}=0$, $\bar{\bm{K}}_1=k_{\theta}\left(\sqrt{3}/2,-1/2\right)^T$, $\bar{\bm{K}}_2=k_{\theta}\left(\sqrt{3}/2,1/2\right)^T$, and $\bar{\bm{M}}=\left(\bar{\bm{K}}_1+\bar{\bm{K}}_2 \right)/2$. $\bm{K}^{\eta}_l$ and $\bar{\bm{K}}_{1,2}$ are related by
\begin{align}
\bm{K}^{\eta}_l = \eta\left(n_m\bm{G}^m_1+n_m\bm{G}^m_2+\bar{\bm{K}}_l\right).\label{eq:Kpoints_atomic}
\end{align}
To simulate the corrugation in $z$-axis, we adopt the empirical height profile \cite{18prx_koshino,23prb_tapw,14prb_corrugation}
\begin{align}
z^l(\bm{r}) = (-1)^{l+1}\left\{\frac{d_0}{2}+d_1\left[\cos(\bm{G}_1^m\cdot \bm{r})+\cos(\bm{G}_2^m\cdot \bm{r})+\cos((\bm{G}_1^m+\bm{G}_2^m)\cdot \bm{r})\right]\right\},
\end{align}
where $d_0 \approx 0.343$ nm, $d_1\approx0.0278$ nm. The height profile preserves the $D_6$ symmetry and translation symmetry [Fig. \ref{fig:atomic_lattice}(a)], and separates the flat bands from remote bands by a gap about $25$ meV [Fig. \ref{fig:H0_band}(a)]. We note that the in-plane relaxation can also be similarly simulated using analytical methods \cite{24a_analytical_relaxtion}, which we will not cover in this study.

\begin{figure}
\includegraphics[width=\textwidth]{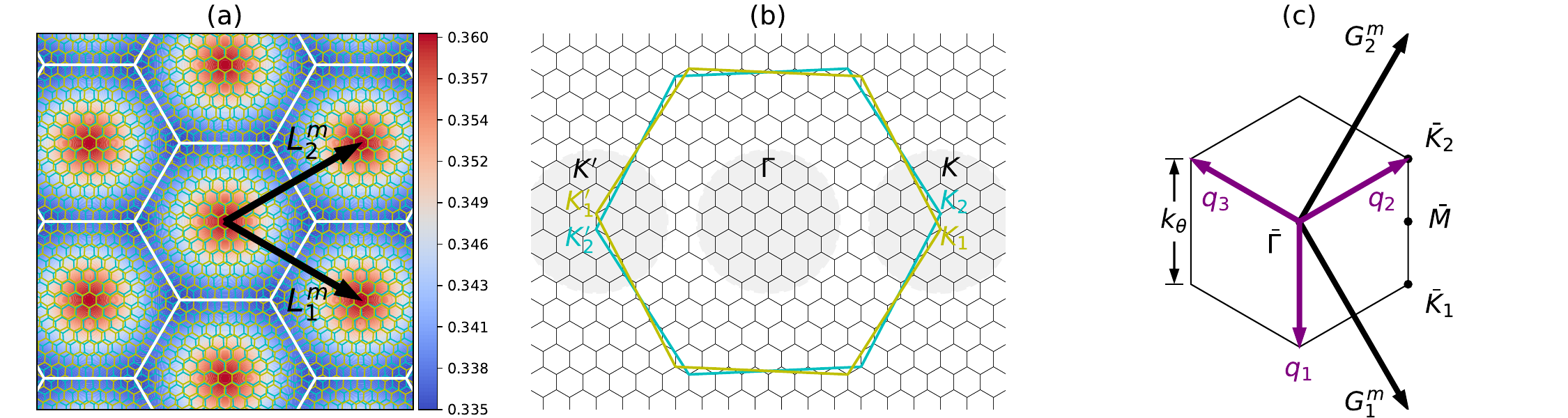}
\caption{\label{fig:atomic_lattice} (a) The moir\'{e} superlattice of TBG. For visual convenience we plot the case with $n_m=6$ ($\theta=5.086^{\circ}$). The light green and light blue hexagons represent the graphene lattice in the top ($l=1$) and bottom ($l=2$) layers, respectively. The color shows the interlayer distance between the two layers, in the unit of nm. (b) The atomic Brillouin zones (aBZ) of the top (green) and bottom (blue) layers, and the nested moir\'{e} Brillouin zone (mBZ, tiny black hexagons). The shadows depict the $K(K')$ valleys and the $\Gamma$ valley. (c) The mBZ with high-symmetry points and vectors defined in detail.}
\end{figure}

\section{Tight-binding model versus TAPW method \label{Appendix_TAPW}}
\subsection{The tight-binding model}
This appendix provides a short review of the tight-binding model (TB) and the truncated atomic plane wave (TAPW) method \cite{23prb_tapw}. The spinless TB Hamiltonian of TBG reads
\begin{align}\label{eq:fullHam}
H^{\text{TB}} = \sum_{I i l\alpha}\sum_{I'i'l'\alpha'} t(\bm{R}_{Iil\alpha}-\bm{R}_{I'i'l'\alpha'}) C^{\dagger}_{Iil\alpha} C_{I'i'l'\alpha'},
\end{align}
where $C^{\dagger}_{I i \alpha}$ ($C_{I i \alpha}$) creates (annihilates) a $p_z$-orbital electron located at $\bm{R}_{Ii\alpha}$. The hopping integral $t$ is determined through the Slater-Koster formula under two-center approximation
\begin{equation}
t(\bm{r}) = V_\pi\left(1-\frac{z^2}{r^2}\right)+V_\sigma\frac{z^2}{r^2},\quad V_\pi = -2.7 e^{-(r-a_c)/r_0} \text{eV}, \quad V_\sigma = 0.48 e^{-(r-d_0)/r_0} \text{eV},
\label{eq:SK}
\end{equation}
where $z = \bm{r} \cdot \bm{e}_z$, $a_c=a_0/\sqrt{3}$ is the nearest-neighbour carbon distance, $r_0 = 0.184a_0$ is the decaying length \cite{12prb_Koshino_Moon}. The moir\'{e} Bloch basis is defined as (suppose we have $N_m$ supercells)
\begin{align}
C_{il\alpha\bar{\bm{k}}}^{\dagger}=\frac{1}{\sqrt{N_m}}\sum_I e^{i\bar{\bm{k}}\cdot (\bm{L}_I+\bm{\tau}_{il\alpha})}C^{\dagger}_{Iil\alpha}, \label{eq:basis_moire_Bloch}
\end{align}
under which the TB Hamiltonian is decomposed as  
\begin{align}
H^{\text{TB}}=\sum_{\bar{\bm{k}}}\sum_{il\alpha,i'l'\alpha'}C^{\dagger}_{il\alpha\bar{\bm{k}}}H^{\text{TB}}_{il\alpha,i'l'\alpha'}(\bar{\bm{k}})C_{i'l'\alpha'\bar{\bm{k}}}, \quad
H^{\text{TB}}_{il\alpha,i'l'\alpha'}(\bar{\bm{k}})=e^{-i\bar{\bm{k}}\cdot\bar{\bm{\tau}}_{il\alpha,i'l'\alpha'}}t(\bar{\bm{\tau}}_{il\alpha,i'l'\alpha'}), \label{eq:Hamk_old}
\end{align}
where $\bar{\bm{\tau}}_{il\alpha,i'l'\alpha'}$ is the nearest relative displacement between sublattice $(i,l,\alpha)$ and $(i',l',\alpha')$, i.e, $\bar{\bm{\tau}}_{il\alpha,i'l'\alpha'}$ takes $\bm{\tau}_{il\alpha}-\bm{\tau}_{i'l'\alpha'}+\bm{L}_I$ with smallest length $|\bm{\tau}_{il\alpha}-\bm{\tau}_{i'l'\alpha'}+\bm{L}_I|$. 

\subsection{Truncated atomic plane wave method}
However, Eq. (\ref{eq:Hamk_old}) is redundant if we care only about the physics near the charge neutrality point (CNP). A better basis set is the monolayer Bloch basis (atomic plane waves) with wave vector $\bm{k}=\bar{\bm{k}}+\bm{G}$ defined in aBZ, which is connected with the moir\'{e} Bloch basis Eq. (\ref{eq:basis_moire_Bloch}) through a unitary transformation
\begin{align}
C^{\dagger}_{l\alpha,\bar{\bm{k}}+\bm{G}}\equiv \frac{1}{\sqrt{N_m N_a}}\sum_{Ii}e^{i(\bar{\bm{k}}+\bm{G})\cdot (\bm{L}_{I}+\bm{\tau}_{il\alpha})}C^{\dagger}_{Iil\alpha}=\sum_{i}C_{il\alpha\bar{\bm{k}}}^{\dagger}X_{i,\bm{G}}(l\alpha), \quad X_{i,\bm{G}}(l\alpha) = \frac{1}{\sqrt{N_a}}e^{i\bm{\tau}_{il\alpha}\cdot\bm{G}}.\label{eq:transform_in}
\end{align}
The unitarity of the transformation can be proved by noticing that
\begin{align}
\delta_{\bar{\bm{k}},0} \delta_{\bm{G},0}=\delta_{\bar{\bm{k}}+\bm{G},0}=\frac{1}{N_m N_a}\sum_{Ii}e^{i\bm{R}_{Iil\alpha}\cdot(\bar{\bm{k}}+\bm{G})}=\frac{1}{N_a}\sum_{i}e^{i\bm{\tau}_{il\alpha}\cdot \bm{G}}\frac{1}{N_m}\sum_{I}e^{i\bm{R}_{Iil\alpha}\cdot\bar{\bm{k}}}=\frac{\delta_{\bar{\bm{k}},0}}{N_a}\sum_{i}e^{i\bm{\tau}_{il\alpha}\cdot \bm{G}},
\end{align}
indicating that $N_a^{-1}\sum_{i}e^{i\bm{\tau}_{il\alpha}\cdot \bm{G}}=\delta_{\bm{G},0}$, thus $\sum_{i}X_{i,\bm{G}}^*(l\alpha)X_{i,\bm{G}'}(l\alpha)=N_a^{-1}\sum_{i}e^{i\bm{\tau}_{i\alpha}\cdot (\bm{G}'-\bm{G})}=\delta_{\bm{G}\bm{G}'}$, or $X^{\dagger}(l\alpha)X(l\alpha)=I_{N_a\times N_a}$. Using the duel relation one may also prove $X(l\alpha)X^{\dagger}(l\alpha)=I_{N_a\times N_a}$. The reverse transformation of Eq. (\ref{eq:transform_in}) $C_{il\alpha\bar{\bm{k}}}^{\dagger}=\sum_{\bm{G}}C^{\dagger}_{l\alpha,\bar{\bm{k}}+\bm{G}}X^*_{i,\bm{G}}(l\alpha)$ transforms the full Hamiltonian (\ref{eq:Hamk_old}) to
\begin{align}
H^{\text{TB}}=\sum_{\bar{\bm{k}}}\sum_{\bm{G}l\alpha}\sum_{\bm{G}'l'\alpha'}C^{\dagger}_{l\alpha,\bar{\bm{k}}+\bm{G}}\left[X^{\dagger}(l\alpha)H^{\text{TB}}_{l\alpha,l'\alpha'}(\bar{\bm{k}})X(l'\alpha')\right]_{\bm{G},\bm{G}'}C_{l'\alpha',\bar{\bm{k}}+\bm{G}'}. \label{eq:projection}
\end{align}
In the above equation the summations are over all $N_a$ commensurate vectors $\bm{G}$. The TAPW approximation is simply the truncation (projection) of the summation to the two low-energy atomic valleys, 
\begin{align}
\begin{split}
&H^{\text{TAPW}}=\sum_{\bar{\bm{k}}}\sum_{\eta l\alpha}\sum_{\eta'l'\alpha'}\sum_{\bm{G}}^{|\bm{G}-\bm{K}_l^{\eta}|<G_{\text{c}}}\sum_{\bm{G}'}^{|\bm{G}'-\bm{K}_{l'}^{\eta'}|<G_{\text{c}}}C^{\dagger}_{\eta l\alpha,\bar{\bm{k}}+\bm{G}}H^{\text{TAPW}}_{\bm{G}l\alpha,\bm{G}'l'\alpha'}(\bar{\bm{k}})C_{\eta'l'\alpha',\bar{\bm{k}} +\bm{G}'},\\
&H^{\text{TAPW}}_{\bm{G}l\alpha,\bm{G}'l'\alpha'}(\bar{\bm{k}})=\left[X^{\dagger}(l\alpha)H^{\text{TB}}_{l\alpha,l'\alpha'}(\bar{\bm{k}})X(l'\alpha')\right]_{\bm{G},\bm{G}'}.
\end{split}
\label{eq:Hamk_tapw}
\end{align}
The valley index $\eta=K,K'$ is put on $C_{\eta l\alpha,\bar{\bm{k}}+\bm{G}}$ to indicate where $\bm{G}$ is located, and the cutoff $G_{\text{c}}$ should be large enough for convergence. It can be proved that the intralayer part of the truncated Hamiltonian reduces to the monolayer Hamiltonian, i.e.,
\begin{align}
H^{\text{TAPW}}_{\bm{G}l\alpha,\bm{G}'l\alpha'}(\bar{\bm{k}})=\delta_{\bm{G}\bm{G}'}H^{l}_{\alpha,\alpha'}(\bar{\bm{k}}+\bm{G})=\delta_{\bm{G}\bm{G}'}\sum_{\bm{R}_l}e^{- i (\bar{\bm{k}}+\bm{G}) \cdot
  (\bm{R}_l+\bm{\tau}_{l\alpha}-\bm{\tau}_{l\alpha'})} t (\bm{R}_l+\bm{\tau}_{l\alpha}-\bm{\tau}_{l\alpha'}),
\end{align}
where $\bm{R}_l$ sums over all the lattice vectors of the $l$-th monolayer.

The truncated Hamiltonian has not only a much simplified matrix structure (same as the BM model), but also the high accuracy (same as the TB model) near CNP. For rigid lattice, $H^{\text{TAPW}}$ is valley-diagonal (intervalley coupling $\sim 10^{-7}$ eV), but the intervalley terms are finite if the $K$-phonon distortion is applied. 

Such projection allows an efficient scheme to calculate the moir\'{e}-scale electron-phonon coupling (EPC) in a frozen-phonon manner. Notice that the truncation method can also be applied on dynamical matrix to strictly calculate the moir\'{e} phonons, see Refs. \cite{23prb_tapw,22prb_csc_model}. However, what we care about here is their couplings with the low-energy electrons. So we start directly from monolayer phonons that are much simpler than the exact moir\'{e} phonons.

\section{Continuum model and heavy-fermion representation \label{Appendix_BM_ZLL}}
\subsection{The BM model and its symmetries}
The BM Hamiltonian can be obtained by applying the following approximations on Eq. (\ref{eq:Hamk_tapw}): (a) use $k\cdot p$ model (same as monolayer graphene) on the intralayer terms, (2) retain only the largest three interlayer scatterings with smallest momentum transfer, and neglect the momentum dependence \cite{23prb_tapw}. The resulting Hamiltonian can be compactly written using plane wave basis as (the summation over $\bm{G}$ is always within the cutoff)
\begin{align}
\begin{split}
& H^{\text{BM}} =\sum_{\bar{\bm{k}}}\sum_{\eta}\sum_{\bm{G}l\alpha}\sum_{\bm{G}'l'\alpha'}C^{\dagger}_{\eta l\alpha,\bar{\bm{k}}+\bm{G}}\langle \eta l\alpha,\bar{\bm{k}}+\bm{G}|H^{\text{BM}}_{\eta}(\bm{p},\bm{r})|\eta l'\alpha',\bar{\bm{k}}+\bm{G}'\rangle C_{\eta l'\alpha',\bar{\bm{k}} +\bm{G}'},\\
& H^{\text{BM}}_{\eta}(\bm{p},\bm{r}) = \left(\begin{array}{cc}
    v_F (\bm{p}-\hbar\bm{K}^{\eta}_1)\cdot (\eta \sigma_x,\sigma_y) & U_{\eta} (\bm{r})\\
    U_{\eta}^{\dag}
    (\bm{r}) & v_F (\bm{p}-\hbar\bm{K}^{\eta}_2)\cdot (\eta \sigma_x,\sigma_y)
  \end{array}\right),\\
& U_{\eta}(\bm{r})=\left(\begin{array}{cc}
    u_0 & u_1\\
    u_1 & u_0
  \end{array}\right) +
  \left(\begin{array}{cc}
    u_0 & u_1 \omega^{\eta}\\
    u_1 \omega^{-\eta} & u_0
  \end{array}\right) e^{i\eta\bm{G}_1^m \cdot \bm{r}} +\left(\begin{array}{cc}
    u_0 & u_1 \omega^{-\eta}\\
    u_1 \omega^{\eta} & u_0
  \end{array}\right) e^{-i\eta\bm{G}_2^m \cdot \bm{r}},
\end{split}
\label{eq:Hamk_BM_full}
\end{align}
where $\bm{p}=-i\hbar\nabla$, $\omega=\exp(i2\pi/3)$. The parameters can be directly read from the truncated Hamiltonian (\ref{eq:Hamk_tapw}): $\hbar v_F=0.52316$ eV$\cdot$nm, $u_0=85.463$ meV, and $u_1=102.85$ meV.

The BM model has the translation symmetry, the $D_6$ rotation symmetry, time-reversal symmetry $\mathcal{T}$, and a particle-hole symmetry $\mathcal{P}$ \cite{11pnas_BM_model,18prb_Adrian_Po,18prb_emergent_D6}. Under a general operation $g$, the 8-component (valley, layer, sublattice) wave function $f(\bm{r})$ changes to $(gf)(\bm{r})=D(g)f(g^{-1}\bm{r})$, where $D(g)$ is the matrix to make $D(g)H^{\text{BM}}(g^{-1}\bm{p},g^{-1}\bm{r})D^{-1}(g)=\pm H^{\text{BM}}(\bm{p},\bm{r})$ satisfied (the ``$-$'' case is only for $g=\mathcal{P}$). The $D(g)$ matrices are found as (our choice of $\mathcal{P}$ does not change the coordinates but switches the valleys)
\begin{align}
\begin{split}
 & D_{\eta'l'\xi',\eta l \xi}(C_{3 z}) = e^{i(\bm{K}^{\eta}_l-C_{3z}\bm{K}^{\eta}_l)\cdot \bm{r}}\left[e^{i\frac{2 \pi}{3}\tau_z \sigma_z }\right]_{\eta'l'\xi',\eta l \xi}, \quad
 D(C_{2 x}) = \varrho_x \sigma_x,\quad
 D(C_{2 z}) = \tau_x \sigma_x, \\
 & D(\mathcal{T}) = \tau_x \mathcal{K}, \quad
D_{\eta' l' \xi', \eta l \xi}(\mathcal{P}) =
  -e^{- i (\bm{K}_1^{\eta} +\bm{K}_2^{\eta}) \cdot \bm{r}}
  [\tau_x \varrho_y \sigma_x]_{\eta' l' \xi', \eta l \xi},
 \label{eq:symmetry_continuum_original}
\end{split}
\end{align}
where $\tau_i$, $\varrho_j$, and $\sigma_k$ are identity and Pauli matrices introduced in valley, layer, and sublattice spaces, with $\eta$, $l$, and $\xi$ being their indexes, respectively. $\mathcal{K}$ is the complex conjugation operator. The momentum-translation prefactor in $D(C_{3z})$ appears because the rotated wave vectors should be translated back to its equivalent valley that we have specially chosen. To gauge out these phases and make symmetry analysis simpler, some authors choose the convention that the momentum is relative to $\bm{K}^{\eta}_l$, which corresponds to a gauge transformation $C_{\eta l\alpha,\bar{\bm{k}}+\bm{G}}\rightarrow C_{\eta l\alpha,\bar{\bm{k}}+\bm{G}-\bm{K}^{\eta}_l}$ ($\bm{G}-\bm{K}^{\eta}_l$ is redefined as $-\bm{Q}_{\eta l}$ in Refs. \cite{22prl_heavy_fermion_Zhida,22prb_Kekule_Fabrizio}). Correspondingly the Hamiltonian looks more symmetric,
\begin{align}
\begin{split}
& H^{\text{BM}}_{\eta}(\bm{p},\bm{r}) = \left(\begin{array}{cc}
    v_F \bm{p}\cdot (\eta \sigma_x,\sigma_y) & U_{\eta} (\bm{r})\\
    U_{\eta}^{\dag}
    (\bm{r}) & v_F \bm{p}\cdot (\eta \sigma_x,\sigma_y)
  \end{array}\right),\\
& U_{\eta}(\bm{r})=\left(\begin{array}{cc}
    u_0 & u_1\\
    u_1 & u_0
  \end{array}\right)e^{-i\eta\bm{q}_1\cdot\bm{r}}  +\left(\begin{array}{cc}
    u_0 & u_1 \omega^{-\eta}\\
    u_1 \omega^{\eta} & u_0
  \end{array}\right) e^{-i\eta\bm{q}_2\cdot\bm{r}} +
  \left(\begin{array}{cc}
    u_0 & u_1 \omega^{\eta}\\
    u_1 \omega^{-\eta} & u_0
  \end{array}\right) e^{-i\eta\bm{q}_3\cdot\bm{r}},
\end{split}
\label{eq:Ham_BM_symmetric}
\end{align}
where $\bm{q}_1=k_{\theta}(0,-1)^T$, $\bm{q}_2=C_{3z}\bm{q}_1$, and $\bm{q}_3=C^2_{3z}\bm{q}_1$, see Fig. \ref{fig:atomic_lattice}(c). Using this gauge the symmetry matrices are
\begin{align}
\begin{split}
 D(C_{3 z}) = e^{i\frac{2 \pi}{3}\tau_z \sigma_z}, \quad
 D(C_{2 x}) = \varrho_x \sigma_x,\quad
 D(C_{2 z}) = \tau_x \sigma_x, \quad
 D(\mathcal{T}) = \tau_x \mathcal{K}, \quad
 D(\mathcal{P}) = -\tau_x \varrho_y \sigma_x.
 \label{eq:symmetry_continuum_symmetric}
\end{split}
\end{align}

\subsection{Pseudo zeroth Landau levels and its Gaussian approximation}
The following part is a brief derivation of zeroth Landau level (ZLL) orbitals \cite{22prb_ZLL_OPW_representation}, which play the central role of $f$ orbitals in determining the moir\'{e} phonons. Focusing on $\eta=K$ valley, we expand the moir\'{e} potential to its linear order in coordinates $U(\bm{r})=3u_0\sigma_0-i3k_{\theta}u_1(-y/2,x/2)\cdot \bm{\sigma}+O[(r/L_{\theta})^2]$. After transformation $\mathcal{U}_y^{\dagger} H\mathcal{U}_y$, where $\mathcal{U}_y$ is the matrix diagonalizing the Pauli matrix $\varrho_y$, the local Hamiltonian near the AA center can be written as
\begin{align}
\tilde{H}^{\text{AA}}=\mathcal{U}_y^{\dagger} \left[ v_F\bm{p}\cdot \bm{\sigma}+3u_0\varrho_x+3k_{\theta}u_1\varrho_y\left(-\frac{y}{2},\frac{x}{2}\right)\cdot\bm{\sigma} \right] \mathcal{U}_y=v_F(\bm{p}+\varrho_ze\bm{A})\cdot\bm{\sigma}+3u_0\varrho_y,\label{eq:Ham_local_AA}
\end{align}
where the pseudo-field $\bm{A}=B_{\theta}(-y/2,x/2)$, $B_{\theta}=3k_{\theta}u_1/(ev_F)$. The local Hamiltonian (\ref{eq:Ham_local_AA}) has localized zero modes protected by the chiral symmetry: $\{\varrho_z\sigma_z,\tilde{H}^{\text{AA}}\}=0$, akin to the ZLL problem of Dirac fermions \cite{19prb_ZLL_jianpeng}. The eigen-equation $\tilde{H}^{\text{AA}}\Phi(\bm{r})=0$ for the zero mode $\Phi=(y_1,y_2,y_3,y_4)^T$ can be decoupled as two sets of equations [$r=\sqrt{x^2+y^2}$, $\phi=\arctan{(y/x)}$]:
\begin{align}
\begin{split}
\left(\frac{\partial}{\partial r}-\frac{r}{2l_B^2}+\frac{i}{r}\frac{\partial}{\partial \phi} \right)y_1=-\frac{t_{\theta}}{l_B}e^{-i\phi}y_4,&\quad 
\left(\frac{\partial}{\partial r}-\frac{r}{2l_B^2}-\frac{i}{r}\frac{\partial}{\partial \phi} \right)y_4 =\frac{t_{\theta}}{l_B}e^{i\phi}y_1,\\
\left(\frac{\partial}{\partial r}+\frac{r}{2l_B^2}-\frac{i}{r}\frac{\partial}{\partial \phi} \right)y_2=-\frac{t_{\theta}}{l_B}e^{i\phi}y_3,&\quad 
\left(\frac{\partial}{\partial r}+\frac{r}{2l_B^2}+\frac{i}{r}\frac{\partial}{\partial \phi} \right)y_3=\frac{t_{\theta}}{l_B}e^{-i\phi}y_2,
\end{split}
\label{eq:y1y2y3y4}
\end{align}
where $l_B=\sqrt{\hbar/(eB_{\theta})}$ is the magnetic length, $t_{\theta}=3u_0l_B/(\hbar v_F)$ (Ref. \cite{22prb_ZLL_OPW_representation} uses $\lambda_{\theta}=t_{\theta}/\sqrt{2}$ instead). Eliminating $y_3$ and $y_4$ results in the following equations for $y_1$ and $y_2$,
\begin{align}
\left(r^2\frac{\partial^2}{\partial r^2}+r\frac{\partial}{\partial r}+t_{\theta}^2\frac{r^2}{l_B^2}+\frac{\partial^2}{\partial \phi^2}\right)e^{-r^2/(4l_B^2)}y_1=0, \quad
\left(r^2\frac{\partial^2}{\partial r^2}+r\frac{\partial}{\partial r}+t_{\theta}^2\frac{r^2}{l_B^2}+\frac{\partial^2}{\partial \phi^2}\right)e^{r^2/(4l_B^2)}y_2=0.
\end{align}
Then use the trick of separation of variables, $e^{\mp r^2/(4l_B^2)}y_{1,2}=R(r)\Theta(\phi)$, we get
\begin{align}
r^2\frac{d^2}{dr^2}R+r\frac{d}{dr}R+\left(t_{\theta}^2\frac{r^2}{l_B^2}-n^2\right)R=0, \quad \frac{d^2}{d\phi^2}\Theta+n^2\Theta=0.
\end{align}
The single-valuedness requires $\Theta(\phi+2\pi)=\Theta(\phi)\sim e^{in\phi}$, so $n$ must be integers. The equation for $R$ is a Bessel equation, leading to $R\sim J_n(t_{\theta}r/l_B)$. We hunt for localized modes, so $y_1=0$ (otherwise $|y_1|\sim e^{r^2/(4l_B^2)}/\sqrt{r}$ for $r\rightarrow \infty$). So the nonzero components are $y_2(\bm{r})\sim e^{-r^2/(4l_B^2)+in\phi}J_n(t_{\theta}r/l_B)$ and $y_3(\bm{r})\sim -e^{-r^2/(4l_B^2)+i(n-1)\phi}J_{n-1}(t_{\theta}r/l_B)$ [Eq. (\ref{eq:y1y2y3y4})]. Then transform back: $\tilde{\Phi}=\mathcal{U}_y(0,y_2,y_3,0)^T=(y_3,y_2,-iy_3,iy_2)^T/\sqrt{2}$, and pick two most localized ones ($n=0,1$) as the $f$ orbitals. Their wave functions are (still call them ZLLs)
\begin{align}
\tilde{\Phi}_{K+} (\bm{r}) &= 
  \frac{1}{\sqrt{2}} \left(\begin{array}{cc}
    e^{i \frac{\pi}{4}} w_0 (\bm{r}) \\
    - e^{i \frac{\pi}{4}} w_1 (\bm{r}) \\
    e^{- i \frac{\pi}{4}} w_0 (\bm{r}) \\
    e^{- i \frac{\pi}{4}} w_1 (\bm{r})
  \end{array}\right), \quad
\tilde{\Phi}_{K-}(\bm{r}) =\frac{1}{\sqrt{2}} \left(\begin{array}{cc}
    e^{- i \frac{\pi}{4}} w_1^* (\bm{r})\\
    e^{- i \frac{\pi}{4}} w_0 (\bm{r})\\
    - e^{i \frac{\pi}{4}} w_1^*(\bm{r})\\
    e^{i \frac{\pi}{4}} w_0 (\bm{r})
  \end{array}\right),\label{eq:ZLL_WF_matrix}
\end{align}
with the sublattice orbitals ($I_n$ is the first-kind modified Bessel function)
\begin{align}
w_n(\bm{r}) &=\mathcal{C}e^{-\frac{r^2}{4l_B^2}}J_n\left(t_{\theta}\frac{r}{l_B}\right)e^{in\phi}, \quad 
\mathcal{C}=\frac{e^{t_{\theta}^2/2}}{\sqrt{2\pi l_B^2\left[ I_0(t_{\theta}^2)+I_1(t_{\theta}^2)\right]}}.
\label{eq:ZLL_WF_appendix}
\end{align}
In the convention (\ref{eq:Hamk_BM_full}) the wave functions $\Phi_{\eta t}$ differ from Eq. (\ref{eq:ZLL_WF_matrix}) by a gauge: $\Phi_{\eta t}(\bm{r},l\alpha)=\exp(i\bm{K}_l^{\eta}\cdot \bm{r})\tilde\Phi_{\eta t}(\bm{r},l\alpha)$. The indexes $\eta$ and $t=\pm$ in $\Phi_{\eta t}$ represent the valley and angular momentum ($C_{3z}$ eigenvalue: $\omega^t$, or the hidden Chern number), respectively. The two ZLLs in the $K'$ valley are obtained by time reversal: $\Phi_{K't}=\Phi^*_{K\bar{t}}$ ($\bar{t}=-t$). The factors $e^{\pm i\pi/4}$ in Eq. (\ref{eq:ZLL_WF_matrix}) are introduced to make the four ZLLs transform as
\begin{align}
  D^f (C_{3 z}) = e^{i \frac{2 \pi}{3} \sigma_z}, \quad
  D^f (C_{2 x}) = \sigma_x, \quad 
  D^f (C_{2 z}) = \tau_x, \quad
  D^f (\mathcal{T}) = \tau_x \sigma_x \mathcal{K}, \quad
  D^f (\mathcal{P}) = \tau_x \sigma_z . 
\label{eq:ZLL_repre_appendix}
\end{align}
$D^f$ is the representation matrix so that $g\Phi_{\eta t}(\bm{r}) = \sum_{\eta't'}\Phi_{\eta't'}D^{f}_{\eta't',\eta t}(g)$.

The ZLLs Eq. (\ref{eq:ZLL_WF_appendix}) captures all typical properties of the $f$ electron Wannier functions in Refs. \cite{22prl_heavy_fermion_Zhida,19prr_8orbital_model}. In the $K$ valley, Ref. \cite{22prl_heavy_fermion_Zhida} proposes the following Gaussian functions to approximate them
\begin{align}
  w_0(\bm{r})=\frac{\alpha_1}{l_1\sqrt{\pi}}e^{-\frac{r^2}{2l_1^2}}, \quad w_1(\bm{r})=\frac{\alpha_2}{l_2^2\sqrt{\pi}}(x+iy)e^{-\frac{r^2}{2l_2^2}},
\label{eq:Wannier_WF_appendix}
\end{align}
which share exactly the same symmetry as Eq. (\ref{eq:ZLL_WF_appendix}) under $D_6$ rotations. Note that in the $K'$ valley, the order of the Gaussian orbitals in Ref. \cite{22prl_heavy_fermion_Zhida} is reversed compared to ours, resulting in a different but equivalent symmetry representation. The representation here is consistent with Refs. \cite{20epjp_JT_BM_Fabrizio,22prb_Kekule_Fabrizio}.

Since the ZLLs Eq. (\ref{eq:ZLL_WF_appendix}) are most localized zero modes of the linearized BM model, we may expect their large overlap with the flat bands. We define the $\bar{\bm{k}}$-resolved and the overall overlap functions
\begin{align}
O_f(\bar{\bm{k}}) =\frac{1}{2}\sum_{t,n=\pm1}|\langle \Phi_{\eta t \bar{\bm{k}}}|\psi_{\eta n\bar{\bm{k}}}\rangle|^2, \quad
O_f = \frac{1}{N_m}\sum_{\bar{\bm{k}}}O_f(\bar{\bm{k}}),
\label{eq:ovlp_function}
\end{align}
where $|\psi_{\eta n\bar{\bm{k}}}\rangle$ are flat-band eigenstates, and $|\Phi_{\eta t\bar{\bm{k}}}\rangle=N_m^{-1/2}\sum_{\bm{R}}e^{i\bar{\bm{k}}\cdot \bm{R}}|\Phi_{\eta t}(\bm{r}-\bm{R})\rangle$ are Bloch sums of ZLLs. A direct calculation using parameters below Eq. (\ref{eq:Hamk_BM_full}) (which gives $l_B\approx2.2936$ nm and $t_{\theta}\approx 1.1240$) yields an overall overlap $O_f\approx 66.94\%$ with the TB flat bands, which is relatively high but lower than the BM case where the non-locality of realistic moir\'{e} potential is discarded. In the 8-band model introduced below, the ZLLs serve as trial $f$ orbitals during the Wannierization, and the overlap is further optimized to $O_f\approx 94.34\%$ .

\subsection{The 8-band model \label{appendix_8band_model}}
An 8-band lattice model for TBG was proposed in Ref. \cite{19prr_8orbital_model}. Although the particle-hole symmetry is locally missing, the single-particle bands are accurately retained. Recently such model has made success in studying the magnetoplasmons \cite{23_jpcm_magnetoplasmon} and explaining the cascades effects \cite{23nc_cascades_dmft}. The model (in each valley/spin) contains a triangular lattice formed by two $p_{\pm}$ orbitals centered at AA sites ($f$ orbitals), a triangular lattice formed by an $s$ orbital centered at AA sites, a hexagonal lattice formed by two $p_z$ orbitals centered at AB/BA sites, and a kagome lattice formed by three $s$ orbitals centered at the domain walls which we denote by DW1, DW2, and DW3. We denote the 8 spinless orbitals as $p_{+}@$AA, $p_{-}@$AA, $s@$AA, $p_z @$AB, $p_z @$BA, $s$@DW1, $s@$DW2, $s@$DW3, with the generating operators 
\begin{align}
d^{\dagger}_{\eta\bm{R}}=(f^{\dagger}_{\eta+,\bm{R}},f^{\dagger}_{\eta-,\bm{R}},{\tau}^{\dagger}_{\eta\bm{R}},{\eta}^{\dagger}_{\eta,\bm{R}+\bm{\tau}_{\text{AB}}},{\eta}^{\dagger}_{\eta,\bm{R}+\bm{\tau}_{\text{BA}}},{\kappa}^{\dagger}_{\eta,\bm{R}+\bm{\tau}_{\text{DW1}}},{\kappa}^{\dagger}_{\eta,\bm{R}+\bm{\tau}_{\text{DW2}}},{\kappa}^{\dagger}_{\eta,\bm{R}+\bm{\tau}_{\text{DW3}}}). \label{eq:d_operator_8band}
\end{align}
Their wave functions $\Phi_{\eta n}(\bm{r}-\bm{\tau}_n-\bm{R})$, $n=1,2,...,8$ are localized at the Wannier centers
\begin{align}
\begin{split}
&\bm{\tau}_1=\bm{\tau}_2=\bm{\tau}_3=\bm{\tau}_{\text{AA}}=(0,0)^T, \\
& \bm{\tau}_4=\bm{\tau}_{\text{AB}}=(\bm{L}_1^m+\bm{L}_2^m)/3,\quad \bm{\tau}_5=\bm{\tau}_{\text{BA}}=-(\bm{L}_1^m+\bm{L}_2^m)/3,\\
& \bm{\tau}_6=\bm{\tau}_{\text{DW1}}=(\bm{L}_2^m-\bm{L}_1^m)/2, \quad \bm{\tau}_7=\bm{\tau}_{\text{DW2}}=-\bm{L}_2^m/2, \quad \bm{\tau}_8=\bm{\tau}_{\text{DW3}}=\bm{L}_1^m/2. 
\end{split}
\end{align}
The shape of these orbitals and the superlattice formed by them are shown in Fig. \ref{fig:8band}. In real-space the $p_{\pm}@$AA orbitals [Figs. \ref{fig:8band}(b)(c)] have the local symmetry
\begin{align}
\begin{split}
& C_{3z}\Phi_{\eta 1}(\bm{r}) = \omega\Phi_{\eta 1}(\bm{r}),\quad C_{3z}\Phi_{\eta 2}(\bm{r}) = \omega\Phi_{\eta 2}(\bm{r}),\quad C_{2x}\Phi_{\eta 1}(\bm{r}) = \Phi_{\eta 2}(\bm{r}),\quad C_{2x}\Phi_{\eta 2}(\bm{r}) = \Phi_{{\eta} 1}(\bm{r}),\\
& C_{2z}\mathcal{T}\Phi_{{\eta} 1}(\bm{r}) = \Phi_{{\eta} 2}(\bm{r})\mathcal{K},\quad C_{2z}\mathcal{T}\Phi_{{\eta} 2}(\bm{r}) = \Phi_{{\eta} 1}(\bm{r})\mathcal{K},
\end{split}
\end{align}
which is exactly the symmetry of ZLLs Eq. (\ref{eq:ZLL_repre_appendix}), so we still call them $f$ orbitals. The remaining six orbitals are $c$ orbitals which all have zero components at AA sites. The $s@$AA orbital [Fig. \ref{fig:8band}(d)] has the symmetry
\begin{align}
& C_{3z}\Phi_{\eta 3}(\bm{r})=\Phi_{\eta 3}(\bm{r}),\quad C_{2x}\Phi_{\eta 3}(\bm{r})=\Phi_{\eta 3}(\bm{r}),\quad C_{2z}\mathcal{T}\Phi_{\eta 3}(\bm{r})=\Phi_{\eta 3}(\bm{r})\mathcal{K}.
\end{align}
The $p_z @$AB and $p_z @$BA orbitals [Figs. \ref{fig:8band}(e)(f)] transform as
\begin{align}
\begin{split}
& C_{3z}\Phi_{\eta 4}(\bm{r}-\bm{\tau}_4)=\Phi_{\eta 4}(\bm{r}-\bm{\tau}_4+\bm{L}_1^m),\quad C_{3z}\Phi_{\eta 5}(\bm{r}-\bm{\tau}_5)=\Phi_{\eta 5}(\bm{r}-\bm{\tau}_5-\bm{L}_1^m),\quad\\
& C_{2x}\Phi_{\eta 4}(\bm{r}-\bm{\tau}_4)=\Phi_{\eta 4}(\bm{r}-\bm{\tau}_4),\quad C_{2x}\Phi_{\eta 5}(\bm{r}-\bm{\tau}_5)=\Phi_{\eta 5}(\bm{r}-\bm{\tau}_5),\quad\\
& C_{2z}\mathcal{T}\Phi_{\eta 4}(\bm{r}-\bm{\tau}_4) = \Phi_{\eta 5}(\bm{r}-\bm{\tau}_5)\mathcal{K},\quad C_{2z}\mathcal{T}\Phi_{\eta 5}(\bm{r}-\bm{\tau}_5) = \Phi_{\eta 4}(\bm{r}-\bm{\tau}_4)\mathcal{K}.
\end{split}
\end{align}
And the last three $c$ orbitals $s@$DWn [Figs. \ref{fig:8band}(g)(h)(i)] transform as
\begin{align}
\begin{split}
& C_{3z}\Phi_{\eta 6}(\bm{r}-\bm{\tau}_6)=\Phi_{\eta 7}(\bm{r}-\bm{\tau}_7),\quad C_{3z}\Phi_{\eta 7}(\bm{r}-\bm{\tau}_7)=\Phi_{\eta 8}(\bm{r}-\bm{\tau}_8),\quad C_{3z}\Phi_{\eta 8}(\bm{r}-\bm{\tau}_8)=\Phi_{\eta 6}(\bm{r}-\bm{\tau}_6),\quad\\
& C_{2x}\Phi_{\eta 6}(\bm{r}-\bm{\tau}_6)=\Phi_{\eta 6}(\bm{r}-\bm{\tau}_6+\bm{L}_2^m-\bm{L}_1^m),\quad C_{2x}\Phi_{\eta 7}(\bm{r}-\bm{\tau}_7)=\Phi_{\eta 8}(\bm{r}-\bm{\tau}_8+\bm{L}_1^m),\\
& C_{2x}\Phi_{\eta 8}(\bm{r}-\bm{\tau}_8)=\Phi_{\eta 7}(\bm{r}-\bm{\tau}_7-\bm{L}_2^m),\\
& C_{2z}\mathcal{T}\Phi_{\eta 6}(\bm{r}-\bm{\tau}_6)=\Phi_{\eta 6}(\bm{r}-\bm{\tau}_6+\bm{L}_2^m-\bm{L}_1^m)\mathcal{K},\quad C_{2z}\mathcal{T}\Phi_{\eta 7}(\bm{r}-\bm{\tau}_7)=\Phi_{\eta 7}(\bm{r}-\bm{\tau}_7-\bm{L}_2^m)\mathcal{K},\\
& C_{2z}\mathcal{T}\Phi_{\eta 8}(\bm{r}-\bm{\tau}_8)=\Phi_{\eta 8}(\bm{r}-\bm{\tau}_8+\bm{L}_1^m)\mathcal{K}.
\end{split}
\end{align}
The time reversal relates the two valleys by
\begin{align}
\mathcal{T}\Phi_{\eta 1}(\bm{r})=\Phi_{\bar{\eta} 2}(\bm{r})\mathcal{K},\quad \mathcal{T}\Phi_{\eta 2}(\bm{r})=\Phi_{\bar{\eta} 1}(\bm{r})\mathcal{K},\quad \mathcal{T}\Phi_{\eta n}(\bm{r}-\bm{\tau}_n)=\Phi_{\bar{\eta} n}(\bm{r}-\bm{\tau}_n)\mathcal{K} \text{ for } n\geq3. \label{eq:time_revsersal}
\end{align}

The $f$ orbitals ($p_{\pm}$@AA) are well-localized around AA centers, and all other six $c$ orbitals have zero components at AA centers $\bm{r}=0$. In $k$-space, the $f$ orbitals are mainly distributed in the mBZ edge of the flat bands ($94.43\%$) and the mBZ center of the nearest remote bands ($5.57\%$), see Fig. \ref{fig:H0_band}. 

\begin{figure}
\includegraphics[width=\textwidth]{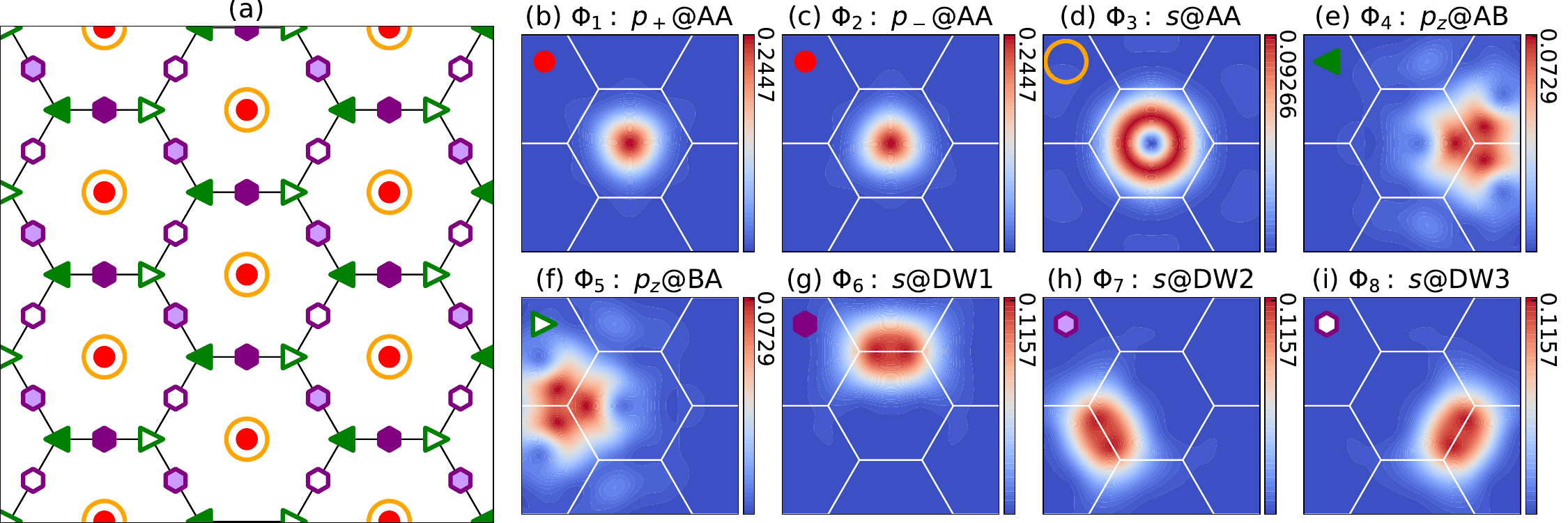}
\caption{\label{fig:8band} (a) The lattice configuration of the 8-band electron model. Different symbols represent different sublattices. Red dots: triangular lattice formed by the $p_{\pm}$@AA orbitals ($f$ orbital). Orange rings: triangular lattice formed by the $s$@AA orbitals. Green triangles: honeycomb lattice formed by the $p_{z}$@AB/BA orbitals. Purple hexagons: kagome lattice formed by the $s$@DWn ($n=1,2,3$) orbitals. The density distribution of these Wannier orbitals in the first supercell, $|\Phi_{n}(\bm{r}-\bm{\tau}_n)|^2=\sum_{l\alpha}|\Phi_{n}(\bm{r}-\bm{\tau}_n,l\alpha)|^2$, are shown in (b)-(i) for the $\eta=K$ valley, in the unit of $k_{\theta}^{2}$.}
\end{figure}

\begin{figure}
\includegraphics[width=\textwidth]{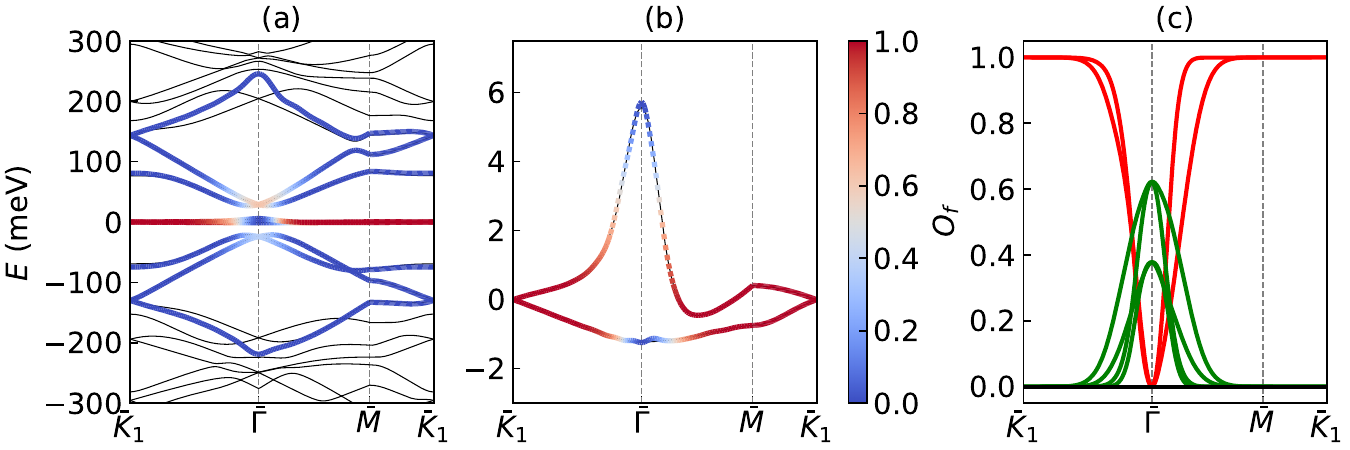}
\caption{\label{fig:H0_band} (a) Energy bands of the original TB model (black lines) and the fitted 8-band model (colored lines) of MATBG, projected onto the $K$ valley. (b) The snapshot of the flat bands. In (a) and (b) the color represents the ratio of $f$ electron components, which is also shown explicitly in (c) with the red, green, and black lines indicating the overlap between $f$ orbitals and the two flat bands, four nearest remote bands, and two farthest remote bands, respectively.}
\end{figure}

We obtain the model following the multistep Wannierization given in Ref. \cite{19prr_8orbital_model}. In the initial step, we use the analytical ZLLs Eq. (\ref{eq:ZLL_WF_matrix}) as the trial states of $f$ orbitals, and use Gaussian functions for the other six $c$ orbitals. The projection parameters are carefully tuned so that the flat bands have $O_f\approx94.34\%$ components of $f$ orbitals. During the Wannierization the integration is always calculated in $k$-space. In numerics the Wannier orbitals are written conveniently using the original plane wave basis,
\begin{align}
|\Phi_{\eta n}(\bm{r}-\bm{\tau}_n-\bm{R})\rangle = \frac{1}{\sqrt{N_m}}\sum_{\bar{\bm{k}}}\sum_{\bm{G}l\alpha}|\eta l\alpha,\bar{\bm{k}}+\bm{G}\rangle e^{-i\bar{\bm{k}}\cdot \bm{R}}\mathcal{W}_{\bm{G}l\alpha,n}^{\eta}(\bar{\bm{k}}). \label{eq:WF_plwv}
\end{align}
The matrix $\mathcal{W}$ is saved with a $27\times 27$ meshgrid of $\bar{\bm{k}}$ points and $61$ $\bm{G}$ vectors for further calculations of the hopping, Coulomb interaction, and the EPC matrices.

Analytically it is more convenient to use Gaussian orbitals (\ref{eq:ZLL_WF_matrix}), (\ref{eq:Wannier_WF_appendix}). The three parameters $\alpha_1$ ($\alpha_2^2=1-\alpha_1^2$), $l_1$ and $l_2$ can be obtained by fitting the numerical Wannier functions. The resultant values are $\alpha_1\approx 0.8096$, $l_1\approx 2.9033$ nm and $l_2\approx 3.6484$ nm, with an optimal overlap $88.45\%$ with the Wannier $f$ orbitals. The Gaussian Wannier functions will help obtain analytical results about moir\'{e} phonons in Appendix \ref{Appendix_epc_continuum}.

\section{Phonon basis, frozen-phonon method, and symmetrization \label{appendix_phonon}}
\subsection{Folded monolayer optical phonon basis}
This part serves as the review of the monolayer optical phonons, which is the starting point of this study. In the $K(K')$ valley we focus on the iTO mode ($K$-phonons), and in the $\Gamma$ valley we focus on the degenerate iTO and iLO modes ($\Gamma$-phonons). First we adopt the Einstein phonon approximation which assumes the constant phonon frequencies $\omega_K(\bm{q})= 150$ meV and $\omega_{\Gamma}(\bm{q})=$ 180 meV for $K$-phonons and $\Gamma$-phonons, respectively. The phase of the eigenmode vector $\bm{\epsilon}^{b}(\bm{q})$ in Eq. (\ref{eq:u_field}) is cumbersome as $\bm{q}$ changes \cite{07prb_constant_plrz}. Since we are only interested in a limited region near $K(K')$ or $\Gamma$, the Hilbert space spanned by $\bm{\epsilon}^{b}(\bm{q})$ should remain stable at various $\bm{q}$. This expectation coincides with the spirit of $k\cdot p$ theory. So we fix the monolayer $\bm{\epsilon}^{K,K'}(\bm{q})$ and $\bm{\epsilon}^{x,y}(\bm{q})$ as (in accordance with Appendix \ref{Appendix_lattice})
\begin{subequations}
\begin{align}
\bm{\epsilon}^{K}_A= \left( \frac{i}{2},-\frac{1}{2} \right)^T, \quad \bm{\epsilon}^{K}_B= \left( \frac{i}{2},\frac{1}{2} \right)^T,\quad 
\bm{\epsilon}^{K'}_A= \left( -\frac{i}{2},-\frac{1}{2} \right)^T, \quad \bm{\epsilon}^{K'}_B= \left( -\frac{i}{2},\frac{1}{2} \right)^T, 
\label{eq:eigenmode_iTO_vector} \\
\bm{\epsilon}^{x}_A= \left( -\frac{1}{\sqrt{2}},0\right)^T, \quad \bm{\epsilon}^{x}_B= \left( \frac{1}{\sqrt{2}}, 0\right)^T,\quad
\bm{\epsilon}^{y}_A= \left( 0,-\frac{1}{\sqrt{2}} \right)^T, \quad \bm{\epsilon}^{y}_B= \left( 0,\frac{1}{\sqrt{2}} \right)^T,
\label{eq:eigenmode_iLO_vector}
\end{align}
\end{subequations}
which are eigenmode vectors exactly at high-symmetry points $K(K')$ and $\Gamma$, i.e., eigenmodes at $\bm{q}=0$. To check the validity of this approximation we numerically calculate the overlaps ($\langle \bm{\epsilon}^1|\bm{\epsilon}^2 \rangle=\bm{\epsilon}^{1*}_A\cdot \bm{\epsilon}^2_A+\bm{\epsilon}^{1*}_B\cdot \bm{\epsilon}^2_B$)
\begin{align}
O^{\eta}_K(\bm{q})=|\langle \bm{\epsilon}^{\eta}(\bm{q})|\bm{\epsilon}^{\eta}\rangle|^2,\quad
O^{\mu}_{\Gamma}(\bm{q})=\frac{1}{2}\sum_{\mu'=x,y}|\langle \bm{\epsilon}^{\mu'}(\bm{q})|\bm{\epsilon}^{\mu}\rangle|^2,
\end{align}
using the phonon model in Refs. \cite{15prl_chiral_phonon,23_liu_ek_phonon}. Within the cutoff $\bm{q}\lesssim 4\sqrt{3}k_{\theta}\approx 0.13 K$, $O^{\eta}_K(\bm{q})$ and $O^{\mu}_{\Gamma}(\bm{q})$ are found always larger than $95.70\%$ and $99.96\%$, respectively. Therefore, neglecting the $\bm{q}$-dependence is an excellent approximation. Note that the $A_1$ and $B_1$ displacements shown in Fig. \ref{fig:monolayer_phonon}(b) are real superposition of Eq. (\ref{eq:eigenmode_iTO_vector}): $\bm{\epsilon}^{A_1}=(\bm{\epsilon}^{K}+\bm{\epsilon}^{K'})/\sqrt{2}$, $\bm{\epsilon}^{B_1}=i(\bm{\epsilon}^{K}-\bm{\epsilon}^{K'})/\sqrt{2}$. In the following we keep using the valley basis Eq. (\ref{eq:eigenmode_iTO_vector}) rather than the $A_1$, $B_1$ basis (they will be naturally recovered once the displacement fields are symmetrized).

In the twisted bilayers, the eigenmodes in the $l$-th layer will be rotated to $\bm{\epsilon}^b_{l\alpha}=R[(-1)^l\theta/2]\bm{\epsilon}^b_{\alpha}$. The displacement fields of $K$-phonons and $\Gamma$-phonons have the form of plane waves,
\begin{subequations}
\begin{align}
\bm{u}_{l\eta\bm{q}}(\bm{r})&=\frac{1}{\sqrt{N_m N_a}}\sum_{\bm{R}_l \alpha}e^{i(\bm{q}+\bm{K}_l^{\eta})\cdot \bm{r}}\delta_{\bm{r},\bm{R}_l+\bm{\tau}_{l\alpha}}\bm{\epsilon}^{\eta}_{l\alpha}, \quad \eta = K,K', \label{eq:u_field_K} \\
\bm{u}_{l\mu\bm{q}}(\bm{r})&=\frac{1}{\sqrt{N_m N_a}}\sum_{\bm{R}_l \alpha}e^{i\bm{q}\cdot \bm{r}}\delta_{\bm{r},\bm{R}_l+\bm{\tau}_{l\alpha}}\bm{\epsilon}^{\mu}_{l\alpha},\quad \mu = x,y. \label{eq:u_field_G} 
\end{align}
\end{subequations}
Here $\delta$ takes value $1$ if its indexes coincide or $0$ otherwise. The summation is over all atomic lattice $\bm{R}_l$ in a specific layer $l$ (the distortion happens only in one monolayer). The above distortion fields can be viewed as ``plane wave basis'' for phonons, and they are orthonormal in the sense ($b_{1,2}=K,K',x,y$)
\begin{align}
\langle \bm{u}_{l_1b_1\bm{q}_1}|\bm{u}_{l_2 b_2 \bm{q}_2}\rangle = \sum_{l'\bm{R}_{l'}\alpha'}\bm{u}^{*}_{l_1 b_1 \bm{q}_1}(\bm{R}_{l'}+\bm{\tau}_{l'\alpha'})\cdot \bm{u}_{l_2 b_2 \bm{q}_2}(\bm{R}_{l'}+\bm{\tau}_{l'\alpha'}) =\delta_{\bm{q}_1\bm{q}_2}\delta_{l_1l_2}\delta_{b_1b_2}.\label{eq:phonon_normalization}
\end{align}
Usually we decompose $\bm{q}=\bar{\bm{q}}+\bm{Q}_l^{\eta}$ in Eq. (\ref{eq:u_field_K}) and $\bm{q}=\bar{\bm{q}}+\bm{G}$ in Eq. (\ref{eq:u_field_G}) to represent that these fields are folded in the supercell. Here $\bar{\bm{q}}\in$mBZ, $\bm{Q}_l^{\eta}$ is the vector connecting $\bm{K}_l^{\eta}$ to its nearby moir\'{e} $\bar{\bm{\Gamma}}$ points, i.e., $\bm{Q}_l^{\eta}+\bm{K}_l$ and $\bm{G}$ are linear combinations of $\bm{G}_1^m$ and $\bm{G}_2^m$.

Now we identify the symmetry of these modes. For a generic distortion field $\bm{u}$, under rotation $g$ it will change to $\bm{u}^g=g\bm{u}$, with $\bm{u}^g(\bm{r})=g\bm{u}(g^{-1}\bm{r})$, while time symmetry $\mathcal{T}$ takes the complex conjugation. For example, 
\begin{align}
\begin{split}
(C_{3z}\bm{u}_{lx\bm{q}})(\bm{r}) =& \frac{1}{\sqrt{N_m N_a}}\sum_{\bm{R}_l \alpha}e^{i(C_{3z}\bm{q})\cdot \bm{r}}\delta_{\bm{r},C_{3z}(\bm{R}_l+\bm{\tau}_{l\alpha})}C_{3z}\bm{\epsilon}^{x}_{l\alpha}\\
=&\frac{1}{\sqrt{N_m N_a}}\sum_{\bm{R}_l \alpha}e^{i(C_{3z}\bm{q})\cdot \bm{r}}\delta_{\bm{r},\bm{R}_l+\bm{\tau}_{l\alpha}}\left(-\frac{1}{2}\bm{\epsilon}^{x}_{l\alpha}+\frac{\sqrt{3}}{2}\bm{\epsilon}^{y}_{l\alpha}\right)\\
=&-\frac{1}{2}\bm{u}_{lx,C_{3z}\bm{q}}+\frac{\sqrt{3}}{2}\bm{u}_{ly,C_{3z}\bm{q}},
\end{split}
\end{align}
where we have used $C_{3z}(\bm{R}_l+\bm{\tau}_{l\alpha})=\bm{R}'_{l}+\bm{\tau}_{l\alpha}$ and the rotation properties of $\bm{\epsilon}_{l\alpha}^{x,y}$. We can directly check that
\begin{align}
C_{3z}\bm{u}_{l\eta\bm{q}}=\bm{u}_{l\eta,C_{3z}\bm{q}}, \quad C_{2x}\bm{u}_{l\eta\bm{q}}=\bm{u}_{\bar{l}\eta,C_{2x}\bm{q}}, \quad C_{2z}\bm{u}_{l\eta\bm{q}}=\bm{u}_{l\bar{\eta},-\bm{q}}, \quad \mathcal{T}\bm{u}_{l\eta\bm{q}}=\bm{u}_{l\bar{\eta},-\bm{q}},\label{eq:u_symmetry_K}
\end{align}
and for $\Gamma$-phonons
\begin{align}
\begin{split}
C_{3z}\bm{u}_{lx\bm{q}}=-\frac{1}{2}\bm{u}_{lx,C_{3z}\bm{q}}+\frac{\sqrt{3}}{2}\bm{u}_{ly,C_{3z}\bm{q}},&\quad C_{3z}\bm{u}_{ly\bm{q}}=-\frac{\sqrt{3}}{2}\bm{u}_{lx,C_{3z}\bm{q}}-\frac{1}{2}\bm{u}_{ly,C_{3z}\bm{q}}, \\
C_{2x}\bm{u}_{lx\bm{q}}=-\bm{u}_{\bar{l}x,C_{2x}\bm{q}}, \quad C_{2x}\bm{u}_{ly\bm{q}}=\bm{u}_{\bar{l}y,C_{2x}\bm{q}}, &\quad 
C_{2z}\bm{u}_{lx(y)\bm{q}}=\bm{u}_{lx(y),-\bm{q}}, \quad \mathcal{T}\bm{u}_{lx(y)\bm{q}}=\bm{u}_{lx(y),-\bm{q}}.
\end{split}
\label{eq:u_symmetry_Gamma}
\end{align}
It is reasonable that $C_{3z}$, $C_{2z}$ and $\mathcal{T}$ do not switch the layers. These operator are the same as the monolayer ones. In contrast, $C_{2x}$ interchanges the layers which is the only symmetry different from monolayer. These phonon basis are in general complex and inconvenient to use in frozen-phonon scheme. They will be symmetrized in the following using Eqs. (\ref{eq:u_symmetry_K}) and (\ref{eq:u_symmetry_Gamma}).

\subsection{Frozen-phonon method, normalization of phonon modes \label{Appendix_frozen_phonon}}
In this appendix we show the details, especially the normalization of the phonons in the frozen-phonon scheme. We start with a generic phonon mode $b$ with momentum $\bm{q}\in$aBZ, which has the displacement field $\bm{u}_{b\bm{q}}(\bm{r})$ with norm $1$ [Eq. (\ref{eq:phonon_normalization})]. Note that $\bm{u}_{b\bm{q}}$ defined here can represent both the $K$-phonons ($|\bm{q}|\approx |K|$) and $\Gamma$-phonons ($|\bm{q}|\approx 0$), i.e., the valley index is implicitly contained by the momentum, which is slightly different from Eq. (\ref{eq:u_field_K}). Now we introduce the generalized coordinate $\mathcal{Q}_{b\bm{q}}$ (dimension: length) to quantify the distortion strength. The electron Hamiltonian with the tiny frozen lattice distortion $\mathcal{Q}_{b\bm{q}}\bm{u}_{b\bm{q}}$ can be expanded as power series of $\mathcal{Q}_{b\bm{q}}$
\begin{align}
H_0(\mathcal{Q}_{b\bm{q}}) = H_0(0) + \mathcal{Q}_{b\bm{q}} H_{b\bm{q}}'+O(\mathcal{Q}_{b\bm{q}}^2),
\end{align}
where $H_{b\bm{q}}'=\partial_{\mathcal{Q}_{b\bm{q}}} H_0(\mathcal{Q}_{b\bm{q}})|_{\mathcal{Q}_{b\bm{q}}=0}$ is the gradient. The EPC Hamiltonian is simply the quantization of $\mathcal{Q}_{b\bm{q}}$ in the linear term above, by introducing the bosonic operators $a_{b\bm{q}}$
\begin{align}
\mathcal{Q}_{b\bm{q}} = \sqrt{\frac{\hbar}{2M_c\omega_b}}(a_{b\bm{q}}+a_{b,-\bm{q}}^{\dagger}), \quad
\mathcal{P}_{b\bm{q}} = i\sqrt{\frac{\hbar\omega_b M_c}{2}}(a_{b\bm{q}}^{\dagger}-a_{b,-\bm{q}}), \label{eq:boson_operator_definition}
\end{align}
where $M_c$ is the carbon atom mass, $\omega_b$ is the phonon frequency, and $\mathcal{P}_{b\bm{q}}$ is the generalized momenta canonically conjugated to $\mathcal{Q}_{b\bm{q}}$. Now the EPC Hamiltonian reads
\begin{align}
H_{\text{epc},b\bm{q}}=&\frac{1}{\sqrt{2N_m}}h_{b\bm{q}}(a_{b\bm{q}}+a^{\dagger}_{b,-\bm{q}}),\\
h_{b\bm{q}}=\sqrt{\frac{N_m\hbar}{M_c\omega_b}}\frac{\partial H_0(\mathcal{Q}_{b\bm{q}})}{\partial \mathcal{Q}_{b\bm{q}}}\bigg|_{\mathcal{Q}_{b\bm{q}}=0}=&\sqrt{\frac{N_m\hbar}{M_c\omega_b}}\lim_{\mathcal{Q}_{b\bm{q}}\rightarrow0}\frac{H_0(\mathcal{Q}_{b\bm{q}})-H_0(0)}{\mathcal{Q}_{b\bm{q}}}\approx H_0\left(\sqrt{\frac{N_m\hbar}{M_c\omega_b}}\right)-H_0(0). \label{eq:normalize_condition_u}
\end{align}
In other words, the EPC matrix $h_{b\bm{q}}$ is just the variance of the Hamiltonian after applying the distortion $\tilde{\bm{u}}_{b\bm{q}}=\sqrt{N_m}l_b\bm{u}_{b\bm{q}}$, whose mean distortion length per supercell is frozen as the phonon length $l_b$,
\begin{align}
\left|\tilde{\bm{u}}_{b\bm{q}}(\bm{r})\right|=\sqrt{\frac{1}{N_m}\sum_{l\bm{R}_l \alpha}\left|\tilde{\bm{u}}_{b\bm{q}}(\bm{R}_l+\bm{\tau}_{l\alpha})\right|^2}=l_b, \quad l_b=\sqrt{\frac{\hbar}{M_c\omega_b}}.\label{eq:phonon_mean_dist}
\end{align}
Using $\omega_K=150$ meV and $\omega_{\Gamma}=180$ meV, we get $l_K\approx 48.17$ m$\text{\AA}$ and $l_{\Gamma}\approx 43.97$ m\AA, which safely locate in the parameter regime where $H_0(\bm{u})\propto \bm{u}$ and Eq. (\ref{eq:normalize_condition_u}) holds. The above formula is also valid for multiple phonons, and the complete EPC Hamiltonian Eq. (\ref{eq:H_epc_full}) is the summation over all $\bm{q}$ and branches $b$ of interest.

\subsection{Symmetrization of folded phonons \label{appendix_frozen_phonon_symmetrized}}
As illustrated in the main text, to an approximation we can focus on phonon basis preserving the moir\'{e} translation symmetry. Then the construction of moir\'{e} phonons reduces to determining the periodic part of the ``Bloch phonons'' Eq. (\ref{eq:moire_phonon_peridic}). Such periodic fields for $K$-phonons in general take the form
\begin{align}
\bm{u}(\bm{r})=\frac{1}{\lambda}\sum_{l\eta\bm{Q}_l^{\eta}}\lambda_{l\eta}(\bm{Q}_l^{\eta})\bm{u}_{l\eta\bm{Q}_l^{\eta}}(\bm{r}),
\end{align}
where $\bm{Q}_l^{\eta}$ are vectors so that $\bm{K}_l^{\eta}+\bm{Q}_l^{\eta}$ locates at $\bar{\bm{\Gamma}}$ ($\equiv n_1\bm{G}_1^m+n_2\bm{G}_2^m$), see Fig. \ref{fig:Q_G}(a). In the main text, we state that for the mode with a specific symmetry, the coefficient $\lambda_{l\eta}(\bm{Q}_l^{\eta})$ is just the Pauli matrices coefficient in the decomposition of the EPC matrix. However, this strategy cannot be directly imposed in the frozen-phonon scheme because such plane wave phonons will in general not be real (this is doable using a simplified layer-decoupled EPC model, see Appendix \ref{Appendix_epc_continuum}). Instead, we directly superpose these plane wave basis to form real and symmetric modes.

From Eq. (\ref{eq:u_symmetry_K}) we know twelve $\bm{u}_{l\eta\bm{Q}_l^{\eta}}$ form a $D_6$-invariant subspace, 3 from each layer and valley. The target is to decompose this subspace into (real) irreducible representations (irreps) of $D_6$. Here we directly give the results. The 12-dimensional representation space is decomposed as
\begin{align}
A_1 \oplus B_1 \oplus A_2 \oplus B_2\oplus 2E_1\oplus 2E_2.
\end{align}
The coefficients $\lambda(\bm{Q}_{l}^{\eta})$ are proportional to the unitary matrix that rotates the original representation to the decoupled irreps. In this case they can be expressed as $s$-type functions $\lambda\sim f(Q)$ ($Q$ is the common norm of the twelve vectors $\bm{Q}_l^{\eta}$) or $p$-type functions $\lambda (\bm{Q}_l^{\eta})\sim (\bm{Q}^{\eta}_l)_{x,y}f(Q)$. The twelve $\bm{Q}_l^{\eta}$ can be grouped into four sets. The first set contains three vectors (connected through $C_{3z}$) in layer $l=1$ and valley $\eta=K$, which we denote by $\{\bm{Q}\}_{1K}$. The $C_{2z}$, $C_{2x}$, and $C_{2z}C_{2x}$ rotations of this set yield another three sets living in other valley and layer, denoted by $\{\bm{Q}\}_{1K'}$, $\{\bm{Q}\}_{2K}$, and $\{\bm{Q}\}_{2K'}$, respectively. Then the twelve real irreps can be expressed as (not normalized)
\begin{subequations}
\begin{align}
& A_1:\quad \sum\nolimits_{\{\bm{Q}\}_{1K}}\bm{u}_{1K\bm{Q}}+\sum\nolimits_{\{\bm{Q}\}_{1K'}}\bm{u}_{1K'\bm{Q}}
            +(1\leftrightarrow 2), \label{eq:uK_a}\\
& A_2:\quad \sum\nolimits_{\{\bm{Q}\}_{1K}}\bm{u}_{1K\bm{Q}}+\sum\nolimits_{\{\bm{Q}\}_{1K'}}\bm{u}_{1K'\bm{Q}}
             -(1\leftrightarrow 2), \label{eq:uK_b}\\
& B_1:\quad i\sum\nolimits_{\{\bm{Q}\}_{1K}}\bm{u}_{1K\bm{Q}}-i\sum\nolimits_{\{\bm{Q}\}_{1K'}}\bm{u}_{1K'\bm{Q}}
              +(1\leftrightarrow 2), \label{eq:uK_c}\\
& B_2:\quad i\sum\nolimits_{\{\bm{Q}\}_{1K}}\bm{u}_{1K\bm{Q}}-i\sum\nolimits_{\{\bm{Q}\}_{1K'}}\bm{u}_{1K'\bm{Q}}
              -(1\leftrightarrow 2), \label{eq:uK_d}\\
& E_1:\quad \begin{cases} 
            i\sum\nolimits_{\{\bm{Q}\}_{1K}}Q_x\bm{u}_{1K\bm{Q}}+i\sum\nolimits_{\{\bm{Q}\}_{1K'}}Q_x\bm{u}_{1K'\bm{Q}}
            +(1\leftrightarrow 2), \\
            i\sum\nolimits_{\{\bm{Q}\}_{1K}}Q_y\bm{u}_{1K\bm{Q}}+i\sum\nolimits_{\{\bm{Q}\}_{1K'}}Q_y\bm{u}_{1K'\bm{Q}}
            +(1\leftrightarrow 2),
            \end{cases} \label{eq:uK_e}\\
& E_1:\quad \begin{cases} 
            i\sum\nolimits_{\{\bm{Q}\}_{1K}}Q_y\bm{u}_{1K\bm{Q}}+i\sum\nolimits_{\{\bm{Q}\}_{1K'}}Q_y\bm{u}_{1K'\bm{Q}}
            -(1\leftrightarrow 2), \\
            -i\sum\nolimits_{\{\bm{Q}\}_{1K}}Q_x\bm{u}_{1K\bm{Q}}-i\sum\nolimits_{\{\bm{Q}\}_{1K'}}Q_x\bm{u}_{1K'\bm{Q}}
            -(1\leftrightarrow 2),
            \end{cases} \label{eq:uK_f}\\
& E_2:\quad \begin{cases} 
            \sum\nolimits_{\{\bm{Q}\}_{1K}}Q_x\bm{u}_{1K\bm{Q}}-\sum\nolimits_{\{\bm{Q}\}_{1K'}}Q_x\bm{u}_{1K'\bm{Q}}
            +(1\leftrightarrow 2), \\
            \sum\nolimits_{\{\bm{Q}\}_{1K}}Q_y\bm{u}_{1K\bm{Q}}-\sum\nolimits_{\{\bm{Q}\}_{1K'}}Q_y\bm{u}_{1K'\bm{Q}}
            +(1\leftrightarrow 2),
            \end{cases} \label{eq:uK_g}\\
& E_2:\quad \begin{cases} 
            \sum\nolimits_{\{\bm{Q}\}_{1K}}Q_y\bm{u}_{1K\bm{Q}}-\sum\nolimits_{\{\bm{Q}\}_{1K'}}Q_y\bm{u}_{1K'\bm{Q}}
            -(1\leftrightarrow 2), \\
            -\sum\nolimits_{\{\bm{Q}\}_{1K}}Q_x\bm{u}_{1K\bm{Q}}+\sum\nolimits_{\{\bm{Q}\}_{1K'}}Q_x\bm{u}_{1K'\bm{Q}}
            -(1\leftrightarrow 2).
            \end{cases} \label{eq:uK_h}
\end{align}
\end{subequations}
It is easy to check the orthogonality of these basis, by noticing that $Q_x+Q'_x+Q''_x=Q_y+Q'_y+Q''_y=Q_xQ_y+Q'_xQ'_y+Q''_xQ''_y=0$ is always true for any $\bm{Q}$, $\bm{Q}'$, and $\bm{Q}''$ connected by $C_{3z}$: $\bm{Q}$, $\bm{Q}'=C_{3z}\bm{Q}$, and $\bm{Q}''=C_{3z}^2\bm{Q}$. The above fields are also real, $\mathcal{T}\bm{u}=\bm{u}$, thus can be imposed into the TB program after normalization Eq. (\ref{eq:phonon_mean_dist}). 

\begin{figure*}
\includegraphics[width=0.7\textwidth]{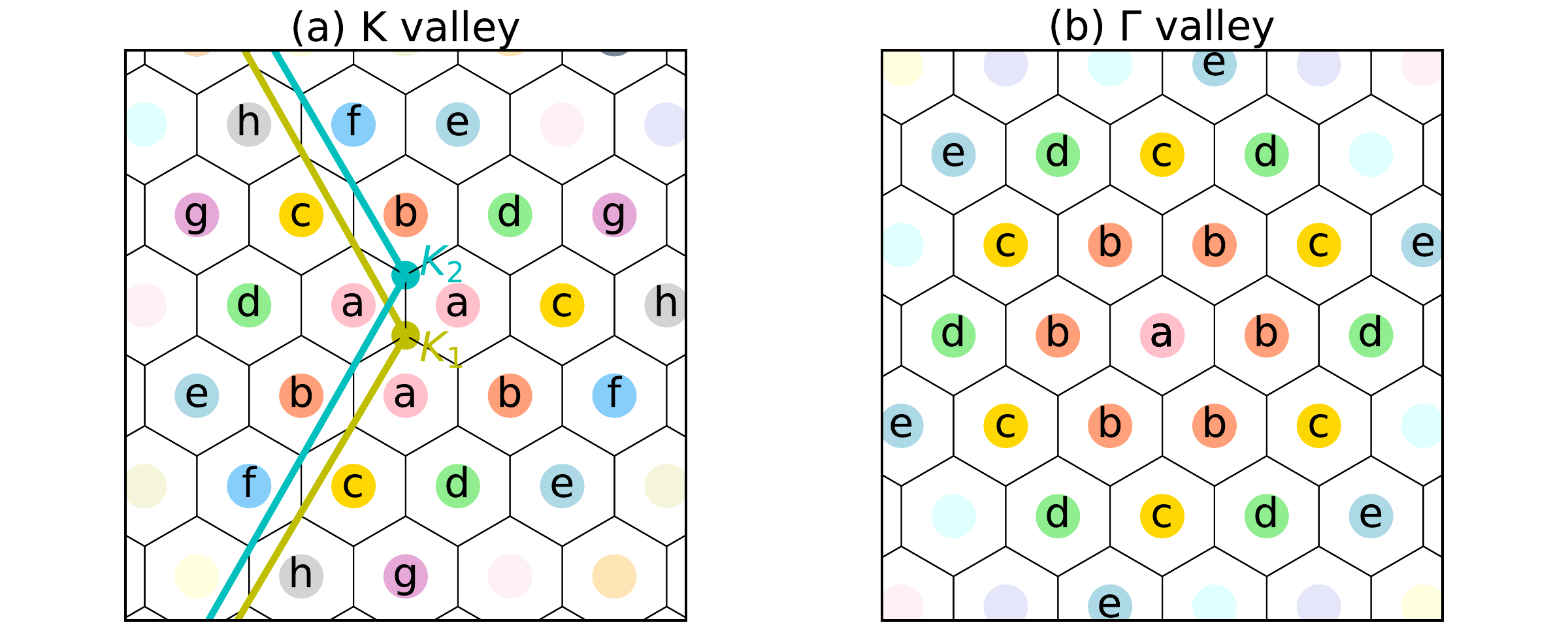}% Here is how to import EPS art
\caption{\label{fig:Q_G}
(a) The colored dots represent the commensurate momentum $\bm{K}_1+\bm{Q}_1^{K}$ in the $K$ valley of the top layer ($l=1$). The green and blue lines show the aBZ boundary of top and bottom layers. (b) The colored dots represent the commensurate $\bm{G}$ in the $\Gamma$ valley of the top layer. In each plot, dots with the same color are connected by $D_6$ rotations, and will be symmetrized together (with the partners in another valley/layer). We label the leading modes using symbols `a', `b', etc., and show the corresponding frozen-phonon bands for the $A_1$, $B_1$ modes in $K$ valley in Fig. \ref{fig:band_A1_Q}, and $E_2$ mode in $\Gamma$ valley in Fig. \ref{fig:band_E2_G}.}
\end{figure*}

\begin{figure*}
\includegraphics[width=0.9\textwidth]{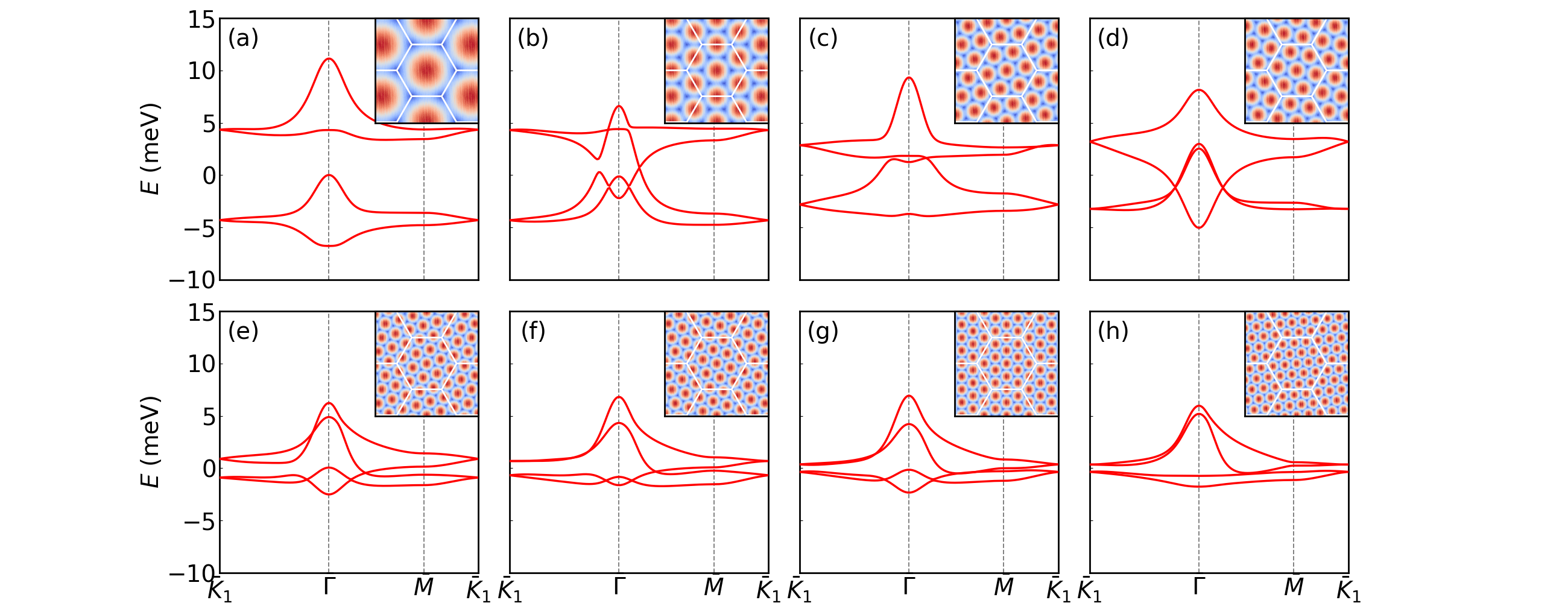}% Here is how to import EPS art
\caption{\label{fig:band_A1_Q}
The frozen-phonon flat bands with the commensurate ($K$-phonon) $A_1$ [Eq. (\ref{eq:uK_a})] or $B_1$ [Eq. (\ref{eq:uK_c})] distortion fields. From (a) to (h), as the distortion field (the inset shows its strength envelope on the top layer) becomes more delocalized, the flat bands gradually returns to its undistorted shape. The first order mode in (a) is qualitatively the same as the moir\'{e} phonon found in Refs. \cite{20epjp_JT_BM_Fabrizio,22prb_Kekule_Fabrizio}.}
\end{figure*}

\begin{figure*}
\includegraphics[width=0.9\textwidth]{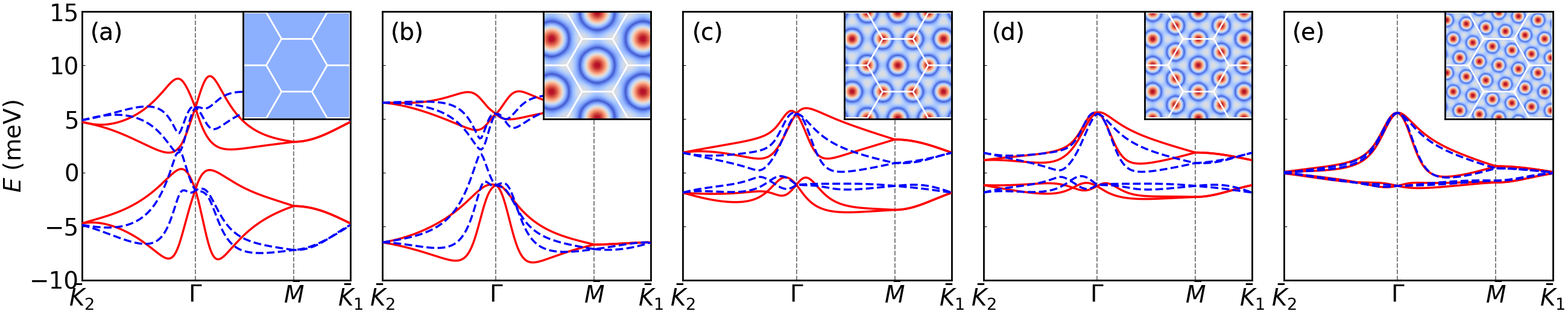}% Here is how to import EPS art
\caption{\label{fig:band_E2_G}
The frozen-phonon flat bands with the commensurate ($\Gamma$-phonon) $E_2$ distortions [Eq. (\ref{eq:uG_n})], in the $x$ direction (red solid lines) and $y$ direction (blue dashed lines). The inset shows the corresponding envelope of the distortion strength on the top layer. The anisotropic correction of $d$-type distortions Eq. (\ref{eq:uG_p}) are not included here.}
\end{figure*}

Since the distortion fields input are symmetrized, the resulting onsite EPC matrices directly shape like $\lambda_{ij}(Q)\tau_{i}\sigma_j$ once projected onto the $f$ orbitals. Numerically we find only the $A_1(\tau_x)$ mode Eq. (\ref{eq:uK_a}), $B_1(\tau_y)$ mode Eq. (\ref{eq:uK_c}), $E_1(\tau_y\sigma_x,\tau_y\sigma_y)$ mode Eq. (\ref{eq:uK_e}), and $E_2(\tau_x\sigma_x,\tau_x\sigma_y)$ mode Eq. (\ref{eq:uK_g}) will strongly couple with $f$ orbitals. These six modes are the moir\'{e} $K$-phonons shown in the main text, which has coupling strength two orders larger than the remaining modes. As an example, in Fig. \ref{fig:band_A1_Q} we show the flat bands splitted by $A_1$ fields ($B_1$ fields give exactly the same bands), when vectors $\bm{Q}_l^{\eta}$ go away from $\bm{K}_l^{\eta}$. We also observe that, all these six moir\'{e} $K$-modes have layer-even displacements so that atoms in the two layers have in-phase local oscillations.

For $\Gamma$-phonons, the Bloch distortion field is the superposition of Eq. (\ref{eq:u_field_G})
\begin{align}
\bm{u}(\bm{r})=\frac{1}{\lambda}\sum_{l\mu\bm{G}}\lambda_{l\mu}(\bm{G})\bm{u}_{l\mu\bm{G}}(\bm{r}),
\end{align}
where the summation is over $\mu=x,y$ and $\bm{G}=n_1\bm{G}_1^m+n_2\bm{G}_2^m$. Eq.(\ref{eq:u_symmetry_Gamma}) indicates that this time the $D_6$-invariant subspace contains $24$ basis carrying $12$ different $\bm{G}$ vectors ($\bm{G}=0$ is an exception): in each layer, there are six $(x,y)$ pairs whose momentum $\bm{G}$ are connected through $C_{6z}=C^{-1}_{3z}C_{2z}$. The decomposed irreps are doubled as well,
\begin{align}
2A_1 \oplus 2A_2 \oplus 2B_1 \oplus 2B_2 \oplus 4E_1 \oplus 4E_2.
\end{align}
Similar to $K$-phonons, we group the $\bm{G}$ vectors by four sets, $\{\bm{G}\}_{1+}$, $\{\bm{G}\}_{1-}$, $\{\bm{G}\}_{2+}$, $\{\bm{G}\}_{2-}$, where $\{\bm{G}\}_{1+}$ contains three $\bm{G}$ vectors connected by $C_{3z}$ in the layer $l=1$, and the other three sets are rotated by $C_{2z}$, $C_{2x}$, $C_{2z}C_{2x}$, respectively. Unlike valley index of $K$-phonons, here we can randomly label ``$\pm$'' on the two $C_{3z}$-invariant sets (different choice only results in a phase of modes after symmetrization). Using these notations the 24 symmetrized modes can be written as (real, but not normalized)
\begin{subequations}
\begin{align}
& A_1:\quad (\sum\nolimits_{\{\bm{G}\}_{1+}} -\sum\nolimits_{\{\bm{G}\}_{1-}} )\left(G_x\bm{u}_{1x\bm{G}}+G_y\bm{u}_{1y\bm{G}}\right)
            -(1\leftrightarrow 2), \label{eq:uG_a}\\
& A_1:\quad (\sum\nolimits_{\{\bm{G}\}_{1+}} -\sum\nolimits_{\{\bm{G}\}_{1-}} )\left(G_y\bm{u}_{1x\bm{G}}-G_x\bm{u}_{1y\bm{G}}\right)
            +(1\leftrightarrow 2), \label{eq:uG_b}\\
& A_2:\quad (\sum\nolimits_{\{\bm{G}\}_{1+}} -\sum\nolimits_{\{\bm{G}\}_{1-}} )\left(G_x\bm{u}_{1x\bm{G}}+G_y\bm{u}_{1y\bm{G}}\right)
            +(1\leftrightarrow 2), \label{eq:uG_c}\\
& A_2:\quad (\sum\nolimits_{\{\bm{G}\}_{1+}} -\sum\nolimits_{\{\bm{G}\}_{1-}} )\left(G_y\bm{u}_{1x\bm{G}}-G_x\bm{u}_{1y\bm{G}}\right)
            -(1\leftrightarrow 2), \label{eq:uG_d}\\
& B_1:\quad i(\sum\nolimits_{\{\bm{G}\}_{1+}} +\sum\nolimits_{\{\bm{G}\}_{1-}} )\left(G_x\bm{u}_{1x\bm{G}}+G_y\bm{u}_{1y\bm{G}}\right)
            -(1\leftrightarrow 2), \label{eq:uG_e}\\
& B_1:\quad i(\sum\nolimits_{\{\bm{G}\}_{1+}} +\sum\nolimits_{\{\bm{G}\}_{1-}} )\left(G_y\bm{u}_{1x\bm{G}}-G_x\bm{u}_{1y\bm{G}}\right)
            +(1\leftrightarrow 2), \label{eq:uG_f}\\
& B_2:\quad i(\sum\nolimits_{\{\bm{G}\}_{1+}} +\sum\nolimits_{\{\bm{G}\}_{1-}} )\left(G_x\bm{u}_{1x\bm{G}}+G_y\bm{u}_{1y\bm{G}}\right)
            +(1\leftrightarrow 2), \label{eq:uG_g}\\
& B_2:\quad i(\sum\nolimits_{\{\bm{G}\}_{1+}} +\sum\nolimits_{\{\bm{G}\}_{1-}} )\left(-G_y\bm{u}_{1x\bm{G}}+G_x\bm{u}_{1y\bm{G}}\right)
            -(1\leftrightarrow 2), \label{eq:uG_h}\\
& E_1:\quad \begin{cases} 
            i(\sum\nolimits_{\{\bm{G}\}_{1+}}-\sum\nolimits_{\{\bm{G}\}_{1-}})\bm{u}_{1y\bm{G}}
            +(1\leftrightarrow 2), \\
            i(\sum\nolimits_{\{\bm{G}\}_{1+}}-\sum\nolimits_{\{\bm{G}\}_{1-}})(-\bm{u}_{1x\bm{G}})
            +(1\leftrightarrow 2),
            \end{cases} \label{eq:uG_i}\\
& E_1:\quad \begin{cases} 
            i(\sum\nolimits_{\{\bm{G}\}_{1+}}-\sum\nolimits_{\{\bm{G}\}_{1-}})\bm{u}_{1x\bm{G}}
            -(1\leftrightarrow 2), \\
            i(\sum\nolimits_{\{\bm{G}\}_{1+}}-\sum\nolimits_{\{\bm{G}\}_{1-}})\bm{u}_{1y\bm{G}}
            -(1\leftrightarrow 2),
            \end{cases} \label{eq:uG_j}\\
& E_1:\quad \begin{cases} 
            i(\sum\nolimits_{\{\bm{G}\}_{1+}}+\sum\nolimits_{\{\bm{G}\}_{1-}})(G_y\bm{u}_{1x\bm{G}}+G_x\bm{u}_{1y\bm{G}})
            +(1\leftrightarrow 2), \\
            i(\sum\nolimits_{\{\bm{G}\}_{1+}}+\sum\nolimits_{\{\bm{G}\}_{1-}})(G_x\bm{u}_{1x\bm{G}}-G_y\bm{u}_{1y\bm{G}})
            +(1\leftrightarrow 2),
            \end{cases} \label{eq:uG_k}\\
& E_1:\quad \begin{cases} 
            i(\sum\nolimits_{\{\bm{G}\}_{1+}}+\sum\nolimits_{\{\bm{G}\}_{1-}})(-G_x\bm{u}_{1x\bm{G}}+G_y\bm{u}_{1y\bm{G}})
            -(1\leftrightarrow 2), \\
            i(\sum\nolimits_{\{\bm{G}\}_{1+}}+\sum\nolimits_{\{\bm{G}\}_{1-}})(G_y\bm{u}_{1x\bm{G}}+G_x\bm{u}_{1y\bm{G}})
            -(1\leftrightarrow 2),
            \end{cases} \label{eq:uG_l}\\
& E_2:\quad \begin{cases} 
            (\sum\nolimits_{\{\bm{G}\}_{1+}}+\sum\nolimits_{\{\bm{G}\}_{1-}})\bm{u}_{1y\bm{G}}
            +(1\leftrightarrow 2), \\
            (\sum\nolimits_{\{\bm{G}\}_{1+}}+\sum\nolimits_{\{\bm{G}\}_{1-}})(-\bm{u}_{1x\bm{G}})
            +(1\leftrightarrow 2),
            \end{cases} \label{eq:uG_m}\\
& E_2:\quad \begin{cases} 
            (\sum\nolimits_{\{\bm{G}\}_{1+}}+\sum\nolimits_{\{\bm{G}\}_{1-}})\bm{u}_{1x\bm{G}}
            -(1\leftrightarrow 2), \\
            (\sum\nolimits_{\{\bm{G}\}_{1+}}+\sum\nolimits_{\{\bm{G}\}_{1-}})\bm{u}_{1y\bm{G}}
            -(1\leftrightarrow 2),
            \end{cases} \label{eq:uG_n}\\
& E_2:\quad \begin{cases} 
            (\sum\nolimits_{\{\bm{G}\}_{1+}}-\sum\nolimits_{\{\bm{G}\}_{1-}})(G_y\bm{u}_{1x\bm{G}}+G_x\bm{u}_{1y\bm{G}})
            +(1\leftrightarrow 2), \\
            (\sum\nolimits_{\{\bm{G}\}_{1+}}-\sum\nolimits_{\{\bm{G}\}_{1-}})(G_x\bm{u}_{1x\bm{G}}-G_y\bm{u}_{1y\bm{G}})
            +(1\leftrightarrow 2),
            \end{cases} \label{eq:uG_o}\\
& E_2:\quad \begin{cases} 
            (\sum\nolimits_{\{\bm{G}\}_{1+}}-\sum\nolimits_{\{\bm{G}\}_{1-}})(-G_x\bm{u}_{1x\bm{G}}+G_y\bm{u}_{1y\bm{G}})
            -(1\leftrightarrow 2), \\
            (\sum\nolimits_{\{\bm{G}\}_{1+}}-\sum\nolimits_{\{\bm{G}\}_{1-}})(G_y\bm{u}_{1x\bm{G}}+G_x\bm{u}_{1y\bm{G}})
            -(1\leftrightarrow 2).
            \end{cases} \label{eq:uG_p}
\end{align}
\end{subequations}
These fields are by construction real and orthogonal. Our calculations show that the layer-odd (the two layers osscilate in opposite directions) modes $E_2(\sigma_x,\sigma_y)$ (\ref{eq:uG_n}) and $B_2(\tau_z\sigma_z)$ (\ref{eq:uG_h}) strongly couple to $f$ orbitals. The $E_2$ mode (\ref{eq:uG_p}) weakly couple to $f$ orbitals, contributing a tiny correction ($\sim 2.5\%$) to the total $E_2$ modes (see Appendix \ref{Appendix_epc_continuum}). In the special case of $\bm{G}=0$, only $E_2$ basis (\ref{eq:uG_m}) and (\ref{eq:uG_n}) exist, corresponding to the layer-even and layer-odd combinations of the two homogeneous optical modes from the two layers.

\subsection{Other branches}
In this study we only discuss about the iTO phonons in the $K(K')$ valley and the iLO/iTO phonons in the $\Gamma$ valley. Recently Ref. \cite{23_liu_ek_phonon} proved that, in the monolayer level, the coupling to $A_1,B_1$ branches is the only efficient optical EPC near $K(K)'$ as a result of the symmetry constraint as well as the two-center approximation. We note that all other monolayer branches can be treated by the same method proposed in this paper.

\section{Moir\'{e} phonons from layer-decoupled phonon model \label{Appendix_epc_continuum}}
In this appendix we reformulate the moir\'{e} phonons analytically. The basic idea is to start with an approximated continuum EPC Hamiltonian with a constant monolayer coupling $\gamma_K$ ($\gamma_{\Gamma}$). Such model is widely used in previous studies \cite{08prb_graphene_symmetry,18prl_SC_from_phonon,23prb_pair_critical_field,23_liu_ek_phonon,23a_EPC_vs_IKS_kwan,23a_off_diagonal_sc}. Although it is less accurate (the interlayer phonon coupling is completely neglected in this model), its simple form suggests more comprehensive analytical studies. The analysis here is the implementation of the theory proposed in the Section \ref{Sec2}.

\subsection{Layer-decoupled EPC model}
For the optical $K$-phonons and $\Gamma$-phonons, the EPC Hamiltonian is approximately contributed by two independent layers, and can be written simply as
\begin{align}
H_{\text{epc}}^K =& \frac{\gamma_{K}}{\sqrt{2N_mN_a}}\sum_{l\eta\bm{q}}\sum_{\alpha\beta\bm{k}}C^{\dagger}_{\bar{\eta} l\alpha,\bm{k}+\bm{q}+\bm{K}^{\bar{\eta}}_l}(\sigma_x)_{\alpha\beta}C_{\eta l\beta,\bm{k}+\bm{K}^{\eta}_l}(a_{l\eta\bm{q}}+a_{l\bar{\eta},-\bm{q}}^{\dagger}),\label{eq:Hepc_T_continuum} \\
H^{\Gamma}_{\text{epc}} =& \frac{\gamma_{\Gamma}}{\sqrt{2N_mN_a}}\sum_{l\bm{q}}\sum_{\eta\alpha\beta\bm{k}}C^{\dagger}_{\eta l\alpha,\bm{k}+\bm{q}+\bm{K}_l^{\eta}}(-\eta\sigma_y)_{\alpha\beta}C_{\eta l\beta,\bm{k}+\bm{K}_l^{\eta}}(a_{lx\bm{q}}+a_{lx,-\bm{q}}^{\dagger})\nonumber \\ 
&+\frac{\gamma_{\Gamma}}{\sqrt{2N_mN_a}}\sum_{l\bm{q}}\sum_{\eta\alpha\beta\bm{k}}C^{\dagger}_{\eta l\alpha,\bm{k}+\bm{q}+\bm{K}_l^{\eta}}(\sigma_x)_{\alpha\beta}C_{\eta l\beta,\bm{k}+\bm{K}_l^{\eta}}(a_{ly\bm{q}}+a_{ly,-\bm{q}}^{\dagger}), \label{eq:Hepc_L_continuum}
\end{align}
where $C_{\eta l\alpha,\bm{k}+\bm{K}^{\eta}_l}$ denotes the spinless plane wave electron with momentum $\bm{k}+\bm{K}_l^{\eta}$ in the layer $l$, valley $\eta$, and sublattice $\alpha$, $a_{l\eta\bm{q}}$ denotes the $K$-phonon with momentum $\bm{q}+\bm{K}_l^{\eta}$ in the layer $l$ and valley $\eta$, $a_{l\mu\bm{q}}$ denotes the $\Gamma$-phonon with momentum $\bm{q}$ and direction $\mu=x,y$ in the layer $l$. Notice here $\bm{k}$ and $\bm{q}$ are not constrained in mBZ. The coupling constants $\gamma_K\approx 814$ meV and $\gamma_{\Gamma}=528$ meV. They are directly read from the frozen-phonon Hamiltonian, but the weak $\bm{q}$-dependence are neglected by taking the averages. The value of $\gamma_{K}$ here are quite close to the one [$17\text{eV}\text{\AA}^{-1}\sqrt{\hbar/(M_c\omega_K)}$] given in Ref. \cite{23_liu_ek_phonon} that are obtained as an integration of the Slater-Koster hopping.

Then we project the EPC Hamiltonian into the $f$ orbitals, and retain only the onsite terms,
\begin{align}
H_{\text{epc}}^{K,f} =& \frac{1}{\sqrt{2N_m}}\sum_{l\eta\bm{q}}\sum_{\bm{R}}f_{\bm{R}}^{\dagger}\Gamma_{\bm{q}}^{l\eta}(\bm{R})f_{\bm{R}}(a_{l\eta\bm{q}}+a_{l\bar{\eta},-\bm{q}}^{\dagger}),\\
H_{\text{epc}}^{\Gamma,f} =& \frac{1}{\sqrt{2N_m}}\sum_{l\mu\bm{q}}\sum_{\bm{R}}f_{\bm{R}}^{\dagger}\Gamma_{\bm{q}}^{l\mu}(\bm{R})f_{\bm{R}}(a_{l\mu\bm{q}}+a_{l\mu,-\bm{q}}^{\dagger}),
\end{align}
where the scattering matrix $\Gamma_{\bm{q}}^{lb}(\bm{R})=e^{i\bar{\bm{q}}\cdot\bm{R}}\Gamma_{\bm{q}}^{lb}$ [see Eq. (\ref{eq:scattering_matrix1})], $b=K,K'$ for $K$-phonons and $b=x,y$ for $\Gamma$-phonons. For $K$-phonons it can be expressed as the following integration of the Wannier orbitals
\begin{align}
\begin{split}
\left[\Gamma_{\bm{q}}^{l\eta}\right]_{\xi't',\xi t}=&\frac{\gamma_K}{\sqrt{N_a}}\sum_{\alpha\beta\bm{k}}\langle \Phi_{\xi' t'}(\bm{r})|\bar{\eta} l\alpha,\bm{k}+\bm{q}+\bm{K}^{\bar{\eta}}_l\rangle (\sigma_x)_{\alpha\beta}\langle \eta l\beta,\bm{k}+\bm{K}^{\eta}_l| \Phi_{\xi t}(\bm{r})\rangle\\
=&\delta_{\xi'\bar{\eta}}\delta_{\xi\eta}\frac{\gamma_K}{\sqrt{N_a}}\int \frac{d^2\bm{k}}{(2\pi)^2}d^2\bm{r}'d^2\bm{r}e^{i\bm{k}\cdot(\bm{r}'-\bm{r})}e^{i(\bm{q}+\bm{K}_{l}^{\bar{\eta}})\cdot\bm{r}'-i\bm{K}_{l}^{\eta}\cdot\bm{r}}\sum_{\alpha\beta}\Phi^{*}_{\bar{\eta}t'}(\bm{r}',l\alpha)(\sigma_x)_{\alpha\beta}\Phi_{\eta t}(\bm{r},l\beta)\\
=&\delta_{\xi'\bar{\eta}}\delta_{\xi\eta}\frac{\gamma_K}{\sqrt{N_a}}\int d^2\bm{r} e^{i\bm{q}\cdot\bm{r}}\sum_{\alpha\beta}\tilde{\Phi}^{*}_{\bar{\eta}t'}(\bm{r},l\alpha)(\sigma_x)_{\alpha\beta}\tilde{\Phi}_{\eta t}(\bm{r},l\beta), \label{eq:EPCmat_onsite_K}
\end{split}
\end{align}
where in the second step the summation over $\bm{k}$ is replaced by the integral, and in the third step we have used $\int d^2\bm{k}\exp(i\bm{k}\cdot\bm{r})=(2\pi)^2\delta(\bm{r})$ and switched $\Phi_{\eta t}$ to the symmetric convention Eq. (\ref{eq:Ham_BM_symmetric}). Similarly, for $\Gamma$-phonons $\Gamma_{\bm{q}}^{l\mu}(\bm{R})$ is integrated in real-space as
\begin{align}
\begin{split}
\left[\Gamma_{\bm{q}}^{lx}\right]_{\xi't',\xi t}=&\delta_{\xi'\xi}\frac{\gamma_{\Gamma}}{\sqrt{N_a}}\int d^2\bm{r} e^{i\bm{q}\cdot\bm{r}}\sum_{\alpha\beta}\tilde{\Phi}^{*}_{\xi t'}(\bm{r},l\alpha)(-\xi\sigma_y)_{\alpha\beta}\tilde{\Phi}_{\xi t}(\bm{r},l\beta),\\
\left[\Gamma_{\bm{q}}^{ly}\right]_{\xi't',\xi t}=&\delta_{\xi'\xi}\frac{\gamma_{\Gamma}}{\sqrt{N_a}}\int d^2\bm{r} e^{i\bm{q}\cdot\bm{r}}\sum_{\alpha\beta}\tilde{\Phi}^{*}_{\xi t'}(\bm{r},l\alpha)(\sigma_x)_{\alpha\beta}\tilde{\Phi}_{\xi t}(\bm{r},l\beta). \label{eq:EPCmat_onsite_G}
\end{split}
\end{align}

In the following we will decompose the spinless (spin-diagonal) EPC matrices by Pauli matrices $\tau_i\sigma_j$,
\begin{align}
\Gamma_{\bm{q}}^{lb}=\sum_{ij}\lambda_{ij}^{lb}(\bm{q})\tau_i\sigma_j,\quad \lambda_{ij}^{lb}(\bm{q})=\frac{1}{4}\Tr(\Gamma_{\bm{q}}^{lb}\tau_i\sigma_j).\label{eq:EPC_decomposition}
\end{align}
The time reversal imposes a constraint on the coefficients
\begin{align}
\lambda_{ij}^{l\eta}(\bm{q})=\lambda_{ij}^{l\bar{\eta}*}(-\bm{q}) \ (\eta=K,K'), \quad \lambda_{ij}^{l\mu}(\bm{q})=\lambda_{ij}^{l\mu*}(-\bm{q}) \ (\mu=x,y).
\end{align}
As illustrated in the main text, the moir\'{e} phonon operators and displacement fields at $\bm{R}=\bm{0}$ are 
\begin{align}
a_{ij\bm{0}}=\frac{1}{\sqrt{N_m}\lambda_{ij}}\sum_{lb\bm{q}}\lambda_{ij}^{lb}(\bm{q})a_{lb\bm{q}},\quad
\bm{u}_{ij\bm{0}}(\bm{r})=\frac{1}{\sqrt{N_m}\lambda_{ij}}\sum_{lb\bm{q}}\lambda^{lb *}_{ij}(\bm{q})\bm{u}_{lb\bm{q}}(\bm{r}),\label{eq:moire_distortion_appendix}
\end{align}
and the basis fields $\bm{u}_{lb\bm{q}}$ are defined in Eqs. (\ref{eq:u_field_K}) and (\ref{eq:u_field_G}). The effective onsite EPC constant is then
\begin{align}
\lambda_{ij}=\sqrt{\frac{1}{N_m} \sum_{lb\bm{q}}|\lambda_{ij}^{lb}(\bm{q})|^2}=\sqrt{\frac{N_a S_0}{(2\pi)^2}\sum_{lb}\int d^2\bm{q}|\lambda_{ij}^{lb}(\bm{q})|^2}, \label{eq:scattering_strict}
\end{align}
where $S_0=\sqrt{3}a_0^2/2$ is the unit cell area of graphene.

\subsection{Analytical formulation of moir\'{e} phonons, Gaussian approximation}
To gain more insight about the explicit form of moir\'{e} phonons, we then adopt the $f$ orbital wave functions Eq. (\ref{eq:ZLL_WF_matrix}). From simple algebra we can explicitly write down the coupling matrices for $K$-phonons [written in the basis ($\Phi_{K+}$, $\Phi_{K-}$, $\Phi_{K'+}$, $\Phi_{K'-}$)]
\begin{align}
\Gamma_{\bm{q}}^{1K}=\Gamma_{\bm{q}}^{2K}=\frac{\gamma_K}{2\sqrt{N_a}}\left(\begin{matrix}
0&0&0&0\\0&0&0&0\\s_0-s_1&p_-&0&0\\p_+&s_0-s_1&0&0 \end{matrix}\right),\quad 
\Gamma_{\bm{q}}^{1K'}=\Gamma_{\bm{q}}^{2K'}=\frac{\gamma_K}{2\sqrt{N_a}}\left(\begin{matrix}
0&0&s_0-s_1&-p_-\\0&0&-p_+&s_0-s_1\\0&0&0&0\\0&0&0&0 \end{matrix}\right),
\end{align}
where the $s$- and $p$-type functions are defined through sublattice orbitals Eq. (\ref{eq:ZLL_WF_appendix}) or (\ref{eq:Wannier_WF_appendix})
\begin{align}
\begin{split}
s_0(\bm{q})=\int d^2\bm{r} e^{i\bm{q}\cdot \bm{r}}w_0^2(\bm{r}),&\quad  s_1(\bm{q})=\int d^2\bm{r} e^{i\bm{q}\cdot \bm{r}}|w_1(\bm{r})|^2,\\
p_+(\bm{q})=-2i\int d^2\bm{r} e^{i\bm{q}\cdot \bm{r}}w_0(\bm{r})w_1(\bm{r}),&\quad
p_-(\bm{q})=-2i\int d^2\bm{r} e^{i\bm{q}\cdot \bm{r}}w_0(\bm{r})w_1^*(\bm{r}).\\
\end{split}
\label{eq:sp_type_function}
\end{align}
The $s$ functions are real and radial, while $p_+$ and $p_-$ transform like $x+iy$ and $x-iy$ under rotations. To form real orbitals we define $p_x$ and $p_y$ functions through $p_+(\bm{q})+p_-(\bm{q})=2p_x(\bm{q})$ and $p_+(\bm{q})-p_-(\bm{q})=2ip_y(\bm{q})$. The decomposition Eq. (\ref{eq:EPC_decomposition}) gives explicitly the coefficients
\begin{subequations}
\begin{align}
&A_1(\tau_x\sigma_0):\quad\lambda_{x0}^{1K}(\bm{q})=\lambda_{x0}^{2K}(\bm{q})=\lambda_{x0}^{1K'}(\bm{q})=\lambda_{x0}^{2K'}(\bm{q})=\frac{\gamma_K}{2\sqrt{N_a}}\frac{s_0(\bm{q})-s_1(\bm{q})}{2},\label{eq:lam_A1_appendix}\\
&B_1(\tau_y\sigma_0):\quad\lambda_{y0}^{1K}(\bm{q})=\lambda_{y0}^{2K}(\bm{q})=-\lambda_{y0}^{1K'}(\bm{q})=-\lambda_{y0}^{2K'}(\bm{q})=-i\frac{\gamma_K}{2\sqrt{N_a}}\frac{s_0(\bm{q})-s_1(\bm{q})}{2},\\
&E_1(\tau_y\sigma_x):\quad\lambda_{yx}^{1K}(\bm{q})=\lambda_{yx}^{2K}(\bm{q})=\lambda_{yx}^{1K'}(\bm{q})=\lambda_{yx}^{2K'}(\bm{q})=-i\frac{\gamma_K}{2\sqrt{N_a}}\frac{p_x(\bm{q})}{2},\\
&E_1(\tau_y\sigma_y):\quad\lambda_{yy}^{1K}(\bm{q})=\lambda_{yy}^{2K}(\bm{q})=\lambda_{yy}^{1K'}(\bm{q})=\lambda_{yy}^{2K'}(\bm{q})=-i\frac{\gamma_K}{2\sqrt{N_a}}\frac{p_y(\bm{q})}{2},\\
&E_2(\tau_x\sigma_x):\quad\lambda_{xx}^{1K}(\bm{q})=\lambda_{xx}^{2K}(\bm{q})=-\lambda_{xx}^{1K'}(\bm{q})=-\lambda_{xx}^{2K'}(\bm{q})=\frac{\gamma_K}{2\sqrt{N_a}}\frac{p_x(\bm{q})}{2},\\
&E_2(\tau_x\sigma_y):\quad\lambda_{yy}^{1K}(\bm{q})=\lambda_{xy}^{2K}(\bm{q})=-\lambda_{xy}^{1K'}(\bm{q})=-\lambda_{xy}^{2K'}(\bm{q})=\frac{\gamma_K}{2\sqrt{N_a}}\frac{p_y(\bm{q})}{2},
\end{align}
\end{subequations}
while all other components are zero. If we retain the translation symmetry of moir\'{e} superlattice, the six modes listed above just reduce to the layer-even modes Eqs. (\ref{eq:uK_a}), (\ref{eq:uK_c}), (\ref{eq:uK_e}), (\ref{eq:uK_g}), respectively.

Similarly, for $\Gamma$-phonons, the onsite EPC matrices read
\begin{align}
\begin{split}
\Gamma_{\bm{q}}^{1x}&=-\Gamma_{\bm{q}}^{2x} = \frac{\gamma_{\Gamma}}{2\sqrt{N_a}}\left(\begin{matrix}ip_y&s_0-d_-&0&0\\s_0-d_+&-ip_y&0&0\\
0&0&-ip_y&s_0-d_-\\0&0&s_0-d_+&ip_y
\end{matrix}\right),\\
\Gamma_{\bm{q}}^{1y}&=-\Gamma_{\bm{q}}^{2y} = i\frac{\gamma_{\Gamma}}{2\sqrt{N_a}}\left(\begin{matrix}-p_x&-s_0-d_-&0&0\\s_0+d_+&p_x&0&0\\0&0&p_x&-s_0-d_-\\0&0&s_0+d_+&-p_x
\end{matrix}\right),
\end{split}
\end{align}
where $s_0$, $s_1$, and $p_{\pm}$ are defined in Eq. (\ref{eq:sp_type_function}), and the $d$-type functions are
\begin{align}
d_+(\bm{q})=-\int d^2\bm{r} e^{i\bm{q}\cdot \bm{r}}w_1^2(\bm{r}),&\quad
d_-(\bm{q})=-\int d^2\bm{r} e^{i\bm{q}\cdot \bm{r}}w_1^{*2}(\bm{r}),
\end{align}
which transform like $(x+iy)^2$ and $(x-iy)^2$, respectively. We can also define the real functions $d_{x^2-y^2}$ and $d_{xy}$ through $d_+(\bm{q})+d_-(\bm{q})=2d_{x^2-y^2}(\bm{q})$ and $d_+(\bm{q})-d_-(\bm{q})=2id_{xy}(\bm{q})$. The resulting nonzero coefficients are
\begin{subequations}
\begin{align}
E_2(\tau_0\sigma_x):\quad &\lambda_{0x}^{1x}(\bm{q})=-\lambda_{0x}^{2x}(\bm{q})=\frac{\gamma_{\Gamma}}{2\sqrt{N_a}}\left[s_0(\bm{q})-d_{x^2-y^2}(\bm{q})\right],\quad
\lambda_{0x}^{1y}(\bm{q})=-\lambda_{0x}^{2y}(\bm{q})=-\frac{\gamma_{\Gamma}}{2\sqrt{N_a}}d_{xy}(\bm{q}),\\
E_2(\tau_0\sigma_y):\quad &\lambda_{0y}^{1x}(\bm{q})=-\lambda_{0y}^{2x}(\bm{q})=-\frac{\gamma_{\Gamma}}{2\sqrt{N_a}}d_{xy}(\bm{q}),\quad
\lambda_{0y}^{1y}(\bm{q})=-\lambda_{0y}^{2y}(\bm{q})=\frac{\gamma_{\Gamma}}{2\sqrt{N_a}}\left[s_0(\bm{q})+d_{x^2-y^2}(\bm{q})\right],\\
B_2(\tau_z\sigma_z):\quad &\lambda_{zz}^{1x}(\bm{q})=-\lambda_{zz}^{2x}(\bm{q})=i\frac{\gamma_{\Gamma}}{2\sqrt{N_a}}p_y(\bm{q}),\quad
\lambda_{zz}^{1y}(\bm{q})=-\lambda_{zz}^{2y}(\bm{q})=-i\frac{\gamma_{\Gamma}}{2\sqrt{N_a}}p_x(\bm{q}).
\end{align}
\end{subequations}
Only three moir\'{e} modes exist, which is consistent with symmetry analysis and frozen-phonon calculations. It is noteworthy that the $E_2(\sigma_x,\sigma_y)$ mode has both $s$ and $d$ components. The $s$ component contributes the dominant part, which inherits the monolayer $E_2$ pattern and correspond to the periodic term Eq. (\ref{eq:uG_n}). The $d$ component, corresponding to Eq. (\ref{eq:uG_p}), is quantitatively smaller. The $B_2(\tau_z\sigma_z)$ mode is the only 1D irrep of $\Gamma$-phonons coupled with $f$ electrons, which has the periodic term Eq. (\ref{eq:uG_h}).

If we then use the approximated Gaussian form of the $f$ orbitals (\ref{eq:Wannier_WF_appendix}), things will become more simple. The resulting form factors are also Gaussian,
\begin{align}
\begin{split}
s_0(\bm{q})&=\alpha_1^2e^{-\frac{1}{4}l_1^2 q^2}, \quad s_1(\bm{q})=\alpha_2^2 e^{-\frac{1}{4}l_2^2 q^2}\left(1-\frac{1}{4}l_2^2q^2\right),\\
p_{\pm}(\bm{q})&=\frac{4\alpha_1\alpha_2 l_1^3l_2^2}{(l_1^2+l_2^2)^2}e^{-\frac{l_1^2l_2^2q^2}{2\left(l_1^2+l_2^2\right)}}(q_x\pm iq_2), \quad 
d_{\pm}(\bm{q})=\frac{1}{4}\alpha_2^2 e^{-\frac{1}{4}l_2^2q^2}l_2^2(q_x\pm iq_y)^2,
\end{split}
\label{eq:spd_function}
\end{align}
meaning that all the moir\'{e} phonons' envelopes are also Gaussian. The shape of the form factors are shown in Fig. \ref{fig:spd_function}(a), plotted using parameters $\alpha_1\approx 0.8096$, $l_1\approx 2.9033$ nm and $l_2\approx 3.6484$ nm which are chosen to fit the numerical Wannier orbitals. From the figure we see $d_{\pm}$ is indeed small compared with the others. Once the form factors are obtained, all EPC coefficients $\lambda_{ij}^{lb}$ can be readily calculated. In Fig. \ref{fig:spd_function}(b) we show the EPC coefficient of the moir\'{e} $A_1$ mode (black solid line). The Gaussian approximation is excellent, which provides EPC coefficient that matches well with the frozen-phonon data (blue dots).

The analytical Gaussian form also makes the moir\'{e} phonon acceptable at every momentum $\bm{q}$, not only limited to commensurate $\bar{\bm{q}}=\bar{\bm{0}}$ cases in the frozen-phonon method. In other words, the approximation Eq. (\ref{eq:scattering_approx}) is no longer needed. The resulting onsite EPC constants are
\begin{subequations}
\begin{align}
&K\text{-}A_1,B_1:\quad \lambda_{x0}=\lambda_{y0}=\frac{\gamma_K}{4\pi}\sqrt{\left( \frac{2\alpha_1^4}{l_1^2}+\frac{\alpha_2^4}{l_2^2}-\frac{8\alpha_1^2\alpha_2^2l_1^2}{(l_1^2+l_2^2)^2}\right)\pi S_0},\label{eq:EPC_Gaussian_lam_a}\\
&K\text{-}E_1,E_2:\quad \lambda_{xx}=\lambda_{xy}=\lambda_{yx}=\lambda_{yy} = \frac{\gamma_K}{2\pi}\frac{\alpha_1\alpha_2 l_1}{l_1^2+l_2^2}\sqrt{2\pi S_0},\label{eq:EPC_Gaussian_lam_b}\\
&\Gamma\text{-}E_2:\quad \lambda_{0x}=\lambda_{0y}=\frac{\gamma_{\Gamma}}{2\pi}\sqrt{\left(\frac{\alpha_1^4}{l_1^2}+\frac{\alpha_2^4}{2l_2^2}\right) \pi S_0},\label{eq:EPC_Gaussian_lam_c}\\
&\Gamma\text{-}B_2:\quad \lambda_{zz} = \frac{\gamma_{\Gamma}}{2\pi}\frac{\alpha_1\alpha_2 l_1}{l_1^2+l_2^2}\sqrt{8\pi S_0}. \label{eq:EPC_Gaussian_lam_d}
\end{align}
\end{subequations}
The EPC constant obtained using fitted Gaussian Wannier functions are given in Table \ref{tab:table1_epc_const}. For comparison we also list the corresponding values calculated using the strict projection Eq. (\ref{eq:scattering_strict}) on the approximated EPC Hamiltonian Eqs. (\ref{eq:Hepc_T_continuum}) and (\ref{eq:Hepc_L_continuum}) and the the frozen-phonon method Eq. (\ref{eq:scattering_approx}). 

\subsection{The strong sensibility of EPC on model parameters \label{appendix_E_sensibility}}
From the above analytical results, we see that the effective coupling constants $\lambda_b$ mainly depend on several factors. First, they are proportional to the monolayer coupling strength $\gamma_{K,\Gamma}$. Various graphene models could yield significantly different values of $\gamma_{K,\Gamma}$. In this study, we adopt a TB model that incorporates hoppings to a broad range of atoms, resulting in a $\gamma_{K}$ that is quite close to the one used in Ref. \cite{23_liu_ek_phonon}. The coupling strength also depends on the spatial range of $f$ orbitals, roughly following an inverse law $\lambda_b \sim l^{-1}_{1,2}$. The practical values are thus sensitive to the specific moir\'{e} potential to generate the local $f$ orbitals. Practically the $f$ orbitals become more localized if we start from the continuum model, use a slightly-smaller twist angle \cite{22prb_ZLL_OPW_representation}, or include the relaxation effects. Therefore, it is not surprising to obtain the effective EPC constants enhanced or suppressed by $\sim50\%$ using different electron models.

\begin{figure*}
\includegraphics[width=1.0\textwidth]{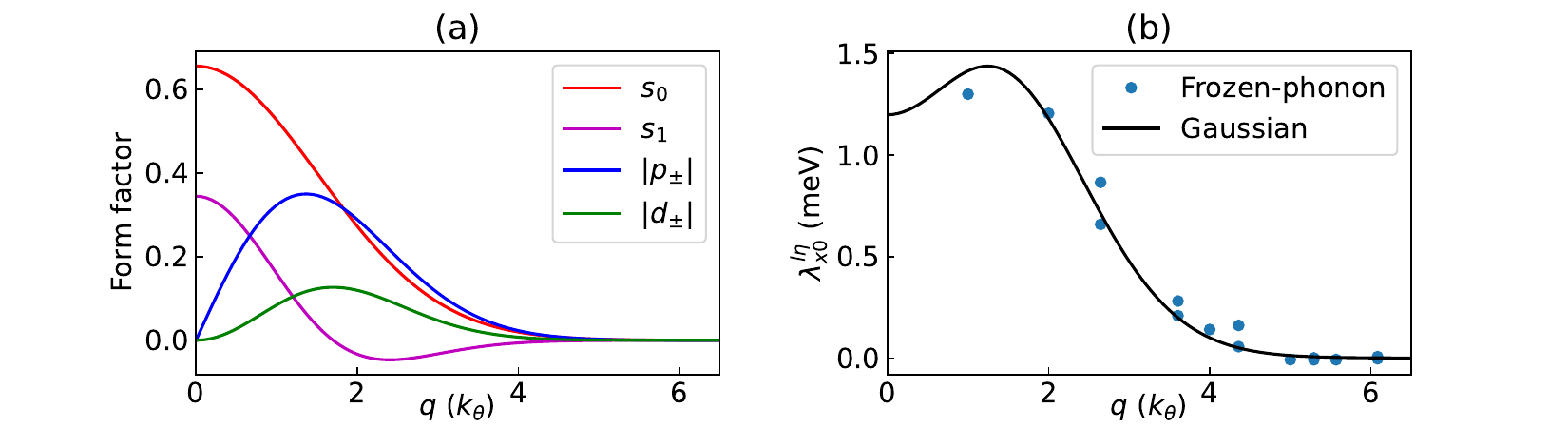}% Here is how to import EPS art
\caption{\label{fig:spd_function}
(a) The shape of the form factors $s_0(\bm{q})$, $s_1(\bm{q})$, $|p_{\pm}(\bm{q})|$, and $|d_{\pm}(\bm{q})|$, calculated using the Gaussian functions [Eqs. (\ref{eq:Wannier_WF_appendix}), (\ref{eq:spd_function})]. (b) The moir\'{e} $A_1$ mode's EPC coefficient $\lambda_{x0}^{l\eta}(\bm{q})$ (\ref{eq:lam_A1_appendix}). The solid line is calculated with the Gaussian functions in (a), while the blue dots are the frozen-phonon data at commensurate momentum ($\bar{\bm{q}}=\bar{\bm{0}}$) [see also Figs. \ref{fig:Q_G}(a) and \ref{fig:band_A1_Q} for the phonon momentum and frozen-phonon bands].}
\end{figure*}

\subsection{Valley-layer locking of moir\'{e} phonons}
By direct decomposition of EPC matrices, we see all the 6 moir\'{e} $K$-phonons are layer-even [$\lambda_{ij}^{1\eta}(\bm{q})=\lambda_{ij}^{2\eta}(\bm{q})$] while all the 3 moir\'{e} $\Gamma$-phonons are layer-odd [$\lambda_{ij}^{1\mu}(\bm{q})=-\lambda_{ij}^{2\mu}(\bm{q})$]. Such kind of valley-layer locking of moir\'{e} phonons is also verified by frozen-phonon calculations. Actually this results from the inner symmetry of the $f$ orbitals. From Eq. (\ref{eq:ZLL_WF_matrix}) one may easily check the following properties of $f$ electrons \cite{22prl_heavy_fermion_Zhida}
\begin{align}
\sum_{l\alpha\beta}\Phi^*_{\bar{\eta} t'}(\bm{r},l\alpha)(\varrho_z\sigma_x)_{l\alpha,l\beta}\Phi_{\eta t}(\bm{r},l\beta)=0, \quad
\sum_{l\alpha\beta}\Phi^*_{\eta t'}(\bm{r},l\alpha)(\varrho_0\sigma_{x,y})_{l\alpha,l\beta}\Phi_{\eta t}(\bm{r},l\beta)=0,
\end{align}
which forbids the layer-odd intervalley intersublattice scatterings in Eq. (\ref{eq:Hepc_T_continuum}) and layer-even intravalley intersublattice scatterings in Eq. (\ref{eq:Hepc_L_continuum}).

\begin{table*}[!htb]
\caption{\label{tab:table1_epc_const}%
The effective onsite EPC constant $\lambda_b$ of various moir\'{e} modes, calculated using different methods. In each method the $f$ orbitals are chosen to be either the Wannier orbitals or the approximated Gaussian functions [Eq. (\ref{eq:Wannier_WF_appendix}), with parameters $\alpha_1\approx 0.8096$, $l_1\approx 2.9033$ nm and $l_2\approx 3.6484$ nm], and the starting EPC Hamiltonian is chosen to be either the frozen-phonon ones [only $\bar{\bm{q}}=0$ cases are summed, see Eq. (\ref{eq:scattering_approx})] or the approximated layer-decoupled ones Eqs. (\ref{eq:Hepc_T_continuum}), (\ref{eq:Hepc_L_continuum}). All numbers are in the unit of meV.}
\begin{ruledtabular}
\begin{tabular}{cccccc}
$f$ orbitals & EPC model & $A_1: \lambda_{x0}$, $B_1: \lambda_{y0}$  &  $E_1: \lambda_{yx},\lambda_{yy}$ \& $E_2:\lambda_{xx},\lambda_{xy}$  & $E_2: \lambda_{0x},\lambda_{0y}$ & $B_2: \lambda_{zz}$ \\
\hline
Gaussian &  layer-decoupled     &$7.371$     &   $4.718$      &   $8.028$     & $6.121$  \\
Wannier  &  layer-decoupled      &$7.364$     &   $4.780$      &   $8.027$     & $6.166$  \\
Wannier  &  frozen-phonon      &$7.347$     &   $4.741$      &   $8.015$     & $6.047$  \\
\end{tabular}
\end{ruledtabular}
\end{table*}

\section{EPC in the 8-band model \label{appendix_epc_lattice_model}}

\subsection{EPC matrix elements}
To construct a complete lattice model containing the EPC, we need to calculate the EPC matrix among all Wannier orbitals. This subsection lists the formula to calculate them. The resulting lattice model is shown in next subsection.

Using Eqs. (\ref{eq:WF_plwv}), the EPC matrices for $K$-phonons [Eq. (\ref{eq:EPCmat_onsite_K})] and $\Gamma$-phonons [Eq. (\ref{eq:EPCmat_onsite_G})] become ($t,t'=\pm$)
\begin{align}
&\left[\Gamma_{\bm{q}}^{l\eta}(\bm{R})\right]_{\xi't',\xi t}=e^{i\bar{\bm{q}}\cdot{\bm{R}}}\delta_{\xi'\bar{\eta}}\delta_{\xi\eta}\frac{\gamma_K}{\sqrt{N_a}}\frac{1}{N_m}\sum_{\bar{\bm{k}}'\bm{G}'}\sum_{\bar{\bm{k}}\bm{G}}\sum_{\alpha\beta}\mathcal{W}^{\bar{\eta}*}_{\bm{G}'l\alpha,t'}(\bar{\bm{k}}')(\sigma_x)_{\alpha\beta}\mathcal{W}^{\eta}_{\bm{G}l\beta,t}(\bar{\bm{k}})\delta_{\bm{q},\bar{\bm{k}}'+\bm{G}'-\bm{K}^{\bar{\eta}}_l-\bar{\bm{k}}-\bm{G}+\bm{K}^{\eta}_l},\\
&\left[\Gamma_{\bm{q}}^{lx}(\bm{R})\right]_{\xi't',\xi t}=e^{i\bar{\bm{q}}\cdot{\bm{R}}}\delta_{\xi'\xi}\frac{\gamma_{\Gamma}}{\sqrt{N_a}}\frac{1}{N_m}\sum_{\bar{\bm{k}}'\bm{G}'}\sum_{\bar{\bm{k}}\bm{G}}\sum_{\alpha\beta}\mathcal{W}^{\xi*}_{\bm{G}'l\alpha,t'}(\bar{\bm{k}}')(-\xi\sigma_y)_{\alpha\beta}\mathcal{W}^{\xi}_{\bm{G}l\beta,t}(\bar{\bm{k}})\delta_{\bm{q},\bar{\bm{k}}'+\bm{G}'-\bar{\bm{k}}-\bm{G}},\\
&\left[\Gamma_{\bm{q}}^{ly}(\bm{R})\right]_{\xi't',\xi t}=e^{i\bar{\bm{q}}\cdot{\bm{R}}}\delta_{\xi'\xi}\frac{\gamma_{\Gamma}}{\sqrt{N_a}}\frac{1}{N_m}\sum_{\bar{\bm{k}}'\bm{G}'}\sum_{\bar{\bm{k}}\bm{G}}\sum_{\alpha\beta}\mathcal{W}^{\xi*}_{\bm{G}'l\alpha,t'}(\bar{\bm{k}}')(\sigma_x)_{\alpha\beta}\mathcal{W}^{\xi}_{\bm{G}l\beta,t}(\bar{\bm{k}})\delta_{\bm{q},\bar{\bm{k}}'+\bm{G}'-\bar{\bm{k}}-\bm{G}}.
\end{align}
The moir\'{e} phonons Eq. (\ref{eq:moire_distortion_appendix}) are then determined by the numerical decomposition Eq. (\ref{eq:EPC_decomposition}). 

Once the moir\'{e} phonons have been obtained, we can calculate the EPC matrices for any two electrons located at any sites of the 8-band model. For the $K$-phonon mode ($b=\tau_x,\tau_y,\tau_y\sigma_x,\tau_y\sigma_y,\tau_x\sigma_x,\tau_x\sigma_y$) with Fourier coefficients $\lambda_{b}^{l\eta}$, and for the $\Gamma$-phonon mode ($b=\sigma_x,\sigma_y,\tau_z\sigma_z$) with coefficients $\lambda_b^{l\mu}$, the complete EPC Hamiltonian in plane wave basis respectively read
\begin{align}
&H_{\text{epc},b}^{K} = \frac{\gamma_{K}}{\sqrt{2N_mN_a}}\sum_{l\eta\bm{q}}\sum_{\alpha\beta\bm{k}}e^{-i\bar{\bm{q}}\cdot\bm{R}}\frac{\lambda_{b}^{l\eta*}(\bm{q})}{\sqrt{N_m}\lambda_b}C^{\dagger}_{\bar{\eta} l\alpha,\bm{k}+\bm{q}+\bm{K}^{\bar{\eta}}_l}(\sigma_x)_{\alpha\beta}C_{\eta l\beta,\bm{k}+\bm{K}^{\eta}_l}(a_{b\bm{R}}+a_{b\bm{R}}^{\dagger}),\\
&H_{\text{epc},b}^{\Gamma} = \frac{\gamma_{\Gamma}}{\sqrt{2N_mN_a}}\sum_{l\bm{q}}\sum_{\eta\alpha\beta\bm{k}}e^{-i\bar{\bm{q}}\cdot\bm{R}}C^{\dagger}_{\eta l\alpha,\bm{k}+\bm{q}+\bm{K}^{\eta}_l}\left[-\eta\sigma_y\frac{\lambda_{b}^{lx*}(\bm{q})}{\sqrt{N_m}\lambda_b}+\sigma_x\frac{\lambda_{b}^{ly*}(\bm{q})}{\sqrt{N_m}\lambda_b}\right]_{\alpha\beta}C_{\eta l\beta,\bm{k}+\bm{K}^{\eta}_l}(a_{b\bm{R}}+a_{b\bm{R}}^{\dagger}).
\end{align}
In the 8-band model, the general EPC Hamiltonian is
\begin{align}
H_{\text{epc}}=\frac{1}{\sqrt{2}}\sum_{b\bm{R}}\sum_{\xi'n'\xi n}\sum_{\bm{L}\bm{T}}d_{\xi'n',\bm{R}+\bm{L}}^{\dagger}\left(\Lambda^{b}_{\xi'\bm{L},\xi\bm{T}}\right)_{n'n} d_{\xi n,\bm{R}+\bm{T}}(a_{b\bm{R}}+a_{b\bm{R}}^{\dagger}),
\end{align}
where the operator $d$ contains both the $f$ electrons ($n=1,2$) and $c$ electrons ($n\geq3$) [Eq. (\ref{eq:d_operator_8band})], and the $\bm{R}$-independent $\Lambda^{b}_{\bm{L},\bm{T}}$ is the scattering matrix between the orbitals $\Phi_{\xi' n'}(\bm{r}-\bm{\tau}_{n'}-\bm{L}-\bm{R})$ and $\Phi_{\xi n}(\bm{r}-\bm{\tau}_{n}-\bm{T}-\bm{R})$ induced the $b$-th moir\'{e} mode localized at $\bm{R}$. For the $K$-phonons, the EPC matrix, written as integrals in $k$-space (for practical calculations) or real-space (for analysis), reads
\begin{align}
\begin{split}
&\left(\Lambda^{b}_{\xi'\bm{L},\xi\bm{T}}\right)_{n'n}\\
=&\frac{\gamma_{K}}{\sqrt{N_a}}\frac{\delta_{\xi'\bar{\xi}}}{N_m^2\lambda_b}\sum_{\bar{\bm{k}}'\bm{G}'}\sum_{\bar{\bm{k}}\bm{G}}\sum_{l\alpha\beta}\sum_{\bm{q}}e^{i\bar{\bm{k}}'\cdot\bm{L}-i\bar{\bm{k}}\cdot\bm{T}}\delta_{\bm{q},\bar{\bm{k}}'+\bm{G}'-\bm{K}_l^{\bar{\xi}}-\bar{\bm{k}}-\bm{G}+\bm{K}_l^{\xi}}\lambda_{b}^{l\xi*}(\bm{q})\mathcal{W}_{\bm{G}'l\alpha,n'}^{\bar{\xi}*}(\bar{\bm{k}}')(\sigma_x)_{\alpha\beta}\mathcal{W}^{\xi}_{\bm{G}l\beta,n}(\bar{\bm{k}})\\
=&\frac{\gamma_{K}}{\sqrt{N_a}}\frac{\delta_{\xi'\bar{\xi}}}{\sqrt{N_m}}\sum_{l\alpha\beta}e^{i\bm{K}_l^{\bar{\xi}}\cdot\bm{L}-i\bm{K}_l^{\xi}\cdot\bm{T}}\int d^2\bm{r}\tilde{\Phi}^{*}_{\bar{\xi} n'}(\bm{r}-\bm{\tau}_{n'}-\bm{L},l\alpha)(\sigma_x)_{\alpha\beta}\tilde{\Phi}_{\xi n}(\bm{r}-\bm{\tau}_n-\bm{T},l\beta)\tilde{u}_b^{l\xi}(\bm{r}),\label{eq:EPC_K_full}
\end{split}
\end{align}
while for the $\Gamma$-phonons,
\begin{align}
\begin{split}
&\left(\Lambda^{b}_{\xi'\bm{L},\xi\bm{T}}\right)_{n'n}\\
=&\frac{\gamma_{\Gamma}}{\sqrt{N_a}}\frac{\delta_{\xi'\xi}}{N_m^2\lambda_b}\sum_{\bar{\bm{k}}'\bm{G}'}\sum_{\bar{\bm{k}}\bm{G}}\sum_{l\alpha\beta}\sum_{\bm{q}}e^{i\bar{\bm{k}}'\cdot\bm{L}-i\bar{\bm{k}}\cdot\bm{T}}\delta_{\bm{q},\bar{\bm{k}}'+\bm{G}'-\bar{\bm{k}}-\bm{G}}\mathcal{W}_{\bm{G}'l\alpha,n'}^{\xi*}(\bar{\bm{k}}')\left[-\xi\sigma_y\lambda_{b}^{lx*}(\bm{q})+\sigma_x\lambda_{b}^{ly*}(\bm{q})\right]_{\alpha\beta}\mathcal{W}^{\xi}_{\bm{G}l\beta,n}(\bar{\bm{k}})\\
=&\frac{\gamma_{\Gamma}}{\sqrt{N_a}}\frac{\delta_{\xi'\xi}}{\sqrt{N_m}}\sum_{l\alpha\beta}e^{i\bm{K}_l^{\xi}\cdot(\bm{L}-\bm{T})}\int d^2\bm{r}\tilde{\Phi}^{*}_{\xi n'}(\bm{r}-\bm{\tau}_{n'}-\bm{L},l\alpha)\left[-\xi\sigma_y \tilde{u}_b^{lx}(\bm{r})+\sigma_x \tilde{u}_b^{ly}(\bm{r})\right]_{\alpha\beta}\tilde{\Phi}_{\xi n}(\bm{r}-\bm{\tau}_n-\bm{T},l\beta), \label{eq:EPC_G_full}
\end{split}
\end{align}
where the envelope functions of electron and phonons [in the gauge of Eqs. (\ref{eq:Ham_BM_symmetric}), (\ref{eq:ZLL_WF_matrix})] are defined as
\begin{align}
&\tilde{\Phi}_{\eta n}(\bm{r}-\bm{\tau}_n-\bm{R},l\alpha)=e^{-i\bm{K}_l^{\eta}\cdot(\bm{r}-\bm{R})}\Phi_{\eta n}(\bm{r}-\bm{\tau}_n-\bm{R},l\alpha),\\
&\tilde{u}_{b}^{l\eta}(\bm{r}-\bm{R})=e^{-i\bm{K}_l^{\eta}\cdot(\bm{r}-\bm{R})}u_{b}^{l\eta}(\bm{r}-\bm{R})=\sum_{\bm{q}}\frac{\lambda_{b}^{l\eta*}(\bm{q})}{\sqrt{N_m}\lambda_b}e^{i\bm{q}\cdot(\bm{r}-\bm{R})},\quad \eta=K,K',\\
&\tilde{u}_{b}^{l\mu}(\bm{r}-\bm{R})=u_{b}^{l\mu}(\bm{r}-\bm{R})=\sum_{\bm{q}}\frac{\lambda_{b}^{l\mu*}(\bm{q})}{\sqrt{N_m}\lambda_b}e^{i\bm{q}\cdot(\bm{r}-\bm{R})},\quad \mu=x,y.
\end{align}
In practical calculations the moir\'{e} phonons and EPC matrices are obtained by performing the integration in $k$-space with the $27\times27$ meshgrid of $\bar{\bm{k}}$ points. The real-space integral form of Eqs. (\ref{eq:EPC_K_full}) and (\ref{eq:EPC_G_full}) is essentially a three-center integration. If the Wannier orbitals and moir\'{e} phonons are well localized (by construction, the localization of moir\'{e} phonons results from the localization of $f$ electrons), we may expect that the coupling will decay fast when their centers deviate far from each other.

\subsection{The effective EPC lattice model}
In this subsection the final EPC lattice model will be presented. For the moir\'{e} $A_1$ ($\tau_x\sigma_0$) and $B_1$ ($\tau_y\sigma_0$) modes of $K$-phonons, the EPC lattice model reads
\begin{align}
\begin{split}
H_{\text{epc},K}^{A_1+B_1}=&\frac{1}{\sqrt{2}}\sum_{\bm{R}}\left(a_{x0,\bm{R}}+a_{x0,\bm{R}}^{\dagger}\right)h_{x0,\bm{R}}+\frac{1}{\sqrt{2}}\sum_{\bm{R}}\left(a_{y0,\bm{R}}+a_{y0,\bm{R}}^{\dagger}\right)h_{y0,\bm{R}}\\
=&\frac{1}{\sqrt{2}}\sum_{\bm{R}}\left(a_{x0,\bm{R}}+a^{\dagger}_{x0,\bm{R}}+ia_{y0\bm{R}}+ia^{\dagger}_{y0,\bm{R}}\right)h_{x0,\bm{R}}^{K'K} + \text{H.c.}, \label{eq:epc_model_8band_K}
\end{split}
\end{align}
where the electron part $h_{x0,\bm{R}}$ and $h_{y0,\bm{R}}$ contains exclusively intervalley scatterings. We decompose them as ($h_{x0,\bm{R}}^{K'K}$ describes the scatterings from $K$ valley to $K'$ valley)
\begin{align}
h_{x0,\bm{R}}=h_{x0,\bm{R}}^{K'K}+h_{x0,\bm{R}}^{K'K\dagger},\quad h_{y0,\bm{R}}=ih_{x0,\bm{R}}^{K'K}-ih_{x0,\bm{R}}^{K'K\dagger}.
\end{align}
Explicitly, the nearest-neighbor $h_{x0,\bm{R}}^{K'K}$ reads [$\omega=\exp(i2\pi/3)$, $\bm{L}=\bm{L}_2^m-\bm{L}^m_1=(0,L_{\theta})^T$, and $g\in \{0,1,2\}$]
\begin{align}
&h_{x0,\bm{R}}^{K'K} \label{eq:epc_model_hK}\\
=& \lambda_{x0}^{ff}\left(f^{\dagger}_{K'+,\bm{R}}f_{K+,\bm{R}}+f^{\dagger}_{K'-,\bm{R}}f_{K-,\bm{R}}\right)\nonumber\\
+&\lambda_{x0}^{\tau\tau}\tau^{\dagger}_{K'\bm{R}}\tau_{K\bm{R}} \nonumber\\
+&\lambda_{x0}^{\eta\eta}\sum_{g}\left[\left(\eta^{\dagger}_{K',\bm{R}+C^{g+1}_{3z}\bm{\tau}_{\text{BA}}}+ \eta^{\dagger}_{K',\bm{R}+C^{g-1}_{3z}\bm{\tau}_{\text{BA}}}\right)\eta_{K,\bm{R}+C_{3z}^{g}\bm{\tau}_{\text{AB}}}+(\text{AB}\leftrightarrow \text{BA})\right] \nonumber\\
+&\lambda_{x0}^{f\tau}\sum_{g}\left[ \omega^{g}\left(f^{\dagger}_{K'+,\bm{R}}\tau_{K,\bm{R}-C_{3z}^g\bm{L}}+\tau^{\dagger}_{K',\bm{R}-C_{3z}^g\bm{L}}f_{K-,\bm{R}}\right)+\omega^{*g}\left(f^{\dagger}_{K'-,\bm{R}}\tau_{K,\bm{R}+C_{3z}^g\bm{L}}+\tau^{\dagger}_{K',\bm{R}+C_{3z}^g\bm{L}}f_{K+,\bm{R}}\right) \right]\nonumber\\
+&\lambda_{x0}^{f\eta}\sum_{g}\left[\omega^{*g}\left(\eta^{\dagger}_{K',\bm{R}+C_{3z}^g\bm{\tau}_{\text{AB}}}f_{K+,\bm{R}}+f^{\dagger}_{K'-,\bm{R}}\eta_{K,\bm{R}+C_{3z}^g\bm{\tau}_{\text{AB}}}\right)-\omega^g\left(\eta^{\dagger}_{K',\bm{R}+C_{3z}^g\bm{\tau}_{\text{AB}}}f_{K-,\bm{R}}+f^{\dagger}_{K'+,\bm{R}}\eta_{K,\bm{R}+C_{3z}^g\bm{\tau}_{\text{AB}}}\right)\right]\nonumber\\
+&\lambda_{x0}^{f\eta*}\sum_g\left[\omega^g\left(\eta^{\dagger}_{K',\bm{R}+C_{3z}^g\bm{\tau}_{\text{BA}}}f_{K-,\bm{R}}+f^{\dagger}_{K'+,\bm{R}}\eta_{K,\bm{R}+C_{3z}^g\bm{\tau}_{\text{BA}}}\right)-\omega^{*g}\left(\eta^{\dagger}_{K',\bm{R}+C_{3z}^g\bm{\tau}_{\text{BA}}}f_{K+,\bm{R}}+f^{\dagger}_{K'-,\bm{R}}\eta_{K,\bm{R}+C_{3z}^g}\bm{\tau}_{\text{BA}}\right)\right]\nonumber \\
+&\lambda_{x0}^{f\kappa}\sum_g\left[\omega^{*g}\left(\kappa^{\dagger}_{K'\bm{R}-C_{3z}^g\bm{\tau}_{\text{DW1}}}f_{K+\bm{R}}+f^{\dagger}_{K'-\bm{R}}\kappa_{K'\bm{R}-C_{3z}^g\bm{\tau}_{\text{DW1}}}\right)+\omega^{g}\left(\kappa^{\dagger}_{K'\bm{R}+C_{3z}^g\bm{\tau}_{\text{DW1}}}f_{K-\bm{R}}+f^{\dagger}_{K'+\bm{R}}\kappa_{K'\bm{R}+C_{3z}^g\bm{\tau}_{\text{DW1}}}\right)\right]\nonumber\\
+& ...  \nonumber
\end{align}
The first term of $h_{x0,\bm{R}}^{K'K}$ with coefficient $\lambda_{x0}^{ff}=\lambda_{x0}$ [Eq. (\ref{eq:EPC_Gaussian_lam_a})] is just the onsite coupling between $f$ orbitals, which has the largest energy scale in this phonon channel. The second term is the onsite coupling among $s@$AA orbitals, which actually governs the opening of the upper inner gap of the flat bands at $\bar{\bm{\Gamma}}$. The third term contains the six nearest-neighbor scatterings among the $p_z@$AB and $p_z@$BA orbitals, which is necessary to open the lower inner gap of flat bands near $\bar{\bm{\Gamma}}$. These three terms already capture faithfully all typical features of the flat band behavior under periodic $A_1$ or $B_1$ distortions. All other terms have only quantitative effects on the detailed dispersion of the frozen-phonon flat band. For instance, the $A_1$ phonon at $\bm{R}$ induces nearest-neighbor scattering from the $f$ orbitals to $p_z$@AB orbitals (the fifth term). Such term will contribute the form factor 
\begin{align}
\lambda_{x0}^{f\eta}\sum_{g}\omega^{*g}\exp(-i\bar{\bm{k}}\cdot C_{3z}^g\bm{\tau}_{\text{AB}}),
\end{align}
where $\lambda_{x0}^{f\eta}=-0.618+0.535i$ meV is in the same scale of $\lambda_{x0}^{\eta\eta}$. Near the mBZ center $\bar{\bm{k}}\approx\bar{\bm{0}}$, where $c$ orbitals enter the flat bands, the form factor almost vanishes (as well as the other terms with prefactors $\omega$, $\omega^*$).

\begin{table*}[!htb]
\caption{\label{tab:table1_epc_lattice}%
The coupling parameters in the lattice EPC Hamiltonian induced by moir\'{e} phonons, calculated using the numerical Wannier orbitals. Here only the leading 6 terms are given for each mode. All numbers are in the energy scale meV.}
\begin{ruledtabular}
\begin{tabular}{cccccc}
$\lambda_{x0}^{ff}=7.364$     & $\lambda_{x0}^{\tau\tau}=-2.684$     &   $\lambda_{x0}^{\eta\eta}=-0.370$      &   $\lambda_{x0}^{f\tau}=0.853$    & $\lambda_{x0}^{f\eta}=-0.618+0.535i$ & $\lambda_{x0}^{f\kappa}=0.617$\\
\hline
$\lambda_{0x}^{ff}=8.027$     &$\lambda_{0x,0}^{f\tau}=1.148$     &   $\lambda_{0x,1}^{f\tau}=0.923$      &   $\lambda_{0x,0}^{f\eta}=-0.598-0.579i$  & $\lambda_{0x,1}^{f\eta}=-0.235-0.287i$ & $\lambda_{0x}^{f\kappa}=0.731$\\
\hline
$\lambda_{xx}^{ff}=4.780$ & $\lambda_{xx}^{f\tau}=2.290$ & $\lambda_{xx}^{f\eta}=0.468-0.572i$ & $\lambda_{xx,0}^{f\kappa}=0.554$ & $\lambda_{xx,1}^{f\kappa}=-0.512$ & $\lambda_{xx}^{\tau\kappa}=-0.585$ \\ 
\hline
$\lambda_{zz}^{ff}=6.166$ & $\lambda_{zz}^{f\tau}=0.683$ &$\lambda_{zz}^{f\eta}=0.412+0.392i$ & $\lambda_{zz}^{f\kappa}=0.545$ & $\lambda_{zz}^{\tau\eta}=0.556+0.868i$ & $\lambda_{zz}^{\tau\kappa}=0.753$
\\
\end{tabular}
\end{ruledtabular}
\end{table*}

For the moir\'{e} $\Gamma$-phonons $E_2(x)$ ($\tau_0\sigma_x$) and $E_2(y)$ ($\tau_0\sigma_y$) modes, the lattice EPC model can be similarly given as
\begin{align}
\begin{split}
H_{\text{epc},\Gamma}^{E_2}=&\frac{1}{\sqrt{2}}\sum_{\bm{R}}\left(a_{0x,\bm{R}}+a_{0x,\bm{R}}^{\dagger}\right)h_{0x,\bm{R}}+\frac{1}{\sqrt{2}}\sum_{\bm{R}}\left(a_{0y,\bm{R}}+a_{0y,\bm{R}}^{\dagger}\right)h_{0y,\bm{R}}\\
=&\frac{1}{\sqrt{2}}\sum_{\bm{R}}\left(a_{0x,\bm{R}}+a^{\dagger}_{0x,\bm{R}}+ia_{0y,\bm{R}}+ia^{\dagger}_{0y,\bm{R}}\right)\left(h^{KK}_{0x,\bm{R}}+h^{K'K'}_{0x,\bm{R}}\right)+\text{H.c.},\label{eq:epc_model_8band_G}
\end{split}
\end{align}
where $h_{0x,\bm{R}}$ and $h_{0y,\bm{R}}$ have been decomposed as
\begin{align}
h_{0x,\bm{R}}=h^{KK}_{0x,\bm{R}}+h^{KK\dagger}_{0x,\bm{R}}+h^{K'K'}_{0x,\bm{R}}+h^{K'K'\dagger}_{0x,\bm{R}},\quad h_{0y,\bm{R}}=i\left(h^{KK}_{0x,\bm{R}}-h^{KK\dagger}_{0x,\bm{R}}+h^{K'K'}_{0x,\bm{R}}-h^{K'K'\dagger}_{0x,\bm{R}}\right).
\end{align}
The explicit expressions for $h^{KK}_{0x,\bm{R}}$ containing the leading terms are ($\bm{L}=\bm{L}_2^m-\bm{L}_1^m$, $g\in \{0,1,2\}$)
\begin{align} 
h^{KK}_{0x,\bm{R}}&=\lambda_{0x}^{ff}f_{K-,\bm{R}}^{\dagger}f_{K+,\bm{R}} \label{eq:epc_model_hG}\\
&+\lambda_{0x,0}^{f\tau} \left({\tau}_{K\bm{R}}^{\dagger}f_{K-,\bm{R}}+f^{\dagger}_{K+,\bm{R}}{\tau}_{K\bm{R}}\right)\nonumber\\
&+\lambda_{0x,1}^{f\tau} \sum_{g}\omega^{g}{\tau}^{\dagger}_{K,\bm{R}+C^g_{3z}\bm{L}}f_{K+,\bm{R}}+\lambda_{0x,1}^{f\tau} \sum_{g}\omega^{g}f^{\dagger}_{K-,\bm{R}}{\tau}_{K,\bm{R}-C^g_{3z}\bm{L}}\nonumber\\
&+\lambda_{0x,0}^{f\eta} \sum_{g}\omega^g\eta^{\dagger}_{K,\bm{R}+C_{3z}^g\bm{\tau}_{\text{AB}}}f_{K+,\bm{R}}+\lambda_{0x,0}^{f\eta}\sum_{g}\omega^g f_{K-,\bm{R}}^{\dagger} \eta_{K,\bm{R}+C_{3z}^g\bm{\tau}_{\text{BA}}}\nonumber\\
&-\lambda_{0x,0}^{f\eta*} \sum_{g} \omega^g\eta^{\dagger}_{K,\bm{R}+C_{3z}^g\bm{\tau}_{\text{BA}}}f_{K+,\bm{R}}-\lambda_{0x,0}^{f\eta*} \sum_{g}\omega^g f^{\dagger}_{K-,\bm{R}}\eta_{K,\bm{R}+C_{3z}^g\bm{\tau}_{\text{AB}}}\nonumber\\
&+\lambda_{0x,1}^{f\eta}\sum_{g}\eta_{K,\bm{R}+C_{3z}^g\bm{\tau}_{\text{AB}}}^{\dagger}f_{K-,\bm{R}}+\lambda_{0x,1}^{f\eta}\sum_{g} f^{\dagger}_{K+,\bm{R}}\eta_{K,\bm{R}+C_{3z}^g\bm{\tau}_{\text{BA}}}\nonumber\\
&-\lambda_{0x,1}^{f\eta*}\sum_g\tau^{\dagger}_{K,\bm{R}+C_{3z}^g\bm{\tau}_{BA}}f_{K-,\bm{R}}-\lambda_{0x,1}^{f\eta*}\sum_{g}f_{K+,\bm{R}}^{\dagger}\eta_{K,\bm{R}+C_{3z}^g\bm{\tau}_{\text{AB}}}\nonumber \\
&+\lambda_{0x}^{f\kappa}\sum_{g}\omega^g f^{\dagger}_{K-,\bm{R}}\kappa_{K,\bm{R}+C^g_{3z}\bm{\tau}_{\text{DW1}}}+\lambda_{0x}^{f\kappa}\sum_{g}\omega^g \kappa^{\dagger}_{K,\bm{R}-C^g_{3z}\bm{\tau}_{\text{DW1}}}f_{K+,\bm{R}}+...\nonumber 
\end{align}
$h^{K'K'}_{0x,\bm{R}}$ can be obtained by time reversal Eq. (\ref{eq:time_revsersal}): $h^{K'K'}_{0x,\bm{R}}=\mathcal{T}h^{KK\dagger}_{0x,\bm{R}}\mathcal{T}^{-1}$. Unlike moir\'{e} $K$-phonons, the above $\Gamma$-phonons will not induce onsite couplings among $c$ orbitals, and couplings between $f$ and $c$ orbtials are more important. The frozen-phonon band can be roughly obtained using the leading terms listed in Eq. (\ref{eq:epc_model_hG}), although more scattering channels with smaller coupling constants are needed in order to accurately repeat the bands.

We then directly list the EPC lattice Hamiltonian for the other modes. For the $E_1$ ($\tau_y\sigma_x,\tau_y\sigma_y$) and $E_2$ ($\tau_x\sigma_x,\tau_x\sigma_y$) modes of $K$-phonons, the model reads
\begin{align}
H_{\text{epc},K}^{E_1+E_2}=\frac{1}{\sqrt{2}}\sum_{\bm{R}}\sum_{i,j=x,y}\left(a_{ij,\bm{R}}+a_{ij,\bm{R}}^{\dagger}\right)h_{ij,\bm{R}},
\end{align}
where for convenience the electron parts can be decomposed as
\begin{align}
\begin{split}
&h_{xx,\bm{R}}=h_{xx,\bm{R}}^{K'K}+h^{KK'}_{xx,\bm{R}}+h_{xx,\bm{R}}^{K'K\dagger}+h_{xx,\bm{R}}^{KK'\dagger},\quad h_{xy,\bm{R}}=ih_{xx,\bm{R}}^{K'K}+ih^{KK'}_{xx,\bm{R}}-ih_{xx,\bm{R}}^{K'K\dagger}-ih_{xx,\bm{R}}^{KK'\dagger}, \\
&h_{yx,\bm{R}}=ih_{xx,\bm{R}}^{K'K}-ih^{KK'}_{xx,\bm{R}}-ih_{xx,\bm{R}}^{K'K\dagger}+ih_{xx,\bm{R}}^{KK'\dagger},\quad
h_{yy,\bm{R}}=-h_{xx,\bm{R}}^{K'K}+h^{KK'}_{xx,\bm{R}}-h_{xx,\bm{R}}^{K'K\dagger}+h_{xx,\bm{R}}^{KK'\dagger}.
\end{split}
\end{align}
The explicit leading terms of $h_{xx,\bm{R}}^{K'K}$ is
\begin{align} 
h^{K'K}_{xx,\bm{R}} &= \lambda_{xx}^{ff} f_{K'-,\bm{R}}^{\dagger} f_{K+,\bm{R}}\\
&+\lambda_{xx}^{f\tau}\left(\tau^{\dagger}_{K'\bm{R}}f_{K-,\bm{R}}+f^{\dagger}_{K'+,\bm{R}}\tau_{K\bm{R}}\right)\nonumber \\
&+\lambda_{xx}^{f\eta}\sum_{g} \left(f_{K'+,\bm{R}}^{\dagger}\eta_{K,\bm{R}+C_{3z}^g\bm{\tau}_{\text{AB}}}+\eta^{\dagger}_{K',\bm{R}+C_{3z}^g\bm{\tau}_{\text{AB}}}f_{K-,\bm{R}}\right) \nonumber\\
&- \lambda_{xx}^{f\eta*}\sum_g\left(f_{K'+,\bm{R}}^{\dagger}\eta_{K,\bm{R}+C_{3z}^g\bm{\tau}_{\text{BA}}}+\eta^{\dagger}_{K',\bm{R}+C_{3z}^g\bm{\tau}_{\text{BA}}}f_{K-,\bm{R}}\right)\nonumber\\
&+\lambda_{xx,0}^{f\kappa}\sum_g \left(f^{\dagger}_{K'+,\bm{R}}\kappa_{K,\bm{R}+C_{3z}^g\bm{\tau}_{\text{DW1}}}\kappa^{\dagger}_{K',\bm{R}+C_{3z}^g\bm{\tau}_{\text{DW1}}}f_{K-,\bm{R}}\right)\nonumber\\
&+\lambda_{xx,1}^{f\kappa}\sum_g \left(f^{\dagger}_{K'+,\bm{R}}\kappa_{K,\bm{R}-C_{3z}^g\bm{\tau}_{\text{DW1}}}+\kappa^{\dagger}_{K',\bm{R}-C_{3z}^g\bm{\tau}_{\text{DW1}}}f_{K-,\bm{R}}\right)\nonumber\\
&+\lambda_{xx}^{\tau\kappa}\sum_g \omega^{*g}\left( \kappa^{\dagger}_{K',\bm{R}+C_{3z}^g\bm{\tau}_{\text{DW1}}}\tau_{K\bm{R}}+\tau_{K'\bm{R}}^{\dagger}\kappa_{K,\bm{R}+C_{3z}^g\bm{\tau}_{\text{DW1}}} \right)+...\nonumber
\end{align}
while $h_{xx,\bm{R}}^{KK'}$ is obtained from $h_{xx,\bm{R}}^{K'K}$ using the $C_{2z}$ rotation: $h_{xx,\bm{R}}^{KK'}=C_{2z}h_{xx,-\bm{R}}^{K'K}C_{2z}^{-1}$. For the $B_2$ ($\tau_z\sigma_z$) mode of $\Gamma$-phonon, the model reads
\begin{align}
H_{\text{epc},\Gamma}^{B_2}=\frac{1}{\sqrt{2}}\sum_{\bm{R}}\left(a_{zz,\bm{R}}+a_{zz,\bm{R}}^{\dagger}\right)h_{zz,\bm{R}},\quad h_{zz,\bm{R}}=h_{zz,\bm{R}}^{KK}+\mathcal{T}h_{zz,\bm{R}}^{KK}\mathcal{T}^{-1},
\end{align}
where 
\begin{align} 
h^{KK}_{zz,\bm{R}}&=\frac{\lambda_{zz}^{ff}}{2}\left(f_{K+,\bm{R}}^{\dagger}f_{K+,\bm{R}}-f_{K-,\bm{R}}^{\dagger}f_{K-,\bm{R}} \right)\label{eq:epc_model_hG_B2}\\
&+\lambda_{zz}^{f\tau}\sum_{g}\left(\omega^{*g}\tau^{\dagger}_{K,\bm{R}-C_{3z}^g\bm{L}}f_{K+,\bm{R}}-\omega^g\tau^{\dagger}_{K,\bm{R}+C_{3z}^g\bm{L}}f_{K-,\bm{R}}\right)\nonumber\\
&+\lambda_{zz}^{f\eta} \sum_{g} \eta^{\dagger}_{K,\bm{R}+C_{3z}^g\bm{\tau}_{\text{AB}}}\left(\omega^g f_{K+,\bm{R}}+\omega^{*g} f_{K-,\bm{R}}\right)-\lambda_{zz}^{f\eta *} \sum_{g} \eta^{\dagger}_{K,\bm{R}+C_{3z}^g\bm{\tau}_{\text{BA}}}\left(\omega^g f_{K+,\bm{R}}+\omega^{*g} f_{K-,\bm{R}}\right) \nonumber\\
&+\lambda_{zz}^{f\kappa} \sum_{g}\left(\omega^{*g}\kappa^{\dagger}_{K,\bm{R}+C_{3z}^g\bm{\tau}_{\text{DW1}}}f_{K+,\bm{R}}-\omega^g \kappa^{\dagger}_{K,\bm{R}-C^g_{3z}\bm{\tau}_{\text{DW1}}}f_{K-,\bm{R}}\right)\nonumber\\
&+\lambda_{zz}^{\tau\eta}\sum_g \eta^{\dagger}_{K,\bm{R}+C_{3z}^g\bm{\tau}_{\text{AB}}}\tau_{K\bm{R}}-\lambda_{zz}^{\tau\eta*}\sum_g \eta^{\dagger}_{K,\bm{R}+C_{3z}^g\bm{\tau}_{\text{BA}}}\tau_{K\bm{R}}\nonumber\\
&+\lambda_{zz}^{\tau\kappa}\sum_g \kappa^{\dagger}_{K,\bm{R}+C_{3z}^g\bm{\tau}_{\text{DW1}}}\tau_{K\bm{R}}-\lambda_{zz}^{\tau\kappa}\sum_g \kappa^{\dagger}_{K,\bm{R}+C_{3z}^g\bm{\tau}_{\text{DW1}}}\tau_{K\bm{R}}+\text{H.c.}+...\nonumber 
\end{align}

Before ending, we comment on the main drawback of our theory. In such projection scheme, although the couplings between the phonons and $f$ orbitals are strictly considered, those with $c$ orbitals are inevitably underestimated due to the cancellation of incoherent phases. Only the symmetry of such scatterings are correctly reflected in our model. 

\subsection{Frozen-phonon scheme in lattice EPC model}
With the lattice EPC Hamiltonian Eqs. (\ref{eq:epc_model_8band_K}) and (\ref{eq:epc_model_8band_G}), the frozen-phonon calculation can be performed as well. Such calculation helps directly verify the validity of the effective lattice model. In this subsection we take moir\'{e} $A_1$ mode $a_{A_1}=a_{x0}$ as an example. First we define the following ``Bloch sum'' of moir\'{e} distortions and phonons,
\begin{align}
\bm{u}_{A_1\bar{\bm{q}}}(\bm{r})=\frac{1}{\sqrt{N_m}}\sum_{\bm{R}}e^{i\bar{\bm{k}}\cdot\bm{R}}\bm{u}(\bm{r}-\bm{R}),\quad a_{A_1\bar{\bm{q}}}=\frac{1}{\sqrt{N_m}}\sum_{\bm{R}}e^{-i\bar{\bm{q}}\cdot\bm{R}}a_{A_1\bm{R}}.
\end{align}
Then the EPC lattice Hamiltonian Eq. (\ref{eq:epc_model_8band_K}) can be written in $k$-space as (hide the electron orbital index)
\begin{align}
\begin{split}
H_{\text{epc}}^{A_1}=&\frac{1}{\sqrt{2}}\sum_{\bm{R}}(a_{A_1\bm{R}}+a_{A_1\bm{R}}^{\dagger})\sum_{\bm{L}\bm{T}}d^{\dagger}_{\bm{R}+\bm{L}}\Lambda_{\bm{L}\bm{T}}^{A_1}d_{\bm{R}+\bm{T}}\\
=&\frac{1}{\sqrt{2N_m}}\sum_{\bar{\bm{q}}\bar{\bm{k}}'\bar{\bm{k}}}\sum_{\bm{L}\bm{T}}\frac{1}{N_m}\sum_{\bm{R}}e^{i(\bar{\bm{q}}+\bar{\bm{k}}-\bar{\bm{k}}')\cdot\bm{R}-i\bar{\bm{k}}'\cdot\bm{L}+i\bar{\bm{k}}\cdot\bm{T}}(a_{A_1\bar{\bm{q}}}+a_{A_1,-\bar{\bm{q}}}^{\dagger})d^{\dagger}_{\bar{\bm{k}}'}\Lambda_{\bm{L}\bm{T}}^{A_1}d_{\bar{\bm{k}}}\\
=&\frac{1}{\sqrt{2N_m}}\sum_{\bar{\bm{k}}'\bar{\bm{k}}}\sum_{\bm{L}\bm{T}}e^{-i\bar{\bm{k}}'\cdot\bm{L}+i\bar{\bm{k}}\cdot\bm{T}}(a_{A_1,[\bar{\bm{k}}'-\bar{\bm{k}}]}+a_{A_1,[\bar{\bm{k}}-\bar{\bm{k}}']}^{\dagger})d^{\dagger}_{\bar{\bm{k}}'}\Lambda_{\bm{L}\bm{T}}^{A_1}d_{\bar{\bm{k}}},
\end{split}
\end{align}
where the symbol $[\bm{q}]$ represents the residue part of $\bm{q}$ in mBZ. The frozen-phonon field in the TB scheme corresponds to the periodic distortion $\bm{u}_{A_1\bar{\bm{0}}}(\bm{r})$. Therefore, in the above equation if we retain only the $\bar{\bm{q}}=0$ terms and recover the generalized coordinate $a_{A_1\bar{\bm{0}}}+a_{A_1\bar{\bm{0}}}^{\dagger}=\sqrt{2M_c\omega_{K}/\hbar}Q_{A_1}$, we get
\begin{align}
H^{A_1}_{\text{epc},\bar{\bm{q}}=\bar{\bm{0}}}=\sqrt{\frac{M_c\omega_K}{N_m\hbar}}Q_{A1}\sum_{\bar{\bm{k}}}\sum_{\bm{L}\bm{T}}e^{i\bar{\bm{k}}\cdot(\bm{T}-\bm{L})}d^{\dagger}_{\bm{k}}\Lambda_{\bm{L}\bm{T}}^{A_1}d_{\bar{\bm{k}}}. \label{eq:frozen_phonon_lattice}
\end{align}
From the discussion of the phonon normalization Eq. (\ref{eq:normalize_condition_u}), when $Q_{A_1}=\sqrt{N_m\hbar/(M_c\omega_K)}$, the resulting Hamiltonian $h_{A_1\bar{\bm{0}}}=\sum_{\bar{\bm{k}}}\sum_{\bm{L}\bm{T}}e^{i\bar{\bm{k}}\cdot(\bm{T}-\bm{L})}d^{\dagger}_{\bm{k}}\Lambda_{\bm{L}\bm{T}}^{A_1}d_{\bar{\bm{k}}}$ is just the electron Hamiltonian under the periodic $A_1$ distortion with the strength frozen at the phonon length $l_b$ [Eq. (\ref{eq:phonon_mean_dist})]. By tuning $Q_{A_1}$ and diagonalizing the corresponding Hamiltonian $H_0+H^{A_1}_{\text{EPC},\bar{\bm{q}}=\bar{\bm{0}}}$, we could obtain the frozen-phonon bands from the lattice model with various distortion strength.

\section{Minimal continuum model from symmetry analysis \label{Appendix_leading_order}}
The symmetry constraints can partially determine the EPC model used in Refs. \cite{20epjp_JT_BM_Fabrizio,22prb_Kekule_Fabrizio}, which contains only three nearest scattering process (in $k$-space) and is qualitatively the same as the leading modes (\ref{eq:leading_order_a}) and (\ref{eq:leading_order_b}).

The $K$-phonon distortions provide a large-momentum intervalley transfer quantified by
\begin{align}
\Delta{\bm{Q}}=\bm{K}_1-\bm{K}'_2=\bm{K}_2-\bm{K}'_1=(2n_m+1)(\bm{G}_1^m+\bm{G}_2^m),\label{eq:Q+-}
\end{align}
while leaving the intravalley part intact. The general influence of them can be described by an intervalley moir\'{e} potential $V(\bm{r})$, so that the full Hamiltonian
\begin{align}
  H = \left(\begin{array}{cc}
    H^{\text{BM}}_K & V\\
    V^{\dag} & H^{\text{BM}}_{K'}
  \end{array}\right), \quad V = \left(\begin{array}{cc}
    v_{11} (\bm{r}) & v_{12} (\bm{r})\\
    v_{21} (\bm{r}) & v_{22} (\bm{r})
  \end{array}\right), \label{eq:generalized_continuum}
\end{align}
where $v_{ll'}(\bm{r})$ are $2\times2$ matrices describing the intralayer ($v_{11}$ and $v_{22}$) and interlayer ($v_{12}$ and $v_{21}$) scatterings. For simplicity we derive the translation-invariant $A_1$ distortions, using the convention (\ref{eq:Hamk_BM_full}) in which the translation is trivial: $V(\bm{r})=V(\bm{r}+\bm{L}_I)$. We could expand $V(\bm{r})$ into Fourier series around $\Delta{\bm{Q}}$ (\ref{eq:Q+-}). For example, the intralayer term can be written as [$V(\bm{r})=V^T(\bm{r})$ is required by $\mathcal{T}$]
\begin{align}
  v_{11} (\bm{r}) = \sum_n e^{i(\bm{G}_n+\Delta{\bm{Q}}) \cdot \bm{r}}
  \left(\begin{array}{cc}
    e_n & f_n\\
    f_n & h_n
  \end{array}\right). \label{eq:v11}
\end{align}
Since all rotation and $\mathcal{T}$ symmetries are preserved for $A_1$ distortions, the conditions $U(g)H(g^{-1}\bm{p},g^{-1}\bm{r})U^{-1}(g)=H(\bm{p},\bm{r})$ are still satisfied. Specifically, $C_{3z}$, $C_{2x}$ and $C_{2z}$ respectively requires
\begin{subequations}
\begin{align}
  v_{11} (\bm{r}) & = \sum_n e^{i (C_{3 z} \bm{G}_n-\bm{G}_2^m
  +\Delta{\bm{Q}}) \cdot
  \bm{r}} \left(\begin{array}{cc}
    \omega^{- 1} e_n & f_n\\
    f_n & \omega h_n
  \end{array}\right),\label{eq:symva}\\
  v_{22} (\bm{r}) & = \sum_n e^{i (C_{2 x} \bm{G}_n+\Delta{\bm{Q}}) \cdot
  \bm{r}} \left(\begin{array}{cc}
    h_n & f_n\\
    f_n & e_n
  \end{array}\right), \label{eq:symvb}\\
  v_{11} (\bm{r}) & = \sum_n e^{i(\bm{G}_n+\Delta{\bm{Q}})\bm{r}}
  \left(\begin{array}{cc}
    h_n^{\ast} & f^{\ast}_n\\
    f^{\ast}_n & e_n^{\ast}
  \end{array}\right) .\label{eq:symvc}
\end{align}
\end{subequations}
From the constraint (\ref{eq:symva}) we have $\{\bm{G}_n \} = \{ C_{3 z} \bm{G}_n -\bm{G}_2^m \}$. The simplest choice with smallest $|\bm{G}_n|$ specifies $\{ \bm{G}_n \} = \{ \bm{0}, \bm{G}_1^m, -\bm{G}_2^m \}$ [i.e., the nearest shell of $\bm{Q}_l^{\eta}$ vectors around $\bm{K}_l^{\eta}$, see Fig. \ref{fig:Q_G}(a) with label (a)], which, together with Eq. (\ref{eq:symvb}), gives (the phase of $g_1$ cannot be determined yet, but we numerically found $g_1\approx0$)
\begin{subequations}
\begin{align}
  v_{11} (\bm{r}) & = e^{i\Delta{\bm{Q}} \cdot \bm{r}}
  \left[ \left(\begin{array}{cc}
    g_1 & g_0\\
    g_0 & g_1^{*}
  \end{array}\right) + e^{i\bm{G}_1^m \cdot \bm{r}}
  \left(\begin{array}{cc}
    \omega g_1 & g_0\\
    g_0 & \omega^{- 1} g_1^{*}
  \end{array}\right) + e^{-i\bm{G}_2^m \cdot \bm{r}}
  \left(\begin{array}{cc}
    \omega^{- 1} g_1 & g_0\\
    g_0 & \omega g_1^{*}
  \end{array}\right) \right], \label{eq:v11_final}\\
  v_{22} (\bm{r}) & = e^{i\Delta{\bm{Q}} \cdot \bm{r}}
  \left[ \left(\begin{array}{cc}
    g_1^{*} & g_0\\
    g_0 & g_1
  \end{array}\right) + e^{-i\bm{G}_1^m \cdot \bm{r}}
  \left(\begin{array}{cc}
    \omega g_1^{*} & g_0\\
    g_0 & \omega^{- 1} g_1
  \end{array}\right) + e^{i\bm{G}_2^m \cdot \bm{r}}
  \left(\begin{array}{cc}
    \omega^{- 1} g_1^{*} & g_0\\
    g_0 & \omega g_1
  \end{array}\right) \right] .\label{eq:v22_final} 
\end{align}
\end{subequations}
A parallel but simpler analysis can be done for the interlayer terms $v_{12}$ and $v_{21}$. Here we directly give the results
\begin{align}
  v_{12} (\bm{r}) = v_{21} (\bm{r}) = e^{i\Delta{\bm{Q}}
  \cdot \bm{r}} \left(\begin{array}{cc}
    & \gamma\\
    \gamma & 
  \end{array}\right). \label{eq:v12_final}
\end{align}
For $A_1$ mode the parameters $g_0$ and $\gamma$ are real, as required by $C_{2z}$ (\ref{eq:symvc}). The solution for $B_1$ distortion is easily obtained by replacing $V$ ($V^{\dag}$) of $A_1$ mode by $-iV$ ($iV^{\dag}$) because it is odd under $C_2z$ rotation. The $B_1$ and $A_1$ potentials are related by a valley SU(2) rotation: $V_{B_1}=\exp(-i\tau_z\pi/4)V_{A_1}\exp(i\tau_z\pi/4)$. The Hamiltonian with such lattice distortions can be written as
\begin{align}
  H &= \left(\begin{array}{cc}
    H^{\text{BM}}_K & (\mathcal{Q}_1-i\mathcal{Q}_2)V\\
    (\mathcal{Q}_1+i\mathcal{Q}_2)V^{\dag} & H^{\text{BM}}_{K'}
  \end{array}\right) 
  =\mathcal{H}_0 +\mathcal{Q}_1 V_{A_1}+\mathcal{Q}_2 V_{B_1},   \label{eq:general_distortion}
\end{align}
where $\mathcal{Q}_1,\mathcal{Q}_2$ are normal coordinates Eq. (\ref{eq:boson_operator_definition}). The above leading order distorted potential are formally obtained purely from the symmetry analysis, rather than from an ``ab-initial'' derivation as done in Ref. \cite{20epjp_JT_BM_Fabrizio}. Once transformed to the gauge Eq. (\ref{eq:Ham_BM_symmetric}), the formula will exactly coincide with Ref. \cite{20epjp_JT_BM_Fabrizio} (except the undetermined phase of $g_1$). Notably, the derivation here can be equally used for higher order distortions (as well as other symmetric branches) by extending the set $\{\bm{G}_n \} = \{ C_{3 z} \bm{G}_n -\bm{G}_2^m \}$ to the outer range.

Not surprisingly, the above model corresponds to the distortion induced by the phonons with $|\bm{Q}_l^{\eta}|=k_{\theta}$. The following artificial field is used in Ref. \cite{20epjp_JT_BM_Fabrizio} to simulate the localized $A_1/B_1$ distortions in the $l$-th layer,
\begin{align}
\bm{u}(\bm{r},l)\sim\sum_{i,j=0,1,2}\left[ \bm{u}^l(\bm{Q}_{ij})e^{i\bm{Q}_{ij}\cdot\bm{r}}\pm c.c.  \right],
\end{align}
where $\bm{Q}_{ij} = C_{3z}^i\bm{K}_{1}-C_{3z}^{j}\bm{K}'_2$ is the momentum transfer between the two valleys. Each $\bm{Q}_{ij}$ represents one specific intervalley scattering which just corresponds to $\{\bm{Q}_l^{\eta}\}$ vectors used in Eqs. (\ref{eq:leading_order_a}), (\ref{eq:leading_order_b}). In our method $g_0$, $g_1$, and $\gamma$ are read from the projected TB Hamiltonian while Refs. \cite{20epjp_JT_BM_Fabrizio,22prb_Kekule_Fabrizio} obtained them by fitting the bands of Eq. (\ref{eq:general_distortion}) to that of the TB model \cite{19prx_JT_TB_Fabrizio}. The latter strategy unnecessarily overestimates the leading term with $|\bm{Q}_l^{\eta}|=k_{\theta}$ and ignores other terms. Our calculation indicates that higher order modes have non-negligible strength and all of them should be honestly reserved.

\section{Phonon-mediated interaction \label{appendix_phonon_interaction}}

\subsection{Numerical full-band mean-field formulation}
The phonon-mediated electron interaction reads ($b$ is the branch index of moir\'{e} phonons)
\begin{align}
H_{\text{P}} = -\frac{1}{2N_m}\sum_{b\bar{\bm{q}}}\frac{1}{\hbar \omega_b}h^{\dagger}_b(\bar{\bm{q}})h_b(\bar{\bm{q}}),\label{eq:epc_int_full}
\end{align}
where the scattering Hamiltonian mediated by moir\'{e} Bloch phonon $a_{\bar{\bm{q}},b}$ Eq. (\ref{eq:moire_phonon_q_space}) is
\begin{align}
h_b(\bar{\bm{q}})=\sum_{s\bar{\bm{k}}}\sum_{\eta\alpha\bm{G}}\sum_{\eta'\alpha'\bm{G}'}\mathcal{H}^{\eta \alpha,\bar{\bm{k}}-\bar{\bm{q}}+\bm{G}}_{\eta'\alpha',\bar{\bm{k}}+\bm{G}'}(b)C^{\dagger}_{s\eta\alpha,\bar{\bm{k}}-\bar{\bm{q}}+\bm{G}}C_{s\eta'\alpha',\bar{\bm{k}}+\bm{G}'}. \label{eq:form_factor}
\end{align}
All complicated umklapp processes are contained in $\mathcal{H}^{\eta \alpha,\bar{\bm{k}}-\bar{\bm{q}}+\bm{G}}_{\eta'\alpha',\bar{\bm{k}}+\bm{G}'}(b)$, including the scatterings among $c$ orbitals. The EPC-induced Hartree and Fock energies are
\begin{align}
E_{\text{P}}^{\text{H}}[\rho] &= -\frac{1}{2N_m}\sum_b \frac{1}{\hbar \omega_b}\left[\sum_{s\bar{\bm{k}}}\sum_{\eta\alpha\bm{G}}\sum_{\eta'\alpha'\bm{G}'}\mathcal{H}^{\eta\alpha,\bar{\bm{k}}+\bm{G}}_{\eta'\alpha',\bar{\bm{k}}+\bm{G}'}(b)\rho^{s'\eta'\alpha'\bm{G}'}_{s\eta\alpha\bm{G}}(\bar{\bm{k}})\right]^2=-\frac{1}{2N_m}\sum_b \frac{1}{\hbar \omega_b}\left[\Tr(\mathcal{H}(b)\rho)\right]^2,\\
E_{\text{P}}^{\text{F}}[\rho] &= \frac{1}{2N_m}\sum_{bs} \frac{1}{\hbar \omega_b}\sum_{\bar{\bm{k}}}\sum_{\eta\alpha\bm{G}}\sum_{\eta'\alpha'\bm{G}'}\sum_{\bar{\bm{p}}}\sum_{\xi\beta\bm{P}}\sum_{\xi'\beta'\bm{P}'}\mathcal{H}^{\eta\alpha,\bar{\bm{p}}+\bm{G}}_{\eta'\alpha',\bar{\bm{k}}+\bm{G}'}(b)\rho_{s\xi\beta\bm{P}}^{s\eta'\alpha'\bm{G}'}(\bar{\bm{k}})\mathcal{H}^{\xi\beta,\bar{\bm{k}}+\bm{P}}_{\xi'\beta',\bar{\bm{p}}+\bm{P}'}(b)\rho_{s\eta\alpha\bm{G}}^{s\xi'\beta'\bm{P}'}(\bar{\bm{p}}) \nonumber\\
& = \frac{1}{2N_m}\sum_{b} \frac{1}{\hbar \omega_b}\Tr(\mathcal{H}(b)\rho\mathcal{H}(b)\rho), \label{eq:Fock_EPC_energy}
\end{align}
where the density matrix $\rho^{s'\eta'\alpha'\bm{G}'}_{s\eta\alpha\bm{G}}(\bar{\bm{k}})=\langle C^{\dagger}_{s\eta\alpha,\bar{\bm{k}}+\bm{G}}C_{s'\eta'\alpha',\bar{\bm{k}}+\bm{G}'}\rangle$. The corresponding Hartree and Fock potentials are
\begin{align}
H_{\text{P}}^{\text{H}}[\rho]=&-\frac{1}{N_m}\sum_{s\bar{\bm{k}}}\sum_{\eta\alpha\bm{G}}\sum_{\eta'\alpha'\bm{G}'}\left[\sum_{b}\frac{1}{\hbar\omega_b}\Tr(\mathcal{H}(b)\rho)\mathcal{H}^{\eta\alpha,\bar{\bm{k}}+\bm{G}}_{\eta'\alpha',\bar{\bm{k}}+\bm{G}'}(b)\right]C^{\dagger}_{s\eta\alpha,\bar{\bm{k}}+\bm{G}}C_{s\eta'\alpha',\bar{\bm{k}}+\bm{G}'}-E_{\text{P}}^{\text{H}}[\rho], \label{eq:Hartree_EPC_potential}\\
H_{\text{P}}^{\text{F}}[\rho]=&\frac{1}{N_m}\sum_{s\bar{\bm{k}}}\sum_{\xi\beta\bm{P}}\sum_{\eta'\alpha'\bm{G}'}\left[\sum_{b}\frac{1}{\hbar\omega_b}\sum_{\bar{\bm{p}}}\sum_{\xi'\beta'\bm{P}'}\sum_{\eta\alpha\bm{G}}\mathcal{H}^{\xi\beta,\bar{\bm{k}}+\bm{P}}_{\xi'\beta',\bar{\bm{p}}+\bm{P}'}(b)\rho_{s\eta\alpha\bm{G}}^{s\xi'\beta'\bm{P}'}(\bar{\bm{p}})\mathcal{H}^{\eta\alpha,\bar{\bm{p}}+\bm{G}}_{\eta'\alpha',\bar{\bm{k}}+\bm{G}'}(b)\right]\nonumber\\&\times C^{\dagger}_{s\xi\beta,\bar{\bm{k}}+\bm{P}}C_{s\eta'\alpha',\bar{\bm{k}}+\bm{G}'}-E_{\text{P}}^{\text{F}}[\rho]. \label{eq:Fock_EPC_potential}
\end{align}
In the frozen-phonon method, only the EPC Hamiltonian Eq. (\ref{eq:form_factor}) at $\bar{\bm{q}}=\bar{\bm{0}}$ are accessible, i.e., only matrices $\mathcal{H}^{\eta\alpha\bar{\bm{k}}+\bm{G}}_{\eta'\alpha'\bar{\bm{k}}+\bm{G}'}(b)$ can be determined. This is sufficient to calculate the Hartree terms, as they only involve scatterings with moir\'{e} momentum conserved. The Fock terms cannot be accurately treated in such scheme and they will not dominate the energy variation. For simplicity the approximation $\mathcal{H}^{\eta\alpha,\bar{\bm{k}}+\bm{G}}_{\xi\beta,\bar{\bm{p}}+\bm{P}}(b)\approx\mathcal{H}^{\eta\alpha\bm{G}}_{\xi\beta\bm{P}}(b)$ is taken since $\mathcal{H}(b)$ is almost independent on $\bar{\bm{k}}$. In practical calculations the density matrix appearing above is replaced by $\rho-\rho_0$, where $\rho_0$ is the non-interacting density matrix (ground state of $H_{0}$ at $\nu=0$). In such double counting scheme the non-symmetry-breaking state $\rho_0$ is always a self-consistent solution. The Hartree-Fock potential Eqs. (\ref{eq:Hartree_EPC_potential}) and (\ref{eq:Fock_EPC_potential}), as well as their Coulomb partners, are iteratively calculated using a $9\times 9$ $\bar{\bm{k}}$-mesh to update the density matrix until the total energy Eq. (\ref{eq:total_energy}) is converged (the convergence criterion $10^{-4}$ meV). 

\subsection{Ordered states}
The calculation is easily converged using the initial states whose symmetries are broken by splitting $f$ orbitals only \cite{22prb_ZLL_OPW_representation}. Specifically, the initial state is chosen as the Slater determinant of the following trial Hamiltonian
\begin{align}
H_{\text{trial}} = H_0+\sum_{ijm\bar{\bm{k}}}U_{ijm}f^{\dagger}_{\bar{\bm{k}}}s_{i}\tau_{j}\sigma_m f_{\bar{\bm{k}}}=H_0+\sum_{ijm\bm{R}}U_{ijm}f^{\dagger}_{\bm{R}}s_{i}\tau_{j}\sigma_m f_{\bm{R}},
\end{align}
where $U_{ijm}s_i\tau_j\sigma_m$ are trial order parameters governing the splitting of $f$ orbitals, whose eigenvalues are in the same order of the Hubbard repulsion among $f$ orbitals $U_0\sim 50$ meV \cite{22prl_heavy_fermion_Zhida,22prb_ZLL_OPW_representation,23nc_cascades_dmft}. Here we list the order parameters with initial strengths, and the corresponding unoccupied $f$ orbitals at fillings $\nu=0,2$. 

At $\nu=0$, the 8 orbitals should be separated into ``$4+4$'' pairs (4 orbitals empty, 4 orbitals occupied). The different orders and the corresponding non-occupied $f$ orbitals can take ($O$ is in the unit of $U_0/2$, and we use $|s\eta t\rangle$ to indicate the $f$ orbitals with spin $s$, valley $\eta$, and angular momentum $t$)
\begin{subequations}
\begin{alignat}{2}
&O^{\text{KIVC}}_{0}=\cos\phi\tau_x\sigma_z+\sin\phi\tau_y\sigma_z: \quad &&\frac{1}{\sqrt{2}}\left(|sK+\rangle+e^{i\phi}|sK'+\rangle\right),\frac{1}{\sqrt{2}}\left(|sK-\rangle-e^{i\phi}|sK'-\rangle\right), \ s=\uparrow,\downarrow,\\
&O^{\text{TIVC}}_{0}=\cos\phi\tau_x+\sin\phi\tau_y: \quad &&\frac{1}{\sqrt{2}}\left(|sKt\rangle+e^{i\phi}|sK't\rangle\right), \ s=\uparrow,\downarrow,\ t=\pm,\label{eq:v0_tivc}\\
&O^{\text{TSM}}_0=\cos\phi\sigma_x+\sin\phi\sigma_y: \quad && \frac{1}{\sqrt{2}}\left(|s\eta+\rangle+e^{i\phi}|s\eta-\rangle\right),\ s=\uparrow,\downarrow,\ \eta=K,K',\\
&O^{\text{VP}}_0=\tau_z: \quad &&|sKt\rangle,\ s=\uparrow,\downarrow,\ t=\pm,\label{eq:v0_vp}\\
&O^{\text{SP}}_0=s_z: \quad &&|\uparrow \eta t\rangle,\ \eta=K,K',\ t=\pm,\\
&O^{\text{SVL}}_0=s_z\tau_z: \quad &&|\uparrow K t\rangle,|\downarrow K' t\rangle, \ t=\pm,\\
&O^{\text{QAH}}_0=\sigma_z: \quad && |s\eta+\rangle,\ s=\uparrow,\downarrow,\ \eta=K,K',\\
&O^{\text{VH}}_0=\tau_z\sigma_z: \quad && |sK+\rangle,|sK'-\rangle,\ s=\uparrow,\downarrow,\\
&O^{\text{SH}}_0=s_z\sigma_z: \quad && |\uparrow \eta+\rangle, |\downarrow\eta-\rangle, \ \eta=K,K'.
\end{alignat}
\end{subequations}
At $\nu=2$, the $f$ orbitals are splitted into ``6+2'' pairs. The orders and unoccupied states can take ($I_0=s_0\tau_0\sigma_0$)
\begin{subequations}
\begin{alignat}{3}
&O^{\text{SP-KIVC}}_2=\left(-I_0+s_z+\tau_x\sigma_z+s_z\tau_x\sigma_z\right)/2: \quad && \frac{1}{\sqrt{2}}\left(|\uparrow K+\rangle+|\uparrow K'+\rangle \right),\frac{1}{\sqrt{2}}\left(|\uparrow K-\rangle-|\uparrow K'-\rangle\right),\\
&O^{\text{SP-TIVC}}_2=\left(-I_0+s_z+\tau_x+s_z\tau_x\right)/2: && \frac{1}{\sqrt{2}}\left(|\uparrow K+\rangle+|\uparrow K'+\rangle \right),\frac{1}{\sqrt{2}}\left(|\uparrow K-\rangle+|\uparrow K'-\rangle\right),\\
&O^{\text{QAH-IVC}}_2=\left(-I_0+\tau_x+\sigma_z+\tau_x\sigma_z\right)/2: \quad && \frac{1}{\sqrt{2}}\left(|\uparrow K+\rangle+|\uparrow K'+\rangle \right),\frac{1}{\sqrt{2}}\left(|\downarrow K+\rangle+|\downarrow K'+\rangle\right),\\
&O^{\text{SH-IVC}}_2=\left(-I_0+\tau_x+s_z\sigma_z+s_z\tau_x\sigma_z\right)/2: \quad && \frac{1}{\sqrt{2}}\left(|\uparrow K+\rangle+|\uparrow K'+\rangle \right),\frac{1}{\sqrt{2}}\left(|\downarrow K-\rangle+|\downarrow K'-\rangle\right),\\
&O^{\text{SVP}}_2=\left(-I_0+s_z+\tau_z+s_z\tau_z\right)/2: \quad && |\uparrow K+\rangle,|\uparrow K-\rangle,\\
&O^{\text{SP-QAH}}_2=\left(-I_0+s_z+\sigma_z+s_z\sigma_z\right)/2: \quad && |\uparrow K+\rangle,|\uparrow K'+\rangle.
\end{alignat}
\end{subequations}

The $f$ orbital with $t=\pm$ will carry Chern number $C=\pm 1$ once hybridized with $c$ orbitals \cite{22prl_heavy_fermion_Zhida}. Using such simple criteria, one can easily identify the topology of various ordered states. The QAH state at $\nu=0$ has $C=\pm4$, the SP-QAH and QAH-IVC states at $\nu=2$ have $C=\pm2$, while all other states are topologically trivial.

\subsection{Effective onsite attraction}
The phonon-mediated electron interaction Eq. (\ref{eq:epc_int_full}) is quite simple if we project onto the $f$ orbitals,
\begin{align}
H_{\text{P}}^f=-\frac{1}{2}\sum_{ij\bm{R}}g_{ij}f_{\bm{R}}^{\dagger}s_0\tau_i\sigma_j f_{\bm{R}}f^{\dagger}_{\bm{R}}s_0\tau_i\sigma_j f_{\bm{R}},\label{eq:epc_int_f}
\end{align}
where $g_{ij}=\lambda_{ij}^2/(\hbar\omega_b)$ quantifies the effective attraction. Using the parameters in Table \ref{tab:epc_const_lam}, we find
\begin{align}
\begin{split}
&K:\quad g_{x0}=g_{y0}\approx 0.363\ \text{meV},\quad g_{xx}=g_{xy}=g_{yx}=g_{yy}\approx 0.152\ \text{meV},\\
&\Gamma:\quad g_{0x}=g_{0y}\approx 0.358\ \text{meV}, \quad g_{zz}\approx 0.211\ \text{meV},
\end{split}
\end{align}
and $g_{ij}=0$ for the other channels. Note that these parameters are strongly dependent on the TB parameters and the moir\'{e} potential, see the discussion in Appendix \ref{appendix_E_sensibility}. Although incomplete, the local attraction Eq. (\ref{eq:epc_int_f}) helps qualitatively understand the numerical results in the main text (and guess possible pairing symmetry \cite{24a_nodal_sc_onsite}). If we only incorporate the $A_1$, $B_1$ modes of $K$-phonons and $E_2$ mode of $\Gamma$-phonons, and take $g_{x0}=g_{y0}=g_{0x}=g_{0y}=g\approx 0.36$ meV, the mean-field energy per supercell contributed by them reads
\begin{align}
\begin{split}
&e_{\text{P}}=e_{\text{P}}^{\text{H}}+e_{\text{P}}^{\text{F}},\\
&e_{\text{P}}^{\text{H}}=-\frac{g}{2}\left[ \Tr^2(\tau_x\tilde{\rho}^f)+\Tr^2(\tau_y\tilde{\rho}^f)+\Tr^2(\sigma_x\tilde{\rho}^f) +\Tr^2(\sigma_y\tilde{\rho}^f)\right],\\
&e_{\text{P}}^{\text{F}}=\frac{g}{2}\left[\Tr(\tau_x\tilde{\rho}^f\tau_x\tilde{\rho}^f)+\Tr(\tau_y\tilde{\rho}^f\tau_y\tilde{\rho}^f)+\Tr(\sigma_x\tilde{\rho}^f\sigma_x\tilde{\rho}^f)+\Tr(\sigma_y\tilde{\rho}^f\sigma_y\tilde{\rho}^f)\right],
\end{split}
\end{align}
where $\tilde{\rho}^f_{s\eta t,s'\eta't'}=\rho^f-\rho^f_0$ is the reduced density matrix for $f$ orbitals, with respect to some reference density $\rho^f_0$. The bare density matrix $\rho^f_{s\eta t,s'\eta't'}=\langle f^{\dagger}_{s'\eta't'\bm{R}}f_{s\eta t\bm{R}}\rangle$ is related to its order matrix $O$ defined in the last subsection by
\begin{align}
\rho^{f}=\frac{1}{2}\left(I_0-O\right),\quad I_0=s_0\tau_0\sigma_0.
\end{align}
It is then straightforward to verify the influence of phonons on various symmetry-breaking orders through simple algebra. We adopt the subtraction scheme $\rho_0^f=I_0/2$ (the energy depends on the subtraction scheme \cite{23a_EPC_vs_IKS_kwan}). At $\nu=0$, the energy change is determined to be
\begin{align}
e^{\text{KIVC}}_{\text{P},0}=-2g,\quad e^{\text{TIVC}}_{\text{P},0}=e^{\text{TSM}}_{\text{P},0}=-6g,\quad e^{\text{SP}}_{\text{P},0}=-e^{\text{VH}}_{\text{P},0}=4g,\quad e^{\text{VP}}_{\text{P},0}=e^{\text{SVL}}_{\text{P},0}=e^{\text{QAH}}_{\text{P},0}=e^{\text{SH}}_{\text{P},0}=0,
\end{align}
and at $\nu=2$, they are found to be
\begin{align}
e^{\text{SP-KIVC}}_{\text{P},2}=e^{\text{SP-TIVC}}_{\text{P},2}=g,\quad e^{\text{SP-QAH}}_{\text{P},2}=e^{\text{SVP}}_{\text{P},2}=2g,\quad e^{\text{QAH-IVC}}_{\text{P},2}=e^{\text{SH-IVC}}_{\text{P},2}=-g.
\end{align}
Note that the above results qualitatively capture the energy trends depicted in Fig. \ref{fig:Evsa}   (except the SP-TIVC state). The quantitative difference is primarily caused by the interaction involving $c$ orbitals, which undeniably influences the condensation energy in the numerical calculations involving all bands.

Compared with the onsite Hubbard repulsion ($U_0>40$ meV), the attraction Eq. (\ref{eq:epc_int_f}) is two orders of magnitude smaller. Fortunately, such attraction just operates with the energy scale of the difference between various ordered states. This enables the TIVC states to prevail over the other states. However, the tiny energy scale also indicates that the Coulomb interaction in the system must be effectively screened for the phonon-mediated channels to play a role in the superconducting physics \cite{23prb_preformed_pair,23a_kondo_song,23_liu_ek_phonon,24a_nodal_sc_onsite}.

\end{widetext}

\bibliography{apssamp}% Produces the bibliography via BibTeX.

\end{document}